\documentclass[prl,twocolumn,preprintnumbers,superscriptaddress,amsmath,amssymb]{revtex4-1}
\usepackage{graphicx}
\usepackage{subfigure}
\usepackage{mathrsfs}
\usepackage{amsfonts}
\usepackage{times}
\usepackage{amsmath}
\usepackage{leftidx}
\usepackage{tikz}
\usepackage{tikz-network}
\usepackage{color}
\usepackage[colorlinks,linkcolor=blue,citecolor=blue]{hyperref}

\newcommand{\Tr}{\operatorname{Tr}}

\usepackage{bbold}
\usepackage{braket}
\usepackage{mathtools}
\usepackage{colortbl}
\usepackage{multirow}

\begin{document}
\title{Topology of Quantum Gaussian States and Operations}
\author{Zongping Gong}
\affiliation{Max-Planck-Institut f\"ur Quantenoptik, Hans-Kopfermann-Stra{\ss}e 1, 85748 Garching, Germany}
\affiliation{Munich Center for Quantum Science and Technology, Schellingstra{\ss}e 4, 80799 M\"unchen, Germany}
\date{\today}
\author{Tommaso Guaita}
\affiliation{Max-Planck-Institut f\"ur Quantenoptik, Hans-Kopfermann-Stra{\ss}e 1, 85748 Garching, Germany}
\affiliation{Munich Center for Quantum Science and Technology, Schellingstra{\ss}e 4, 80799 M\"unchen, Germany}
\date{\today}

\begin{abstract}
As is well-known in the context of topological insulators and superconductors, short-range-correlated fermionic pure Gaussian states with fundamental symmetries are systematically classified by the periodic table. We revisit this topic from a quantum-information-inspired operational perspective without referring to any Hamiltonians, and apply the formalism to bosonic Gaussian states as well as (both fermionic and bosonic) locality-preserving unitary Gaussian operations. We find that while bosonic Gaussian states are all trivial, there exist nontrivial bosonic Gaussian operations that cannot be continuously deformed into the identity under the locality and symmetry constraint. Moreover, we unveil unexpectedly complicated relations between fermionic Gaussian states and operations, pointing especially out that some of the former can be disentangled by the latter under the same symmetry constraint, while some cannot. In turn, we find that some topological operations are genuinely dynamical, in the sense that they cannot create any topological states from a trivial one, yet they are not connected to the identity. The notions of disentanglability and genuinely dynamical topology can be unified in a general picture of unitary-to-state homomorphism, and apply equally to interacting topological phases and quantum cellular automata.
\end{abstract}
\maketitle

\emph{Introduction.---} Classifying topological phases of quantum matter is a central topic in modern condensed matter physics \cite{Ryu2016}. The arguably most well-established paradigm is the classification of free fermions described by band theory, which are expected to be relevant to the majority of natural materials with weak electron interactions \cite{Kane2010,Qi2011}. In the presence of fundamental two-fold Altland-Zirnbauer (AZ) \cite{Altland1997} symmetries, the result is well-known as the periodic table \cite{Ryu2008,Ryu2010,Kitaev2009,Teo2010}.

In contrast, one popular way of classifying strongly interacting topological phases follows a quantum-information-inspired viewpoint \cite{Chen2010,Chen2011,Chen2013,Pollmann2010,Pollmann2011,Schuch2010,Schuch2011}. That is, instead of Hamiltonians, one focuses directly on short-range correlated quantum many-body states, typically in the tensor-network representations \cite{Verstraete2008,Orus2014,Cirac2020}, to see whether one can be transformed into another by a finite-depth quantum circuit of local unitaries with symmetries (if any). For example, topologically ordered states cannot be disentangled by local quantum circuits and thus exhibit long-range entanglement \cite{Chen2010}. Such an operational perspective has recently been used to study the topology of locality-preserving unitaries themselves \cite{Gross2012,Cirac2017,Chen2018,Gong2020,Piroli2021,Ranard2020,Bols2021}, which naturally generalize quantum circuits and are usually called quantum cellular automata (QCA) \cite{Schumacher2004,Arrighi2019,Farrelly2020}. This topic is closely related to Floquet topological phases \cite{Kitagawa2010,Jiang2011,Lindner2013,Else2016,Po2016,Potter2017,Harper2017b,Roy2017,Fidkowski2019,Zhang2021}. 

The operational formalism is certainly applicable also to noninteracting systems, whose ground states are Gaussian. While this point has been tacitly mentioned in the literature \cite{Fidkowski2010,Wen2012,Hastings2013}, to our knowledge, an explicit formalism based on Gaussian states (GSs) alone is still missing. More importantly, while the classification of fermionic Gaussian states (fGSs) should be given by the periodic table, that of unitary Gaussian operations (GOs) remains unclear. We also note that there is considerable recent interest in free-boson topological phases \cite{Clerk2018,Ozawa2019,Ohashi2020}, although the classifications seem to differ a lot depending on the setups \cite{He2021,Clerk2021}. Hence, it is also worthwhile to clarify the classification of bosonic Gaussian states (bGSs) in the operation-based framework.

In this work, we fill the gap between the operational formalism and the classification of GSs, and apply the former to GOs. We find consistent classifications for fGSs, and only one trivial phase for bGSs. Nevertheless, there exist nontrivial bosonic Gaussian operations (bGOs). Remarkably, fGSs and fermionic Gaussian operations (fGOs) are related in a highly nontrivial way --- not all the topological fGSs can be disentangled by fGOs with the same symmetries, and not all the topological fGOs can create topological fGSs from trivial ones. This observation allows us to refine both topological fGSs and fGOs into two types --- disentanglable vs. non-disentanglable, and state-like vs. genuinely dynamical, respectively. These notions are applicable to interacting systems and shed new light on topological phases both in and out of equilibrium.

\emph{Quantum Gaussian states and operations.---} Consider a $d$-dimensional ($d$D) lattice $\Lambda=\mathbb{Z}^d$ with $n$ internal states at each unit cell labeled by the elements in $I$. In terms of the fermion / boson modes $\hat c_{\boldsymbol{r}s}$ / $\hat a_{\boldsymbol{r}s}$ ($\boldsymbol{r}\in\Lambda$, $s\in I$), we can define a set of Majorana fermions $\hat\gamma_{\boldsymbol{r}+s}\equiv \hat c_{\boldsymbol{r}s}^\dag+ \hat c_{\boldsymbol{r}s}$, $\hat \gamma_{\boldsymbol{r}-s}\equiv i(\hat c_{\boldsymbol{r}s}^\dag- \hat c_{\boldsymbol{r}s})$ / quadratures $\hat\xi_{\boldsymbol{r}+s}\equiv \hat a_{\boldsymbol{r}s}^\dag+ \hat a_{\boldsymbol{r}s}$, $\hat\xi_{\boldsymbol{r}-s}\equiv i(\hat a_{\boldsymbol{r}s}^\dag-\hat a_{\boldsymbol{r}s})$. A pure fGS/bGS $|\Psi \rangle$ on $\Lambda$ is fully \cite{bG0} characterized by the covariance matrix $\Gamma $:
\begin{equation}
\begin{split}
(\Gamma_{\rm f})_{\boldsymbol{r}S,\boldsymbol{r}'S'} = \frac{i}{2}\langle\Psi_{\rm f}|[\hat\gamma_{\boldsymbol{r}S}, \hat\gamma_{\boldsymbol{r}'S'}]|\Psi_{\rm f}\rangle,\\
(\Gamma_{\rm b})_{\boldsymbol{r}S,\boldsymbol{r}'S'} = \frac{1}{2}\langle\Psi_{\rm b}|\{\hat\xi_{\boldsymbol{r}S}, \hat\xi_{\boldsymbol{r}'S'}\}|\Psi_{\rm b}\rangle,
\end{split}
\end{equation}
where $S\equiv \pm s$ and the subscript $\rm f/b$ stands for ``fermion/boson". By fully we mean that all the observables and reduced density matrices are calculable in terms of $\Gamma$. We focus on short-range correlated GSs, whose covariance matrices satisfy $|(\Gamma)_{\boldsymbol{r}S,\boldsymbol{r}'S'}|\le \Gamma_0 e^{-|\boldsymbol{r}-\boldsymbol{r}'|/\ell}$ for some $\mathcal{O}(1)$ constants $\Gamma_0$ and $\ell$. 

A unitary fGO/bGO $\hat U $, defined as transforming GSs into GSs, is fully characterized by the representation matrix $V$ determined by the linear action on individual modes: 
\begin{equation}
\begin{split}
\hat U_{\rm f}^\dag\hat \gamma_{\boldsymbol{r}S}\hat U_{\rm f} = \sum_{\boldsymbol{r}',S'}(V_{\rm f})_{\boldsymbol{r}S,\boldsymbol{r}'S'}\hat \gamma_{\boldsymbol{r}'S'}, \\
\hat U_{\rm b}^\dag\hat \xi_{\boldsymbol{r}S}\hat U_{\rm b}= \sum_{\boldsymbol{r}',S'}(V_{\rm b})_{\boldsymbol{r}S,\boldsymbol{r}'S'}\hat \xi_{\boldsymbol{r}'S'}. 
\end{split}
\end{equation}
We focus on locality-preserving GOs, for which, just like $\Gamma $, the entries in $V $ decay exponentially in terms of $|\boldsymbol{r} - \boldsymbol{r}'|$. Note that it is the locality that allows us to distinguish different spatial dimensions. Without locality, all the GSs/GOs can be considered to be of zero dimension.

For simplicity, we assume the lattice translation symmetry such that $(\Gamma)_{\boldsymbol{r}S,\boldsymbol{r}'S'}=(\Gamma)_{\boldsymbol{r}-\boldsymbol{r}',SS'}$, which enables us to perform the Fourier transformation $(\Gamma (\boldsymbol{k}))_{SS'}= \sum_{\delta\boldsymbol{r}\in\Lambda} (\Gamma)_{\delta\boldsymbol{r},SS'} e^{-i\boldsymbol{k}\cdot\delta\boldsymbol{r}}$, where $\boldsymbol{k}\in T^d$ is a wave vector in the Brillouin zone. One can check that $\Gamma (\boldsymbol{k})^*= \Gamma (-\boldsymbol{k})$ and
\begin{align}
\Gamma_{\rm f}(\boldsymbol{k})^\dag &=-\Gamma_{\rm f}(\boldsymbol{k}),\;\;\;\;
\Gamma_{\rm b}(\boldsymbol{k})^\dag =\Gamma_{\rm b}(\boldsymbol{k})>0,\\
\Gamma_{\rm f}(\boldsymbol{k})^2&=-\openone_{2n},\;\;\;\;\;\;\;
\Gamma_{\rm b}(\boldsymbol{k})\sigma\Gamma_{\rm b}(\boldsymbol{k})=\sigma,
\label{flat}
\end{align}
where $\sigma\equiv i\sigma_y\otimes\openone_n$ is the symplectic matrix, $\sigma_{x,y,z}$ is the Pauli matrix and $\openone_n$ is the $n\times n$ identity matrix. Similarly, we can define $V(\boldsymbol{k})$ and confirm the following inherent properties: $V (\boldsymbol{k})^*=V (-\boldsymbol{k})$ and 
\begin{equation}
V_{\rm f}(\boldsymbol{k})V_{\rm f}(\boldsymbol{k})^\dag=\openone_{2n},\;\;\;\;
V_{\rm b}(\boldsymbol{k})\sigma V_{\rm b}(\boldsymbol{k})^\dag=\sigma.
\label{unisym}
\end{equation}
The short-range (exponential decay) nature of $|\Psi \rangle$/$\hat U $ turns out to be equivalent to the analyticity of $\Gamma (\boldsymbol{k})$/$V (\boldsymbol{k})$ in $\boldsymbol{k}$ \cite{Ashida2021}.

To impose a symmetry on a GS/GO, we only have to require the symmetry operator commute \cite{UUd} with $|\Psi \rangle\langle\Psi |$/$\hat U $ and then identify its action on the covariance matrix $\Gamma $/representation matrix $V $. This is different from the way of imposing symmetries to Floquet unitaries \cite{Nakagawa2019}, which is not compatible with our operational framework due to the dynamical breaking of anti-unitary symmetries \cite{McGinley2018}.

\emph{Topological equivalence.---} We say two GSs/GOs are \emph{strictly equivalent} if they can be interpolated by a continuous (more precisely, smooth) path along which both symmetries (if any) and locality are respected. Mathematically, this is nothing but the homotopy equivalence \cite{Whitehead1978} for $\Gamma (\boldsymbol{k})$ and $V (\boldsymbol{k})$, which are smooth maps from $T^d$ to some matrix spaces. First defining strict equivalence for GOs, we can alternatively define that two GSs are strictly equivalent if one can be transformed into the other by a trivial GO strictly equivalent to the identity. Obviously, this  implies the original definition. To see the converse, suppose that $\Gamma (\boldsymbol{k};\lambda)$ ($\lambda\in[0,1]$) interpolates between $\Gamma (\boldsymbol{k};0)$ and $\Gamma (\boldsymbol{k};1)$, then the trivial GO $V (\boldsymbol{k};1)$ can be determined by solving $\partial_\lambda V (\boldsymbol{k};\lambda)=K (\boldsymbol{k};\lambda)V (\boldsymbol{k};\lambda)$ with $V (\boldsymbol{k};0)=\openone_{2n}$ and
\begin{equation}
\begin{split}
K_{\rm f}(\boldsymbol{k};\lambda)&=  \frac{1}{2}\Gamma_{\rm f}(\boldsymbol{k};\lambda)\partial_\lambda\Gamma_{\rm f}(\boldsymbol{k};\lambda),\\
K_{\rm b}(\boldsymbol{k};\lambda)&=  \frac{1}{2}\Gamma_{\rm b}(\boldsymbol{k};\lambda)\sigma\partial_\lambda\Gamma_{\rm b}(\boldsymbol{k};\lambda)\sigma.
\end{split}
\label{Kfb}
\end{equation}
Note that $K_{\rm f}(\boldsymbol{k})=-K_{\rm f}(\boldsymbol{k})^\dag$ and $\sigma K_{\rm b}(\boldsymbol{k})\sigma=K_{\rm b}(\boldsymbol{k})^\dag$, so $V_{\rm f}(\boldsymbol{k})$ is unitary and $V_{\rm b}(\boldsymbol{k})$ is symplectic. This construction is compatible with any additional symmetries \cite{SM}.

Due to 
the physical feasibility of introducing ``catalysts" \cite{Anshu2018}, a more useful definition is 
the following weaker version\textcolor{magenta}{:} 
two GSs/GOs are said to be \emph{equivalent} if they are strictly equivalent after some trivial ancillas are added. Mathematically, this definition is fully captured by the $K$-theory \cite{Karoubi2008}, which concerns essentially the stable homotopy in the presence of additional degrees of freedom. Similar to the case of strict equivalence, we can alternatively define two GSs to be equivalent if one can be transformed into the other by a trivial \cite{tri} GO and assisted by some ancillas. We define trivial GSs to be those without correlations, i.e., $\Gamma (\boldsymbol{k})$ is $\boldsymbol{k}$-independent or $(\Gamma)_{\boldsymbol{r}S,\boldsymbol{r}'S'}\propto \delta_{\boldsymbol{r}\boldsymbol{r}'}$, and of course their equivalent states.

\begin{table}[tbp]
\caption{Dictionary for translating between the physical symmetries of the state (columns 2-4), the emergent symmetries of $i\Gamma_{\rm f}(\boldsymbol{k})$ labelled according to the Hamiltoniam formalism (columns 5-7) and the corresponding AZ classes (column 1).
PHS/SLS stand for particle-hole/sublattice symmetries. SU(2) symmetry marked by $z$/``other" is imposed only in $z$-direciton/other degree of freedom than spin.
Strong topology  of $d$D fGSs/fGOs are classified by $\pi_d(\mathcal{F})$, where $\pi_d$ is the $d$th homotopy group and $\mathcal{F}$ is the classifying space in the last column.
Cells in blue are equivalent to the classification of bGOs without or with involutory/anti-involutory TRS.} 
\begin{center}
\begin{tabular}{c|ccc|ccc|cc}
\hline\hline
\multirow{2}{*}{\;AZ} & \multicolumn{3}{c|}{Physical}   & \multicolumn{3}{c|}{Emergent} & \multicolumn{2}{c}{\;Classifying space}\\
  & \;TRS\; & \;U(1)\; & \;SU(2)\; & \;TRS\; & PHS & \;SLS\; & \; fGSs & \; fGOs \\
\hline
\colorbox{blue!10!white}{A} &  \colorbox{blue!10!white}{0} & 1 & 0 & 0 & 0 & 0 & $\mathcal{C}_0$ &  \colorbox{blue!10!white}{$\mathcal{C}_1$} \\
AIII & $-$ & 0 & $z$ & 0 & 0 & 1 & $\mathcal{C}_1$ & $\mathcal{C}_1^2$ \\
\hline
\colorbox{blue!10!white}{AI} & \colorbox{blue!10!white}{$+$} & 1 & 0 & $+$ & 0 & 0 & $\mathcal{R}_0$ & \colorbox{blue!10!white}{$\mathcal{R}_1$} \\
BDI & $+$ & 0 & 0 & $+$ & $+$ & 1 & $\mathcal{R}_1$ & $\mathcal{R}_1^2$ \\
D & 0 & 0 & 0 & 0 & $+$ & 0 & $\mathcal{R}_2$ & $\mathcal{R}_1$  \\
DIII & $-$ & 0 & 0 & $-$ & $+$ & 1 & $\mathcal{R}_3$ & $\mathcal{C}_1$  \\
\colorbox{blue!10!white}{AII} & \colorbox{blue!10!white}{$-$} & 1 & 0 & $-$ & 0 & 0 & $\mathcal{R}_4$ & \colorbox{blue!10!white}{$\mathcal{R}_5$} \\
CII & $-$ & 0 & other & $-$ & $-$ & 1 & $\mathcal{R}_5$ & $\mathcal{R}_5^2$ \\
C & 0 & 0 & 1 & 0 & $-$ & 0 & $\mathcal{R}_6$ & $\mathcal{R}_5$  \\
CI & $-$ & 0 & 1 & $+$ & $-$ & 1 & $\mathcal{R}_7$ & $\mathcal{C}_1$  \\
\hline\hline
\end{tabular}
\end{center}
\label{table1}
\end{table}

\emph{Classifications.---}
We compute the $K$-theory classification of fGSs by applying the standard Clifford algebra technique \cite{Kitaev2009} directly to $\Gamma_{\rm f}(\boldsymbol{k})$, rather than to the Hamiltonian~\footnote{Note that $i\Gamma_{\rm f}(\boldsymbol{k})$ is Hermitian and 
therefore the same treatment can be applied to it as to a Hamiltonian~\cite{SM}}. The results reproduce the well-known periodic table \cite{Ryu2008,Kitaev2009,Ryu2010} [cf. Table~\ref{table1}]. Crucially, to obtain \emph{emergent} symmetries on $i\Gamma_{\rm f}(\boldsymbol{k})$, which reproduce the standard symmetries of the Hamiltonian formalism~\cite{Ryu2016},
it is necessary to impose different \emph{physical} symmetries to the state $\ket{\Psi_{\rm f}}$.
For example, even without any physical symmetry, $i\Gamma_{\rm f}(\boldsymbol{k})$ exhibits the particle-hole symmetry; a physical spinless time-reversal symmetry (TRS) implies an emergent sublattice symmetry $\{\sigma_z\otimes\openone_n, i\Gamma_{\rm f}(\boldsymbol{k})\}=0$.
Overall, at the physical level it suffices to impose a TRS, a ${\rm U}(1)$ symmetry (which generates a phase to $c_{\boldsymbol{r}s}$'s / $a_{\boldsymbol{r}s}$'s) or a ${\rm SU}(2)$ spin-rotation symmetry to obtain all the AZ classes~\cite{Altland1997,Ryu2008}. See Table~\ref{table1} for the complete dictionary.

\begin{figure}
	\begin{center}
		\includegraphics[width=8.5cm, clip]{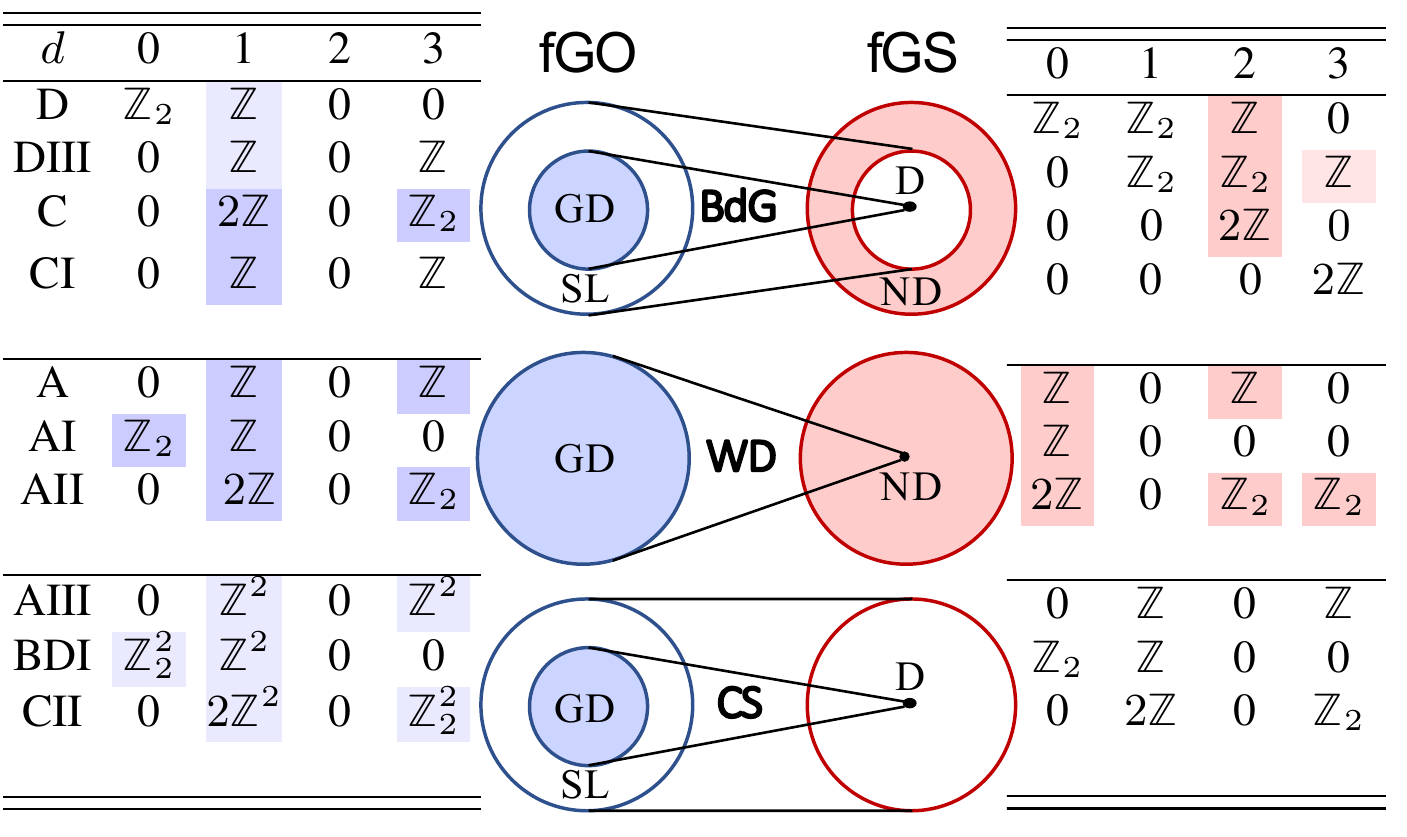}
	\end{center}
	\caption{Homomorphism from the $K$-groups for fGOs to that for fGSs induced by applying fGOs to trivial reference fGSs. The kernel and its complement are identified as genuinely dynamical (GD) and state-like (SL), respectively. The image and its complement are defined as disentanglable (D) and non-disentanglable (ND). For the Bogoliubov-de Gennes (BdG) classes, we have the most general situation. For the Wigner-Dyson (WD) classes, the homomorphism is all-to-trivial and thus all the topological fGOs are genuinely dynamical. For the chiral symmetry (CS) classes, the homomorphism is surjective and thus all the topological fGSs are disentanglable. Cells in light (dark) blue / red indicate that only a subgroup (all) of the topological fGOs / fGSs are genuinely dynamical / non-disentanglable.} 
	\label{fig1}
\end{figure}

To classify fGOs, we utilize the Hermitianization technique developed in the context of Floquet topological phases \cite{Harper2017} and more recent non-Hermitian topological phases \cite{Gong2018,Zhou2019,Kawabata2019}. More in detail, given any unitary $V_{\rm f}(\boldsymbol{k})$, it turns out that $\sigma_+\otimes V_{\rm f}(\boldsymbol{k})+\sigma_-\otimes V_{\rm f}(\boldsymbol{k})^\dag$ ($\sigma_\pm\equiv(\sigma_x\pm i\sigma_y)/2$) is Hermitian and involutory. It can therefore be treated with the same Clifford algebra techniques mentioned before, the results of which are summarized in the right half of the last column in Table~\ref{table1}. The classifying spaces coincide with those obtained through unitarization for non-Hermitian topological phases \cite{Gong2018}. Explicit classification results for fGOs (left) and fGSs (right) with $d\le 3$ are presented in Fig.~\ref{fig1}.

In contrast, all the short-range correlated bGSs are (strictly) trivial. To see this, first noting that $\Gamma_{\rm b}(\boldsymbol{k})>0$, we can uniquely define its logarithm $\log\Gamma_{\rm b}(\boldsymbol{k})$  \cite{Higham2008}, which is Hermitian and anti-commutes with $\sigma$ (cf. Eq.~(\ref{flat})). One can thus continuously deform $\Gamma_{\rm b}(\boldsymbol{k};0)\equiv\Gamma_{\rm b}(\boldsymbol{k})$ into the identity $\Gamma_{\rm b}(\boldsymbol{k};1)\equiv\openone_{2n}$, which corresponds to the vacuum, via
\begin{equation}
\Gamma_{\rm b}(\boldsymbol{k};\lambda) = e^{(1-\lambda)\log\Gamma_{\rm b}(\boldsymbol{k}) },\;\;\;\;\lambda\in[0,1].
\label{exptri}
\end{equation}
This construction is also compatible with additional symmetries. Note that the triviality of bGSs does not contradict the possible nontrivial excited bands \cite{Raghu2008,Hafezi2011,Cooper2019} or dynamical phases \cite{Clerk2018} in stable or unstable free-boson systems, since short-range correlated bGSs correspond to the ground states of stable and gapped free-boson systems \cite{Schuch2006}.

Nevertheless, bGOs may be topologically nontrivial. To see this, we can polar decompose the representation matrix into $V_{\rm b}(\boldsymbol{k})=W_{\rm b}(\boldsymbol{k})P_{\rm b}(\boldsymbol{k})$, where $W_{\rm b}(\boldsymbol{k})$ is unitary and $P_{\rm b}(\boldsymbol{k})$ is Hermitian and positive-definite, both of which satisfy the right relation in Eq.~(\ref{unisym}) \cite{SM}. We can trivialize $P_{\rm b}(\boldsymbol{k})$ into the identity following Eq.~(\ref{exptri}) and accordingly unitarize $V_{\rm b}(\boldsymbol{k})$ into $W_{\rm b}(\boldsymbol{k})$, which satisfies $[W_{\rm b}(\boldsymbol{k}),\sigma]=0$. The classification of bGOs thus turns out to be that of a unitary $W_{\rm b}(\boldsymbol{k})$ that commutes with $\sigma$, \textit{i.e.} equivalent to fGOs in the presence of the ${\rm U}(1)$ symmetry given by $\sigma$. The corresponding rows (with or without TRS) are marked in blue in Table~\ref{table1}.

\emph{Relations between Gaussian states and operations.---} As mentioned previously, topologically equivalent GSs can be related to each other by trivial GOs. Taking the dual of this statement, we know that GOs generating topological GSs from trivial reference states are necessarily topological, and we will refer to them as \emph{state-like} topological GOs. 

However, topological GOs may not always change the topological class of a GS. In this case we refer to them as \emph{genuinely dynamical}. This happens for bosons, and actually also some symmetry classes of fermions. One simple example is class A  or AI in 1D,  where there is no topological fGS while fGOs are classified by $\mathbb{Z}$, which corresponds to the winding number of $V_{\rm f}(k)$ \cite{Kitagawa2010} and is exemplified by lattice translations \cite{Gross2012}. Indeed, lattice translations leave translation-invariant fGSs unchanged, and cannot alter any topological feature even in the presence of disorder.

Topological fGSs that can be generated by acting with a fGO on a trivial state in the same symmetry class are known as \emph{disentanglable}. Notably, there exist also \emph{non-disentaglable} fGSs. This clearly happens in 2D, where fGOs are all trivial while fGSs are not (see Fig.~\ref{fig1}). In particular, class D without any symmetry is classified by $\mathbb{Z}$, which corresponds to the Chern number and is exemplified by the ground state of a chiral superconductor \cite{Sato2009,Qi2010}. 

In general, one can define a group homomorphism from the $K$-group of GOs to that of GSs induced by the functor $F: \hat U\to \hat U|\Psi_0\rangle$, where  $|\Psi_0\rangle$ is a trivial reference GS. This group homomorphism is obviously trivial for bosons, but turns out to be unexpectedly complicated for fermions. From the above examples, where either fGSs or fGOs are always trivial, we already know that this group homomorphism may be neither injective nor surjective. What remains unclear is, when both fGSs and fGOs have nontrivial classifications, whether the former can be disentangled by the latter.

As illustrated in Fig.~\ref{fig1}, the complete answer to the above question reads: fGSs in the Dyson-Wigner classes \cite{Dyson1962}, including A, AI and AII, are fully non-disentanglable. In contrast, the chiral-symmetry classes \cite{Verbaarschot1994}, including AIII, BDI and CII, are fully disentanglable. The remaining four Bogoliubov-de Gennes classes \cite{Altland1997}, including D, DIII, C and CI, are partially disentanglable for specific spatial dimensions. 

\begin{table*}[tbp]
\caption{Examples of topological fGSs that can and cannot be disentangled by fGOs, as well as genuinely dynamical topological fGOs. We can always obtain examples of state-like topological fGOs from the disentanglers for disentanglable topological fGSs.} 
\begin{center}
\begin{tabular}{cccc}
\hline\hline
\;\;\;\;\;\;\;\;Gaussian topological order\;\;\;\;\;\;\;\; & \;\;\;\;\;\;\;\;\;\;\;\;Intrinsic\;\;\;\;\;\;\;\;\;\;\;\; & \;\;\;\;\;\;\;\;\;\;\;\;Symmetry-enriched\;\;\;\;\;\;\;\;\;\;\;\; & \;\;\;\;\;\;\;\;\;\;\;\;Symmetry-protected\;\;\;\;\;\;\;\;\;\;\;\;  \\
\hline
\multirow{2}{*}{Disentanglable fGS (State-like fGO)} & class D, $d=1$ & class BDI, $d=1$ & class BDI, $d=1$ \\
& \multicolumn{2}{c}{Kitaev chain $\times$ $(2m+1)$} & Kitaev chain $\times$ $2m$ \\
\hline
\multirow{2}{*}{Non-disentanglable fGS} & class D, $d=2$ & class A, $d=2$ & class AII, $d=2,3$ \\
& Chiral superconductor & Quantum Hall insulator & TRS topological insulator \\
\hline
\multirow{2}{*}{Genuinely dynamical fGO} & class D, $d=1$ & class AI, $d=1$ & class AII, $d=3$ \\
&\multicolumn{2}{c}{Lattice translation} & TRS $\mathbb{Z}_2$ operation \\
\hline\hline
\end{tabular}
\end{center}
\label{table2}
\end{table*}

Understanding the full disentanglability of chiral-symmetry classes is the easiest. In fact, one can show straightforwardly that any fGSs in these classes can be disentangled by certain fGOs with the same symmetries \cite{SM}. Turning to the Wigner-Dyson classes with particle number conservation, the full non-disentanglability can be understood from the absence of exponentially localized (and compatible with the TRS, if any) Wannier functions \cite{Marzari2007,Vanderbilt2011}. 

For class D/C, the fGSs and fGOs are both nontrivial only in $d=0,1/4,5$, where class BDI/CII also has nontrivial fGSs and can be surjectively (in the sense of topological equivalence) included into class D/C by forgetting the TRS. Recalling the full disentanglability for class BDI/CII, we know that these topological fGSs in class D/C are also disentanglable. 

The remaining classes DIII and CI are the most complicated. By explicitly calculating the winding number or the Chern-Simons form \cite{Ryu2016}, we find that topological fGSs are fully disentanglable in  $d=1,7$ (DIII) or $d=3,5$ (CI), while only a subgroup $2\mathbb{Z}$ out of $\mathbb{Z}$ is disentanglable in $d=3$ (DIII) or $d=7$ (CI). The latter result is consistent with the fact that classes AII and AI are non-disentanglable: let us temporarily use the Hamiltonian formalism. For topological insulators in class DIII/CI, we can surjectively include them into classes AII/AI by forgetting the particle-hole symmetry. Therefore, the former cannot be fully disentanglable since otherwise it would contradict the non-disentanglable nature of the latter.

\emph{Discussions.---} It is well-known that fGSs (similarly to general topological quantum states) can be categorized into intrinsic \cite{Wen1990}, symmetry-protected \cite{Haldane1983} and symmetry-enriched topological order \cite{Ran2013}: nontrivial fGSs in class D are intrinsically topological, while nontrivial symmetric fGSs that are still nontrivial (become trivial) in the absence of symmetries exhibit Gaussian symmetry-protected (symmetry-enriched) order \cite{symc}. Here, we can further refine each category into two classes based on the disentanglability. Similarly, we can also categorize topological fGOs and refine them to be state-like or genuinely dynamical, depending on whether they can generate topological fGSs. See some examples in Table~\ref{table2}.


Furthermore, fGOs are relevant also to free Floquet systems described by time-periodic Bloch Hamiltonians $h(\boldsymbol{k},t+T)=h(\boldsymbol{k},t)$. Although $d$D topological fGOs cannot be written as Floquet unitaries $V_{\rm F}(\boldsymbol{k})\equiv\overrightarrow{\rm T} e^{-i\int^T_0dt \, h(\boldsymbol{k},t)}$ in $d$D, we expect that they can be embedded as the edge dynamics of $(d+1)$D intrinsic Floquet topological phases without equilibrium counterparts. There are some 1D examples in the literature \cite{Lindner2013,Fidkowski2019}, and it would be interesting to consider explicit constructions in higher dimensions. 

Finally, we note that the unitary-to-state homomorphism [cf. Fig.~\ref{fig1}] applies to general locally interacting systems. In particular, disentanglability can be defined for general short-range correlated quantum states by simply extending disentanglers from Gaussian operations to arbitrary QCA. 
For example, the Kitaev chain \cite{Kitaev2001} still exhibits disentanglable intrinsic topological order in the presence of interactions \cite{Wen2017}, since it can be disentangled by a nontrivial fermionic QCA \cite{Fidkowski2019,Piroli2021}. In fact, all the symmetry-protected topological phases classified by (super)cohomology \cite{Chen2013,Gu2014,Tantivasadakarn2018,Ellison2019,Chen2021} are known to be disentanglable. In contrast, the toric code \cite{Kitaev2003}, which is also topologically ordered, is not disentanglable due to the triviality of 2D bosonic (spin) QCA \cite{Freedman2020}. 
Also, the notions of state-like and genuinely dynamical topology apply equally to general QCA. For example, the nontrivial 1D fermionic QCA mentioned and a recently discovered 3D bosonic QCA \cite{Haah2018} that disentangles the Walker-Wang model \cite{Walker2012} are state-like, while the (symmetry-protected) index is a genuinely dynamical topological invariant  in 1D \cite{Gross2012,Cirac2017,Chen2018,Gong2020,Ranard2020}.

\emph{Summary and outlook.---} We have revisited the classification problem of free fermions from an operational perspective, which in turn inspires us to consider the boson counterpart and the classification of GOs. We have found that while bGSs are all trivial, bGOs are not. Most importantly, we have unveiled the complicated relations between fGSs and fGOs, which allow us to refine topological fGSs based on the disentanglability and in turn distinguish state-like topological fGOs from genuinely dynamical ones. The underlying unitary-to-state homomorphism applies equally to interacting systems and provides new insights into both equilibrium and nonequilibrium topological phases.

One obvious open problem is the generalization to generic quantum states and operations. Two specific questions could be finding non-disentanglable symmetry-protected topological phases and genuinely dynamical topological QCA in $d\ge2$D. Even within GSs/GOs, we can consider many generalizations such as to crystalline \cite{Fu2011,Slager2013,Shiozaki2014,Slager2017,Khalaf2018} and higher-order \cite{Benalcazar2017,Schindler2018,Trifunovic2019} topological phases or/and mixed Gaussian states and channels \cite{Diehl2013,Budich2015,Mink2019,Altland2021}. Last but not the least, it might also be interesting to study the implications of topological obstructions for Gaussian variational methods with local ans\"atze \cite{Shi2018,Guaita2019,Hackl2020}.

We thank Ignacio Cirac, Alex Turzillo, Lorenzo Piroli and Nat Tantivasadakarn for very helpful discussions. Z.G. is supported by the Max-Planck-Harvard Research Center for Quantum Optics (MPHQ). T.G. is supported by the Deutsche Forschungs- gemeinschaft (DFG, German Research Foundation) under Germany’s Excellence Strategy – EXC-2111 – 39081486.

\bibliography{GZP_references}

\begin{thebibliography}{102}%
\makeatletter
\providecommand \@ifxundefined [1]{%
 \@ifx{#1\undefined}
}%
\providecommand \@ifnum [1]{%
 \ifnum #1\expandafter \@firstoftwo
 \else \expandafter \@secondoftwo
 \fi
}%
\providecommand \@ifx [1]{%
 \ifx #1\expandafter \@firstoftwo
 \else \expandafter \@secondoftwo
 \fi
}%
\providecommand \natexlab [1]{#1}%
\providecommand \enquote  [1]{``#1''}%
\providecommand \bibnamefont  [1]{#1}%
\providecommand \bibfnamefont [1]{#1}%
\providecommand \citenamefont [1]{#1}%
\providecommand \href@noop [0]{\@secondoftwo}%
\providecommand \href [0]{\begingroup \@sanitize@url \@href}%
\providecommand \@href[1]{\@@startlink{#1}\@@href}%
\providecommand \@@href[1]{\endgroup#1\@@endlink}%
\providecommand \@sanitize@url [0]{\catcode `\\12\catcode `\$12\catcode
  `\&12\catcode `\#12\catcode `\^12\catcode `\_12\catcode `\%12\relax}%
\providecommand \@@startlink[1]{}%
\providecommand \@@endlink[0]{}%
\providecommand \url  [0]{\begingroup\@sanitize@url \@url }%
\providecommand \@url [1]{\endgroup\@href {#1}{\urlprefix }}%
\providecommand \urlprefix  [0]{URL }%
\providecommand \Eprint [0]{\href }%
\providecommand \doibase [0]{http://dx.doi.org/}%
\providecommand \selectlanguage [0]{\@gobble}%
\providecommand \bibinfo  [0]{\@secondoftwo}%
\providecommand \bibfield  [0]{\@secondoftwo}%
\providecommand \translation [1]{[#1]}%
\providecommand \BibitemOpen [0]{}%
\providecommand \bibitemStop [0]{}%
\providecommand \bibitemNoStop [0]{.\EOS\space}%
\providecommand \EOS [0]{\spacefactor3000\relax}%
\providecommand \BibitemShut  [1]{\csname bibitem#1\endcsname}%
\let\auto@bib@innerbib\@empty
\bibitem [{\citenamefont {Chiu}\ \emph {et~al.}(2016)\citenamefont {Chiu},
  \citenamefont {Teo}, \citenamefont {Schnyder},\ and\ \citenamefont
  {Ryu}}]{Ryu2016}%
  \BibitemOpen
  \bibfield  {author} {\bibinfo {author} {\bibfnamefont {C.-K.}\ \bibnamefont
  {Chiu}}, \bibinfo {author} {\bibfnamefont {J.~C.~Y.}\ \bibnamefont {Teo}},
  \bibinfo {author} {\bibfnamefont {A.~P.}\ \bibnamefont {Schnyder}}, \ and\
  \bibinfo {author} {\bibfnamefont {S.}~\bibnamefont {Ryu}},\ }\href {\doibase
  10.1103/RevModPhys.88.035005} {\bibfield  {journal} {\bibinfo  {journal}
  {Rev. Mod. Phys.}\ }\textbf {\bibinfo {volume} {88}},\ \bibinfo {pages}
  {035005} (\bibinfo {year} {2016})}\BibitemShut {NoStop}%
\bibitem [{\citenamefont {Hasan}\ and\ \citenamefont {Kane}(2010)}]{Kane2010}%
  \BibitemOpen
  \bibfield  {author} {\bibinfo {author} {\bibfnamefont {M.~Z.}\ \bibnamefont
  {Hasan}}\ and\ \bibinfo {author} {\bibfnamefont {C.~L.}\ \bibnamefont
  {Kane}},\ }\href {\doibase 10.1103/RevModPhys.82.3045} {\bibfield  {journal}
  {\bibinfo  {journal} {Rev. Mod. Phys.}\ }\textbf {\bibinfo {volume} {82}},\
  \bibinfo {pages} {3045} (\bibinfo {year} {2010})}\BibitemShut {NoStop}%
\bibitem [{\citenamefont {Qi}\ and\ \citenamefont {Zhang}(2011)}]{Qi2011}%
  \BibitemOpen
  \bibfield  {author} {\bibinfo {author} {\bibfnamefont {X.-L.}\ \bibnamefont
  {Qi}}\ and\ \bibinfo {author} {\bibfnamefont {S.-C.}\ \bibnamefont {Zhang}},\
  }\href {\doibase 10.1103/RevModPhys.83.1057} {\bibfield  {journal} {\bibinfo
  {journal} {Rev. Mod. Phys.}\ }\textbf {\bibinfo {volume} {83}},\ \bibinfo
  {pages} {1057} (\bibinfo {year} {2011})}\BibitemShut {NoStop}%
\bibitem [{\citenamefont {Altland}\ and\ \citenamefont
  {Zirnbauer}(1997)}]{Altland1997}%
  \BibitemOpen
  \bibfield  {author} {\bibinfo {author} {\bibfnamefont {A.}~\bibnamefont
  {Altland}}\ and\ \bibinfo {author} {\bibfnamefont {M.~R.}\ \bibnamefont
  {Zirnbauer}},\ }\href {\doibase 10.1103/PhysRevB.55.1142} {\bibfield
  {journal} {\bibinfo  {journal} {Phys. Rev. B}\ }\textbf {\bibinfo {volume}
  {55}},\ \bibinfo {pages} {1142} (\bibinfo {year} {1997})}\BibitemShut
  {NoStop}%
\bibitem [{\citenamefont {Schnyder}\ \emph {et~al.}(2008)\citenamefont
  {Schnyder}, \citenamefont {Ryu}, \citenamefont {Furusaki},\ and\
  \citenamefont {Ludwig}}]{Ryu2008}%
  \BibitemOpen
  \bibfield  {author} {\bibinfo {author} {\bibfnamefont {A.~P.}\ \bibnamefont
  {Schnyder}}, \bibinfo {author} {\bibfnamefont {S.}~\bibnamefont {Ryu}},
  \bibinfo {author} {\bibfnamefont {A.}~\bibnamefont {Furusaki}}, \ and\
  \bibinfo {author} {\bibfnamefont {A.~W.~W.}\ \bibnamefont {Ludwig}},\ }\href
  {\doibase 10.1103/PhysRevB.78.195125} {\bibfield  {journal} {\bibinfo
  {journal} {Phys. Rev. B}\ }\textbf {\bibinfo {volume} {78}},\ \bibinfo
  {pages} {195125} (\bibinfo {year} {2008})}\BibitemShut {NoStop}%
\bibitem [{\citenamefont {Ryu}\ \emph {et~al.}(2010)\citenamefont {Ryu},
  \citenamefont {Schnyder}, \citenamefont {Furusaki},\ and\ \citenamefont
  {Ludwig}}]{Ryu2010}%
  \BibitemOpen
  \bibfield  {author} {\bibinfo {author} {\bibfnamefont {S.}~\bibnamefont
  {Ryu}}, \bibinfo {author} {\bibfnamefont {A.~P.}\ \bibnamefont {Schnyder}},
  \bibinfo {author} {\bibfnamefont {A.}~\bibnamefont {Furusaki}}, \ and\
  \bibinfo {author} {\bibfnamefont {A.~W.~W.}\ \bibnamefont {Ludwig}},\ }\href
  {http://stacks.iop.org/1367-2630/12/i=6/a=065010} {\bibfield  {journal}
  {\bibinfo  {journal} {New J. Phys.}\ }\textbf {\bibinfo {volume} {12}},\
  \bibinfo {pages} {065010} (\bibinfo {year} {2010})}\BibitemShut {NoStop}%
\bibitem [{\citenamefont {Kitaev}(2009)}]{Kitaev2009}%
  \BibitemOpen
  \bibfield  {author} {\bibinfo {author} {\bibfnamefont {A.}~\bibnamefont
  {Kitaev}},\ }\href {https://doi.org/10.1063/1.3149495} {\bibfield  {journal}
  {\bibinfo  {journal} {AIP Conf. Proc.}\ }\textbf {\bibinfo {volume} {1134}},\
  \bibinfo {pages} {22} (\bibinfo {year} {2009})}\BibitemShut {NoStop}%
\bibitem [{\citenamefont {Teo}\ and\ \citenamefont {Kane}(2010)}]{Teo2010}%
  \BibitemOpen
  \bibfield  {author} {\bibinfo {author} {\bibfnamefont {J.~C.~Y.}\
  \bibnamefont {Teo}}\ and\ \bibinfo {author} {\bibfnamefont {C.~L.}\
  \bibnamefont {Kane}},\ }\href {\doibase 10.1103/PhysRevB.82.115120}
  {\bibfield  {journal} {\bibinfo  {journal} {Phys. Rev. B}\ }\textbf {\bibinfo
  {volume} {82}},\ \bibinfo {pages} {115120} (\bibinfo {year}
  {2010})}\BibitemShut {NoStop}%
\bibitem [{\citenamefont {Chen}\ \emph {et~al.}(2010)\citenamefont {Chen},
  \citenamefont {Gu},\ and\ \citenamefont {Wen}}]{Chen2010}%
  \BibitemOpen
  \bibfield  {author} {\bibinfo {author} {\bibfnamefont {X.}~\bibnamefont
  {Chen}}, \bibinfo {author} {\bibfnamefont {Z.-C.}\ \bibnamefont {Gu}}, \ and\
  \bibinfo {author} {\bibfnamefont {X.-G.}\ \bibnamefont {Wen}},\ }\href
  {\doibase 10.1103/PhysRevB.82.155138} {\bibfield  {journal} {\bibinfo
  {journal} {Phys. Rev. B}\ }\textbf {\bibinfo {volume} {82}},\ \bibinfo
  {pages} {155138} (\bibinfo {year} {2010})}\BibitemShut {NoStop}%
\bibitem [{\citenamefont {Chen}\ \emph {et~al.}(2011)\citenamefont {Chen},
  \citenamefont {Gu},\ and\ \citenamefont {Wen}}]{Chen2011}%
  \BibitemOpen
  \bibfield  {author} {\bibinfo {author} {\bibfnamefont {X.}~\bibnamefont
  {Chen}}, \bibinfo {author} {\bibfnamefont {Z.-C.}\ \bibnamefont {Gu}}, \ and\
  \bibinfo {author} {\bibfnamefont {X.-G.}\ \bibnamefont {Wen}},\ }\href
  {\doibase 10.1103/PhysRevB.83.035107} {\bibfield  {journal} {\bibinfo
  {journal} {Phys. Rev. B}\ }\textbf {\bibinfo {volume} {83}},\ \bibinfo
  {pages} {035107} (\bibinfo {year} {2011})}\BibitemShut {NoStop}%
\bibitem [{\citenamefont {Chen}\ \emph {et~al.}(2013)\citenamefont {Chen},
  \citenamefont {Gu}, \citenamefont {Liu},\ and\ \citenamefont
  {Wen}}]{Chen2013}%
  \BibitemOpen
  \bibfield  {author} {\bibinfo {author} {\bibfnamefont {X.}~\bibnamefont
  {Chen}}, \bibinfo {author} {\bibfnamefont {Z.-C.}\ \bibnamefont {Gu}},
  \bibinfo {author} {\bibfnamefont {Z.-X.}\ \bibnamefont {Liu}}, \ and\
  \bibinfo {author} {\bibfnamefont {X.-G.}\ \bibnamefont {Wen}},\ }\href
  {\doibase 10.1103/PhysRevB.87.155114} {\bibfield  {journal} {\bibinfo
  {journal} {Phys. Rev. B}\ }\textbf {\bibinfo {volume} {87}},\ \bibinfo
  {pages} {155114} (\bibinfo {year} {2013})}\BibitemShut {NoStop}%
\bibitem [{\citenamefont {Pollmann}\ \emph {et~al.}(2010)\citenamefont
  {Pollmann}, \citenamefont {Turner}, \citenamefont {Berg},\ and\ \citenamefont
  {Oshikawa}}]{Pollmann2010}%
  \BibitemOpen
  \bibfield  {author} {\bibinfo {author} {\bibfnamefont {F.}~\bibnamefont
  {Pollmann}}, \bibinfo {author} {\bibfnamefont {A.~M.}\ \bibnamefont
  {Turner}}, \bibinfo {author} {\bibfnamefont {E.}~\bibnamefont {Berg}}, \ and\
  \bibinfo {author} {\bibfnamefont {M.}~\bibnamefont {Oshikawa}},\ }\href
  {\doibase 10.1103/PhysRevB.81.064439} {\bibfield  {journal} {\bibinfo
  {journal} {Phys. Rev. B}\ }\textbf {\bibinfo {volume} {81}},\ \bibinfo
  {pages} {064439} (\bibinfo {year} {2010})}\BibitemShut {NoStop}%
\bibitem [{\citenamefont {Turner}\ \emph {et~al.}(2011)\citenamefont {Turner},
  \citenamefont {Pollmann},\ and\ \citenamefont {Berg}}]{Pollmann2011}%
  \BibitemOpen
  \bibfield  {author} {\bibinfo {author} {\bibfnamefont {A.~M.}\ \bibnamefont
  {Turner}}, \bibinfo {author} {\bibfnamefont {F.}~\bibnamefont {Pollmann}}, \
  and\ \bibinfo {author} {\bibfnamefont {E.}~\bibnamefont {Berg}},\ }\href
  {\doibase 10.1103/PhysRevB.83.075102} {\bibfield  {journal} {\bibinfo
  {journal} {Phys. Rev. B}\ }\textbf {\bibinfo {volume} {83}},\ \bibinfo
  {pages} {075102} (\bibinfo {year} {2011})}\BibitemShut {NoStop}%
\bibitem [{\citenamefont {Schuch}\ \emph {et~al.}(2010)\citenamefont {Schuch},
  \citenamefont {Cirac},\ and\ \citenamefont {P\'erez-Garc\'ia}}]{Schuch2010}%
  \BibitemOpen
  \bibfield  {author} {\bibinfo {author} {\bibfnamefont {N.}~\bibnamefont
  {Schuch}}, \bibinfo {author} {\bibfnamefont {I.}~\bibnamefont {Cirac}}, \
  and\ \bibinfo {author} {\bibfnamefont {D.}~\bibnamefont {P\'erez-Garc\'ia}},\
  }\href {\doibase https://doi.org/10.1016/j.aop.2010.05.008} {\bibfield
  {journal} {\bibinfo  {journal} {Ann. Phys.}\ }\textbf {\bibinfo {volume}
  {325}},\ \bibinfo {pages} {2153} (\bibinfo {year} {2010})}\BibitemShut
  {NoStop}%
\bibitem [{\citenamefont {Schuch}\ \emph {et~al.}(2011)\citenamefont {Schuch},
  \citenamefont {P\'erez-Garc\'{\i}a},\ and\ \citenamefont
  {Cirac}}]{Schuch2011}%
  \BibitemOpen
  \bibfield  {author} {\bibinfo {author} {\bibfnamefont {N.}~\bibnamefont
  {Schuch}}, \bibinfo {author} {\bibfnamefont {D.}~\bibnamefont
  {P\'erez-Garc\'{\i}a}}, \ and\ \bibinfo {author} {\bibfnamefont
  {I.}~\bibnamefont {Cirac}},\ }\href {\doibase 10.1103/PhysRevB.84.165139}
  {\bibfield  {journal} {\bibinfo  {journal} {Phys. Rev. B}\ }\textbf {\bibinfo
  {volume} {84}},\ \bibinfo {pages} {165139} (\bibinfo {year}
  {2011})}\BibitemShut {NoStop}%
\bibitem [{\citenamefont {Verstraete}\ \emph {et~al.}(2008)\citenamefont
  {Verstraete}, \citenamefont {Murg},\ and\ \citenamefont
  {Cirac}}]{Verstraete2008}%
  \BibitemOpen
  \bibfield  {author} {\bibinfo {author} {\bibfnamefont {F.}~\bibnamefont
  {Verstraete}}, \bibinfo {author} {\bibfnamefont {V.}~\bibnamefont {Murg}}, \
  and\ \bibinfo {author} {\bibfnamefont {J.~I.}\ \bibnamefont {Cirac}},\ }\href
  {\doibase 10.1080/14789940801912366} {\bibfield  {journal} {\bibinfo
  {journal} {Adv. Phys.}\ }\textbf {\bibinfo {volume} {57}},\ \bibinfo {pages}
  {143} (\bibinfo {year} {2008})}\BibitemShut {NoStop}%
\bibitem [{\citenamefont {Or\'us}(2014)}]{Orus2014}%
  \BibitemOpen
  \bibfield  {author} {\bibinfo {author} {\bibfnamefont {R.}~\bibnamefont
  {Or\'us}},\ }\href {\doibase https://doi.org/10.1016/j.aop.2014.06.013}
  {\bibfield  {journal} {\bibinfo  {journal} {Ann. Phys.}\ }\textbf {\bibinfo
  {volume} {349}},\ \bibinfo {pages} {117} (\bibinfo {year}
  {2014})}\BibitemShut {NoStop}%
\bibitem [{\citenamefont {Cirac}\ \emph {et~al.}(2020)\citenamefont {Cirac},
  \citenamefont {P\'erez-Garc\'ia}, \citenamefont {Schuch},\ and\ \citenamefont
  {Verstraete}}]{Cirac2020}%
  \BibitemOpen
  \bibfield  {author} {\bibinfo {author} {\bibfnamefont {J.~I.}\ \bibnamefont
  {Cirac}}, \bibinfo {author} {\bibfnamefont {D.}~\bibnamefont
  {P\'erez-Garc\'ia}}, \bibinfo {author} {\bibfnamefont {N.}~\bibnamefont
  {Schuch}}, \ and\ \bibinfo {author} {\bibfnamefont {F.}~\bibnamefont
  {Verstraete}},\ }\href@noop {} {\enquote {\bibinfo {title} {Matrix product
  states and projected entangled pair states: Concepts, symmetries, and
  theorems},}\ } (\bibinfo {year} {2020}),\ \bibinfo {note}
  {arXiv:2011.12127}\BibitemShut {NoStop}%
\bibitem [{\citenamefont {Gross}\ \emph {et~al.}(2012)\citenamefont {Gross},
  \citenamefont {Nesme}, \citenamefont {Vogts},\ and\ \citenamefont
  {Werner}}]{Gross2012}%
  \BibitemOpen
  \bibfield  {author} {\bibinfo {author} {\bibfnamefont {D.}~\bibnamefont
  {Gross}}, \bibinfo {author} {\bibfnamefont {V.}~\bibnamefont {Nesme}},
  \bibinfo {author} {\bibfnamefont {H.}~\bibnamefont {Vogts}}, \ and\ \bibinfo
  {author} {\bibfnamefont {R.~F.}\ \bibnamefont {Werner}},\ }\href {\doibase
  10.1007/s00220-012-1423-1} {\bibfield  {journal} {\bibinfo  {journal}
  {Commun. Math. Phys.}\ }\textbf {\bibinfo {volume} {310}},\ \bibinfo {pages}
  {419} (\bibinfo {year} {2012})}\BibitemShut {NoStop}%
\bibitem [{\citenamefont {Cirac}\ \emph {et~al.}(2017)\citenamefont {Cirac},
  \citenamefont {P\'{e}rez-Garc\'{i}a}, \citenamefont {Schuch},\ and\
  \citenamefont {Verstraete}}]{Cirac2017}%
  \BibitemOpen
  \bibfield  {author} {\bibinfo {author} {\bibfnamefont {J.~I.}\ \bibnamefont
  {Cirac}}, \bibinfo {author} {\bibfnamefont {D.}~\bibnamefont
  {P\'{e}rez-Garc\'{i}a}}, \bibinfo {author} {\bibfnamefont {N.}~\bibnamefont
  {Schuch}}, \ and\ \bibinfo {author} {\bibfnamefont {F.}~\bibnamefont
  {Verstraete}},\ }\href {http://stacks.iop.org/1742-5468/2017/i=8/a=083105}
  {\bibfield  {journal} {\bibinfo  {journal} {J. Stat. Mech.}\ ,\ \bibinfo
  {pages} {083105}} (\bibinfo {year} {2017})}\BibitemShut {NoStop}%
\bibitem [{\citenamefont {\ifmmode \mbox{\c{S}}\else
  \c{S}\fi{}ahino\ifmmode~\breve{g}\else \u{g}\fi{}lu}\ \emph
  {et~al.}(2018)\citenamefont {\ifmmode \mbox{\c{S}}\else
  \c{S}\fi{}ahino\ifmmode~\breve{g}\else \u{g}\fi{}lu}, \citenamefont {Shukla},
  \citenamefont {Bi},\ and\ \citenamefont {Chen}}]{Chen2018}%
  \BibitemOpen
  \bibfield  {author} {\bibinfo {author} {\bibfnamefont {M.~B.}\ \bibnamefont
  {\ifmmode \mbox{\c{S}}\else \c{S}\fi{}ahino\ifmmode~\breve{g}\else
  \u{g}\fi{}lu}}, \bibinfo {author} {\bibfnamefont {S.~K.}\ \bibnamefont
  {Shukla}}, \bibinfo {author} {\bibfnamefont {F.}~\bibnamefont {Bi}}, \ and\
  \bibinfo {author} {\bibfnamefont {X.}~\bibnamefont {Chen}},\ }\href {\doibase
  10.1103/PhysRevB.98.245122} {\bibfield  {journal} {\bibinfo  {journal} {Phys.
  Rev. B}\ }\textbf {\bibinfo {volume} {98}},\ \bibinfo {pages} {245122}
  (\bibinfo {year} {2018})}\BibitemShut {NoStop}%
\bibitem [{\citenamefont {Gong}\ \emph {et~al.}(2020)\citenamefont {Gong},
  \citenamefont {S\"underhauf}, \citenamefont {Schuch},\ and\ \citenamefont
  {Cirac}}]{Gong2020}%
  \BibitemOpen
  \bibfield  {author} {\bibinfo {author} {\bibfnamefont {Z.}~\bibnamefont
  {Gong}}, \bibinfo {author} {\bibfnamefont {C.}~\bibnamefont {S\"underhauf}},
  \bibinfo {author} {\bibfnamefont {N.}~\bibnamefont {Schuch}}, \ and\ \bibinfo
  {author} {\bibfnamefont {J.~I.}\ \bibnamefont {Cirac}},\ }\href {\doibase
  10.1103/PhysRevLett.124.100402} {\bibfield  {journal} {\bibinfo  {journal}
  {Phys. Rev. Lett.}\ }\textbf {\bibinfo {volume} {124}},\ \bibinfo {pages}
  {100402} (\bibinfo {year} {2020})}\BibitemShut {NoStop}%
\bibitem [{\citenamefont {Piroli}\ \emph {et~al.}(2021)\citenamefont {Piroli},
  \citenamefont {Turzillo}, \citenamefont {Shukla},\ and\ \citenamefont
  {Cirac}}]{Piroli2021}%
  \BibitemOpen
  \bibfield  {author} {\bibinfo {author} {\bibfnamefont {L.}~\bibnamefont
  {Piroli}}, \bibinfo {author} {\bibfnamefont {A.}~\bibnamefont {Turzillo}},
  \bibinfo {author} {\bibfnamefont {S.~K.}\ \bibnamefont {Shukla}}, \ and\
  \bibinfo {author} {\bibfnamefont {J.~I.}\ \bibnamefont {Cirac}},\ }\href
  {\doibase 10.1088/1742-5468/abd30f} {\bibfield  {journal} {\bibinfo
  {journal} {J. Stat. Mech.}\ }\textbf {\bibinfo {volume} {2021}},\ \bibinfo
  {pages} {013107} (\bibinfo {year} {2021})}\BibitemShut {NoStop}%
\bibitem [{\citenamefont {Ranard}\ \emph {et~al.}(2020)\citenamefont {Ranard},
  \citenamefont {Walter},\ and\ \citenamefont {Witteveen}}]{Ranard2020}%
  \BibitemOpen
  \bibfield  {author} {\bibinfo {author} {\bibfnamefont {D.}~\bibnamefont
  {Ranard}}, \bibinfo {author} {\bibfnamefont {M.}~\bibnamefont {Walter}}, \
  and\ \bibinfo {author} {\bibfnamefont {F.}~\bibnamefont {Witteveen}},\
  }\href@noop {} {\enquote {\bibinfo {title} {A converse to lieb-robinson
  bounds in one dimension using index theory},}\ } (\bibinfo {year} {2020}),\
  \bibinfo {note} {arXiv:2012.00741}\BibitemShut {NoStop}%
\bibitem [{\citenamefont {Bols}(2021)}]{Bols2021}%
  \BibitemOpen
  \bibfield  {author} {\bibinfo {author} {\bibfnamefont {A.}~\bibnamefont
  {Bols}},\ }\href@noop {} {\enquote {\bibinfo {title} {Classification of
  equivariant approximately locality-preserving unitaries on spin chains},}\ }
  (\bibinfo {year} {2021}),\ \bibinfo {note} {arXiv:2106.02145}\BibitemShut
  {NoStop}%
\bibitem [{\citenamefont {Schumacher}\ and\ \citenamefont
  {Werner}(2004)}]{Schumacher2004}%
  \BibitemOpen
  \bibfield  {author} {\bibinfo {author} {\bibfnamefont {B.}~\bibnamefont
  {Schumacher}}\ and\ \bibinfo {author} {\bibfnamefont {R.~F.}\ \bibnamefont
  {Werner}},\ }\href@noop {} {\bibfield  {journal} {\bibinfo  {journal} {arXiv
  quant-ph/0405174}\ } (\bibinfo {year} {2004})}\BibitemShut {NoStop}%
\bibitem [{\citenamefont {Arrighi}(2019)}]{Arrighi2019}%
  \BibitemOpen
  \bibfield  {author} {\bibinfo {author} {\bibfnamefont {P.}~\bibnamefont
  {Arrighi}},\ }\href {\doibase 10.1007/s11047-019-09762-6} {\bibfield
  {journal} {\bibinfo  {journal} {Natural Comp.}\ }\textbf {\bibinfo {volume}
  {18}},\ \bibinfo {pages} {885} (\bibinfo {year} {2019})}\BibitemShut
  {NoStop}%
\bibitem [{\citenamefont {Farrelly}(2020)}]{Farrelly2020}%
  \BibitemOpen
  \bibfield  {author} {\bibinfo {author} {\bibfnamefont {T.}~\bibnamefont
  {Farrelly}},\ }\href {\doibase 10.22331/q-2020-11-30-368} {\bibfield
  {journal} {\bibinfo  {journal} {{Quantum}}\ }\textbf {\bibinfo {volume}
  {4}},\ \bibinfo {pages} {368} (\bibinfo {year} {2020})}\BibitemShut {NoStop}%
\bibitem [{\citenamefont {Kitagawa}\ \emph {et~al.}(2010)\citenamefont
  {Kitagawa}, \citenamefont {Berg}, \citenamefont {Rudner},\ and\ \citenamefont
  {Demler}}]{Kitagawa2010}%
  \BibitemOpen
  \bibfield  {author} {\bibinfo {author} {\bibfnamefont {T.}~\bibnamefont
  {Kitagawa}}, \bibinfo {author} {\bibfnamefont {E.}~\bibnamefont {Berg}},
  \bibinfo {author} {\bibfnamefont {M.}~\bibnamefont {Rudner}}, \ and\ \bibinfo
  {author} {\bibfnamefont {E.}~\bibnamefont {Demler}},\ }\href {\doibase
  10.1103/PhysRevB.82.235114} {\bibfield  {journal} {\bibinfo  {journal} {Phys.
  Rev. B}\ }\textbf {\bibinfo {volume} {82}},\ \bibinfo {pages} {235114}
  (\bibinfo {year} {2010})}\BibitemShut {NoStop}%
\bibitem [{\citenamefont {Jiang}\ \emph {et~al.}(2011)\citenamefont {Jiang},
  \citenamefont {Kitagawa}, \citenamefont {Alicea}, \citenamefont {Akhmerov},
  \citenamefont {Pekker}, \citenamefont {Refael}, \citenamefont {Cirac},
  \citenamefont {Demler}, \citenamefont {Lukin},\ and\ \citenamefont
  {Zoller}}]{Jiang2011}%
  \BibitemOpen
  \bibfield  {author} {\bibinfo {author} {\bibfnamefont {L.}~\bibnamefont
  {Jiang}}, \bibinfo {author} {\bibfnamefont {T.}~\bibnamefont {Kitagawa}},
  \bibinfo {author} {\bibfnamefont {J.}~\bibnamefont {Alicea}}, \bibinfo
  {author} {\bibfnamefont {A.~R.}\ \bibnamefont {Akhmerov}}, \bibinfo {author}
  {\bibfnamefont {D.}~\bibnamefont {Pekker}}, \bibinfo {author} {\bibfnamefont
  {G.}~\bibnamefont {Refael}}, \bibinfo {author} {\bibfnamefont {J.~I.}\
  \bibnamefont {Cirac}}, \bibinfo {author} {\bibfnamefont {E.}~\bibnamefont
  {Demler}}, \bibinfo {author} {\bibfnamefont {M.~D.}\ \bibnamefont {Lukin}}, \
  and\ \bibinfo {author} {\bibfnamefont {P.}~\bibnamefont {Zoller}},\ }\href
  {\doibase 10.1103/PhysRevLett.106.220402} {\bibfield  {journal} {\bibinfo
  {journal} {Phys. Rev. Lett.}\ }\textbf {\bibinfo {volume} {106}},\ \bibinfo
  {pages} {220402} (\bibinfo {year} {2011})}\BibitemShut {NoStop}%
\bibitem [{\citenamefont {Rudner}\ \emph {et~al.}(2013)\citenamefont {Rudner},
  \citenamefont {Lindner}, \citenamefont {Berg},\ and\ \citenamefont
  {Levin}}]{Lindner2013}%
  \BibitemOpen
  \bibfield  {author} {\bibinfo {author} {\bibfnamefont {M.~S.}\ \bibnamefont
  {Rudner}}, \bibinfo {author} {\bibfnamefont {N.~H.}\ \bibnamefont {Lindner}},
  \bibinfo {author} {\bibfnamefont {E.}~\bibnamefont {Berg}}, \ and\ \bibinfo
  {author} {\bibfnamefont {M.}~\bibnamefont {Levin}},\ }\href {\doibase
  10.1103/PhysRevX.3.031005} {\bibfield  {journal} {\bibinfo  {journal} {Phys.
  Rev. X}\ }\textbf {\bibinfo {volume} {3}},\ \bibinfo {pages} {031005}
  (\bibinfo {year} {2013})}\BibitemShut {NoStop}%
\bibitem [{\citenamefont {Else}\ and\ \citenamefont {Nayak}(2016)}]{Else2016}%
  \BibitemOpen
  \bibfield  {author} {\bibinfo {author} {\bibfnamefont {D.~V.}\ \bibnamefont
  {Else}}\ and\ \bibinfo {author} {\bibfnamefont {C.}~\bibnamefont {Nayak}},\
  }\href {\doibase 10.1103/PhysRevB.93.201103} {\bibfield  {journal} {\bibinfo
  {journal} {Phys. Rev. B}\ }\textbf {\bibinfo {volume} {93}},\ \bibinfo
  {pages} {201103(R)} (\bibinfo {year} {2016})}\BibitemShut {NoStop}%
\bibitem [{\citenamefont {Po}\ \emph {et~al.}(2016)\citenamefont {Po},
  \citenamefont {Fidkowski}, \citenamefont {Morimoto}, \citenamefont {Potter},\
  and\ \citenamefont {Vishwanath}}]{Po2016}%
  \BibitemOpen
  \bibfield  {author} {\bibinfo {author} {\bibfnamefont {H.~C.}\ \bibnamefont
  {Po}}, \bibinfo {author} {\bibfnamefont {L.}~\bibnamefont {Fidkowski}},
  \bibinfo {author} {\bibfnamefont {T.}~\bibnamefont {Morimoto}}, \bibinfo
  {author} {\bibfnamefont {A.~C.}\ \bibnamefont {Potter}}, \ and\ \bibinfo
  {author} {\bibfnamefont {A.}~\bibnamefont {Vishwanath}},\ }\href {\doibase
  10.1103/PhysRevX.6.041070} {\bibfield  {journal} {\bibinfo  {journal} {Phys.
  Rev. X}\ }\textbf {\bibinfo {volume} {6}},\ \bibinfo {pages} {041070}
  (\bibinfo {year} {2016})}\BibitemShut {NoStop}%
\bibitem [{\citenamefont {Potter}\ and\ \citenamefont
  {Morimoto}(2017)}]{Potter2017}%
  \BibitemOpen
  \bibfield  {author} {\bibinfo {author} {\bibfnamefont {A.~C.}\ \bibnamefont
  {Potter}}\ and\ \bibinfo {author} {\bibfnamefont {T.}~\bibnamefont
  {Morimoto}},\ }\href {\doibase 10.1103/PhysRevB.95.155126} {\bibfield
  {journal} {\bibinfo  {journal} {Phys. Rev. B}\ }\textbf {\bibinfo {volume}
  {95}},\ \bibinfo {pages} {155126} (\bibinfo {year} {2017})}\BibitemShut
  {NoStop}%
\bibitem [{\citenamefont {Harper}\ and\ \citenamefont
  {Roy}(2017)}]{Harper2017b}%
  \BibitemOpen
  \bibfield  {author} {\bibinfo {author} {\bibfnamefont {F.}~\bibnamefont
  {Harper}}\ and\ \bibinfo {author} {\bibfnamefont {R.}~\bibnamefont {Roy}},\
  }\href {\doibase 10.1103/PhysRevLett.118.115301} {\bibfield  {journal}
  {\bibinfo  {journal} {Phys. Rev. Lett.}\ }\textbf {\bibinfo {volume} {118}},\
  \bibinfo {pages} {115301} (\bibinfo {year} {2017})}\BibitemShut {NoStop}%
\bibitem [{\citenamefont {Roy}\ and\ \citenamefont
  {Harper}(2017{\natexlab{a}})}]{Roy2017}%
  \BibitemOpen
  \bibfield  {author} {\bibinfo {author} {\bibfnamefont {R.}~\bibnamefont
  {Roy}}\ and\ \bibinfo {author} {\bibfnamefont {F.}~\bibnamefont {Harper}},\
  }\href {\doibase 10.1103/PhysRevB.95.195128} {\bibfield  {journal} {\bibinfo
  {journal} {Phys. Rev. B}\ }\textbf {\bibinfo {volume} {95}},\ \bibinfo
  {pages} {195128} (\bibinfo {year} {2017}{\natexlab{a}})}\BibitemShut
  {NoStop}%
\bibitem [{\citenamefont {Fidkowski}\ \emph {et~al.}(2019)\citenamefont
  {Fidkowski}, \citenamefont {Po}, \citenamefont {Potter},\ and\ \citenamefont
  {Vishwanath}}]{Fidkowski2019}%
  \BibitemOpen
  \bibfield  {author} {\bibinfo {author} {\bibfnamefont {L.}~\bibnamefont
  {Fidkowski}}, \bibinfo {author} {\bibfnamefont {H.~C.}\ \bibnamefont {Po}},
  \bibinfo {author} {\bibfnamefont {A.~C.}\ \bibnamefont {Potter}}, \ and\
  \bibinfo {author} {\bibfnamefont {A.}~\bibnamefont {Vishwanath}},\ }\href
  {\doibase 10.1103/PhysRevB.99.085115} {\bibfield  {journal} {\bibinfo
  {journal} {Phys. Rev. B}\ }\textbf {\bibinfo {volume} {99}},\ \bibinfo
  {pages} {085115} (\bibinfo {year} {2019})}\BibitemShut {NoStop}%
\bibitem [{\citenamefont {Zhang}\ and\ \citenamefont
  {Levin}(2021)}]{Zhang2021}%
  \BibitemOpen
  \bibfield  {author} {\bibinfo {author} {\bibfnamefont {C.}~\bibnamefont
  {Zhang}}\ and\ \bibinfo {author} {\bibfnamefont {M.}~\bibnamefont {Levin}},\
  }\href {\doibase 10.1103/PhysRevB.103.064302} {\bibfield  {journal} {\bibinfo
   {journal} {Phys. Rev. B}\ }\textbf {\bibinfo {volume} {103}},\ \bibinfo
  {pages} {064302} (\bibinfo {year} {2021})}\BibitemShut {NoStop}%
\bibitem [{\citenamefont {Fidkowski}(2010)}]{Fidkowski2010}%
  \BibitemOpen
  \bibfield  {author} {\bibinfo {author} {\bibfnamefont {L.}~\bibnamefont
  {Fidkowski}},\ }\href {\doibase 10.1103/PhysRevLett.104.130502} {\bibfield
  {journal} {\bibinfo  {journal} {Phys. Rev. Lett.}\ }\textbf {\bibinfo
  {volume} {104}},\ \bibinfo {pages} {130502} (\bibinfo {year}
  {2010})}\BibitemShut {NoStop}%
\bibitem [{\citenamefont {Wen}(2012)}]{Wen2012}%
  \BibitemOpen
  \bibfield  {author} {\bibinfo {author} {\bibfnamefont {X.-G.}\ \bibnamefont
  {Wen}},\ }\href {\doibase 10.1103/PhysRevB.85.085103} {\bibfield  {journal}
  {\bibinfo  {journal} {Phys. Rev. B}\ }\textbf {\bibinfo {volume} {85}},\
  \bibinfo {pages} {085103} (\bibinfo {year} {2012})}\BibitemShut {NoStop}%
\bibitem [{\citenamefont {Hastings}(2013)}]{Hastings2013}%
  \BibitemOpen
  \bibfield  {author} {\bibinfo {author} {\bibfnamefont {M.~B.}\ \bibnamefont
  {Hastings}},\ }\href {\doibase 10.1103/PhysRevB.88.165114} {\bibfield
  {journal} {\bibinfo  {journal} {Phys. Rev. B}\ }\textbf {\bibinfo {volume}
  {88}},\ \bibinfo {pages} {165114} (\bibinfo {year} {2013})}\BibitemShut
  {NoStop}%
\bibitem [{\citenamefont {McDonald}\ \emph {et~al.}(2018)\citenamefont
  {McDonald}, \citenamefont {Pereg-Barnea},\ and\ \citenamefont
  {Clerk}}]{Clerk2018}%
  \BibitemOpen
  \bibfield  {author} {\bibinfo {author} {\bibfnamefont {A.}~\bibnamefont
  {McDonald}}, \bibinfo {author} {\bibfnamefont {T.}~\bibnamefont
  {Pereg-Barnea}}, \ and\ \bibinfo {author} {\bibfnamefont {A.~A.}\
  \bibnamefont {Clerk}},\ }\href {\doibase 10.1103/PhysRevX.8.041031}
  {\bibfield  {journal} {\bibinfo  {journal} {Phys. Rev. X}\ }\textbf {\bibinfo
  {volume} {8}},\ \bibinfo {pages} {041031} (\bibinfo {year}
  {2018})}\BibitemShut {NoStop}%
\bibitem [{\citenamefont {Ozawa}\ \emph {et~al.}(2019)\citenamefont {Ozawa},
  \citenamefont {Price}, \citenamefont {Amo}, \citenamefont {Goldman},
  \citenamefont {Hafezi}, \citenamefont {Lu}, \citenamefont {Rechtsman},
  \citenamefont {Schuster}, \citenamefont {Simon}, \citenamefont {Zilberberg},\
  and\ \citenamefont {Carusotto}}]{Ozawa2019}%
  \BibitemOpen
  \bibfield  {author} {\bibinfo {author} {\bibfnamefont {T.}~\bibnamefont
  {Ozawa}}, \bibinfo {author} {\bibfnamefont {H.~M.}\ \bibnamefont {Price}},
  \bibinfo {author} {\bibfnamefont {A.}~\bibnamefont {Amo}}, \bibinfo {author}
  {\bibfnamefont {N.}~\bibnamefont {Goldman}}, \bibinfo {author} {\bibfnamefont
  {M.}~\bibnamefont {Hafezi}}, \bibinfo {author} {\bibfnamefont
  {L.}~\bibnamefont {Lu}}, \bibinfo {author} {\bibfnamefont {M.~C.}\
  \bibnamefont {Rechtsman}}, \bibinfo {author} {\bibfnamefont {D.}~\bibnamefont
  {Schuster}}, \bibinfo {author} {\bibfnamefont {J.}~\bibnamefont {Simon}},
  \bibinfo {author} {\bibfnamefont {O.}~\bibnamefont {Zilberberg}}, \ and\
  \bibinfo {author} {\bibfnamefont {I.}~\bibnamefont {Carusotto}},\ }\href
  {\doibase 10.1103/RevModPhys.91.015006} {\bibfield  {journal} {\bibinfo
  {journal} {Rev. Mod. Phys.}\ }\textbf {\bibinfo {volume} {91}},\ \bibinfo
  {pages} {015006} (\bibinfo {year} {2019})}\BibitemShut {NoStop}%
\bibitem [{\citenamefont {Ohashi}\ \emph {et~al.}(2020)\citenamefont {Ohashi},
  \citenamefont {Kobayashi},\ and\ \citenamefont {Kawaguchi}}]{Ohashi2020}%
  \BibitemOpen
  \bibfield  {author} {\bibinfo {author} {\bibfnamefont {T.}~\bibnamefont
  {Ohashi}}, \bibinfo {author} {\bibfnamefont {S.}~\bibnamefont {Kobayashi}}, \
  and\ \bibinfo {author} {\bibfnamefont {Y.}~\bibnamefont {Kawaguchi}},\ }\href
  {\doibase 10.1103/PhysRevA.101.013625} {\bibfield  {journal} {\bibinfo
  {journal} {Phys. Rev. A}\ }\textbf {\bibinfo {volume} {101}},\ \bibinfo
  {pages} {013625} (\bibinfo {year} {2020})}\BibitemShut {NoStop}%
\bibitem [{\citenamefont {He}\ and\ \citenamefont {Chien}(2021)}]{He2021}%
  \BibitemOpen
  \bibfield  {author} {\bibinfo {author} {\bibfnamefont {Y.}~\bibnamefont
  {He}}\ and\ \bibinfo {author} {\bibfnamefont {C.-C.}\ \bibnamefont {Chien}},\
  }\href@noop {} {\enquote {\bibinfo {title} {Comparison of topological
  classifications of quadratic bosonic excitations with examples},}\ }
  (\bibinfo {year} {2021}),\ \bibinfo {note} {arXiv:2103.15200}\BibitemShut
  {NoStop}%
\bibitem [{\citenamefont {Chaudhary}\ \emph {et~al.}(2021)\citenamefont
  {Chaudhary}, \citenamefont {Levin},\ and\ \citenamefont {Clerk}}]{Clerk2021}%
  \BibitemOpen
  \bibfield  {author} {\bibinfo {author} {\bibfnamefont {G.}~\bibnamefont
  {Chaudhary}}, \bibinfo {author} {\bibfnamefont {M.}~\bibnamefont {Levin}}, \
  and\ \bibinfo {author} {\bibfnamefont {A.~A.}\ \bibnamefont {Clerk}},\ }\href
  {\doibase 10.1103/PhysRevB.103.214306} {\bibfield  {journal} {\bibinfo
  {journal} {Phys. Rev. B}\ }\textbf {\bibinfo {volume} {103}},\ \bibinfo
  {pages} {214306} (\bibinfo {year} {2021})}\BibitemShut {NoStop}%
\bibitem [{bG0()}]{bG0}%
  \BibitemOpen
  \href@noop {} {}\bibinfo {note} {Here we assume $\langle\Psi_{\rm
  b}|\xi_{\boldsymbol{r}S}|\Psi_{\rm b}\rangle=0$ $\forall \boldsymbol{r},S$
  for bGSs, which does not alter the topological classification \cite{SM}.
  Similarly, we drop the displacement for bGOs.}\BibitemShut {Stop}%
\bibitem [{\citenamefont {Ashida}\ \emph {et~al.}(2021)\citenamefont {Ashida},
  \citenamefont {Gong},\ and\ \citenamefont {Ueda}}]{Ashida2021}%
  \BibitemOpen
  \bibfield  {author} {\bibinfo {author} {\bibfnamefont {Y.}~\bibnamefont
  {Ashida}}, \bibinfo {author} {\bibfnamefont {Z.}~\bibnamefont {Gong}}, \ and\
  \bibinfo {author} {\bibfnamefont {M.}~\bibnamefont {Ueda}},\ }\href {\doibase
  10.1080/00018732.2021.1876991} {\bibfield  {journal} {\bibinfo  {journal}
  {Adv. Phys.}\ }\textbf {\bibinfo {volume} {69}},\ \bibinfo {pages} {249}
  (\bibinfo {year} {2021})}\BibitemShut {NoStop}%
\bibitem [{UUd()}]{UUd}%
  \BibitemOpen
  \href@noop {} {}\bibinfo {note} {More generally, we may require the
  superoperator $\hat U \cdot \hat U ^\dag$ to commute with $\hat U_{\rm
  s}\cdot \hat U_{\rm s}^\dag$, where $\hat U_{\rm s}$ is the symmetry
  operator. However, if $\hat U_{\rm s}$ is unitary and not traceless, we can
  show that the only possibility is $[\hat U ,\hat U_{\rm s}]=0$. Also, if
  $\hat U_{\rm s}$ is anti-unitary, we can again obtain the commutation
  relation by adding a phase into $\hat U $, which does not alter the
  physics.}\BibitemShut {Stop}%
\bibitem [{\citenamefont {Higashikawa}\ \emph {et~al.}(2019)\citenamefont
  {Higashikawa}, \citenamefont {Nakagawa},\ and\ \citenamefont
  {Ueda}}]{Nakagawa2019}%
  \BibitemOpen
  \bibfield  {author} {\bibinfo {author} {\bibfnamefont {S.}~\bibnamefont
  {Higashikawa}}, \bibinfo {author} {\bibfnamefont {M.}~\bibnamefont
  {Nakagawa}}, \ and\ \bibinfo {author} {\bibfnamefont {M.}~\bibnamefont
  {Ueda}},\ }\href {\doibase 10.1103/PhysRevLett.123.066403} {\bibfield
  {journal} {\bibinfo  {journal} {Phys. Rev. Lett.}\ }\textbf {\bibinfo
  {volume} {123}},\ \bibinfo {pages} {066403} (\bibinfo {year}
  {2019})}\BibitemShut {NoStop}%
\bibitem [{\citenamefont {McGinley}\ and\ \citenamefont
  {Cooper}(2018)}]{McGinley2018}%
  \BibitemOpen
  \bibfield  {author} {\bibinfo {author} {\bibfnamefont {M.}~\bibnamefont
  {McGinley}}\ and\ \bibinfo {author} {\bibfnamefont {N.~R.}\ \bibnamefont
  {Cooper}},\ }\href {\doibase 10.1103/PhysRevLett.121.090401} {\bibfield
  {journal} {\bibinfo  {journal} {Phys. Rev. Lett.}\ }\textbf {\bibinfo
  {volume} {121}},\ \bibinfo {pages} {090401} (\bibinfo {year}
  {2018})}\BibitemShut {NoStop}%
\bibitem [{\citenamefont {Whitehead}(1978)}]{Whitehead1978}%
  \BibitemOpen
  \bibfield  {author} {\bibinfo {author} {\bibfnamefont {G.~W.}\ \bibnamefont
  {Whitehead}},\ }\href@noop {} {\emph {\bibinfo {title} {Elements of Homotopy
  Theory}}}\ (\bibinfo  {publisher} {Springer, New York},\ \bibinfo {year}
  {1978})\BibitemShut {NoStop}%
\bibitem [{SM()}]{SM}%
  \BibitemOpen
  \href@noop {} {}\bibinfo {note} {See Supplemental Material for
  details.}\BibitemShut {Stop}%
\bibitem [{\citenamefont {Anshu}\ \emph {et~al.}(2018)\citenamefont {Anshu},
  \citenamefont {Hsieh},\ and\ \citenamefont {Jain}}]{Anshu2018}%
  \BibitemOpen
  \bibfield  {author} {\bibinfo {author} {\bibfnamefont {A.}~\bibnamefont
  {Anshu}}, \bibinfo {author} {\bibfnamefont {M.-H.}\ \bibnamefont {Hsieh}}, \
  and\ \bibinfo {author} {\bibfnamefont {R.}~\bibnamefont {Jain}},\ }\href
  {\doibase 10.1103/PhysRevLett.121.190504} {\bibfield  {journal} {\bibinfo
  {journal} {Phys. Rev. Lett.}\ }\textbf {\bibinfo {volume} {121}},\ \bibinfo
  {pages} {190504} (\bibinfo {year} {2018})}\BibitemShut {NoStop}%
\bibitem [{\citenamefont {Karoubi}(2008)}]{Karoubi2008}%
  \BibitemOpen
  \bibfield  {author} {\bibinfo {author} {\bibfnamefont {M.}~\bibnamefont
  {Karoubi}},\ }\href@noop {} {\emph {\bibinfo {title} {K-theory: An
  Introduction}}}\ (\bibinfo  {publisher} {Springer, Berlin},\ \bibinfo {year}
  {2008})\BibitemShut {NoStop}%
\bibitem [{tri()}]{tri}%
  \BibitemOpen
  \href@noop {} {}\bibinfo {note} {Here the triviality of GO can be either
  strict or weak, since in the latter case one can add more ancillas to make
  the extended GO strictly trivial.}\BibitemShut {Stop}%
\bibitem [{Note1()}]{Note1}%
  \BibitemOpen
  \bibinfo {note} {Note that $i\Gamma _{\protect \rm f}(\protect \boldsymbol
  {k})$ is Hermitian and therefore the same treatment can be applied to it as
  to a Hamiltonian~\cite {SM}}\BibitemShut {NoStop}%
\bibitem [{\citenamefont {Roy}\ and\ \citenamefont
  {Harper}(2017{\natexlab{b}})}]{Harper2017}%
  \BibitemOpen
  \bibfield  {author} {\bibinfo {author} {\bibfnamefont {R.}~\bibnamefont
  {Roy}}\ and\ \bibinfo {author} {\bibfnamefont {F.}~\bibnamefont {Harper}},\
  }\href {\doibase 10.1103/PhysRevB.96.155118} {\bibfield  {journal} {\bibinfo
  {journal} {Phys. Rev. B}\ }\textbf {\bibinfo {volume} {96}},\ \bibinfo
  {pages} {155118} (\bibinfo {year} {2017}{\natexlab{b}})}\BibitemShut
  {NoStop}%
\bibitem [{\citenamefont {Gong}\ \emph {et~al.}(2018)\citenamefont {Gong},
  \citenamefont {Ashida}, \citenamefont {Kawabata}, \citenamefont {Takasan},
  \citenamefont {Higashikawa},\ and\ \citenamefont {Ueda}}]{Gong2018}%
  \BibitemOpen
  \bibfield  {author} {\bibinfo {author} {\bibfnamefont {Z.}~\bibnamefont
  {Gong}}, \bibinfo {author} {\bibfnamefont {Y.}~\bibnamefont {Ashida}},
  \bibinfo {author} {\bibfnamefont {K.}~\bibnamefont {Kawabata}}, \bibinfo
  {author} {\bibfnamefont {K.}~\bibnamefont {Takasan}}, \bibinfo {author}
  {\bibfnamefont {S.}~\bibnamefont {Higashikawa}}, \ and\ \bibinfo {author}
  {\bibfnamefont {M.}~\bibnamefont {Ueda}},\ }\href {\doibase
  10.1103/PhysRevX.8.031079} {\bibfield  {journal} {\bibinfo  {journal} {Phys.
  Rev. X}\ }\textbf {\bibinfo {volume} {8}},\ \bibinfo {pages} {031079}
  (\bibinfo {year} {2018})}\BibitemShut {NoStop}%
\bibitem [{\citenamefont {Zhou}\ and\ \citenamefont {Lee}(2019)}]{Zhou2019}%
  \BibitemOpen
  \bibfield  {author} {\bibinfo {author} {\bibfnamefont {H.}~\bibnamefont
  {Zhou}}\ and\ \bibinfo {author} {\bibfnamefont {J.~Y.}\ \bibnamefont {Lee}},\
  }\href {\doibase 10.1103/PhysRevB.99.235112} {\bibfield  {journal} {\bibinfo
  {journal} {Phys. Rev. B}\ }\textbf {\bibinfo {volume} {99}},\ \bibinfo
  {pages} {235112} (\bibinfo {year} {2019})}\BibitemShut {NoStop}%
\bibitem [{\citenamefont {Kawabata}\ \emph {et~al.}(2019)\citenamefont
  {Kawabata}, \citenamefont {Shiozaki}, \citenamefont {Ueda},\ and\
  \citenamefont {Sato}}]{Kawabata2019}%
  \BibitemOpen
  \bibfield  {author} {\bibinfo {author} {\bibfnamefont {K.}~\bibnamefont
  {Kawabata}}, \bibinfo {author} {\bibfnamefont {K.}~\bibnamefont {Shiozaki}},
  \bibinfo {author} {\bibfnamefont {M.}~\bibnamefont {Ueda}}, \ and\ \bibinfo
  {author} {\bibfnamefont {M.}~\bibnamefont {Sato}},\ }\href {\doibase
  10.1103/PhysRevX.9.041015} {\bibfield  {journal} {\bibinfo  {journal} {Phys.
  Rev. X}\ }\textbf {\bibinfo {volume} {9}},\ \bibinfo {pages} {041015}
  (\bibinfo {year} {2019})}\BibitemShut {NoStop}%
\bibitem [{\citenamefont {Higham}(2008)}]{Higham2008}%
  \BibitemOpen
  \bibfield  {author} {\bibinfo {author} {\bibfnamefont {N.~J.}\ \bibnamefont
  {Higham}},\ }\href@noop {} {\emph {\bibinfo {title} {Functions of Matrices:
  Theory and Computation}}}\ (\bibinfo  {publisher} {SIAM, Philadelphia},\
  \bibinfo {year} {2008})\BibitemShut {NoStop}%
\bibitem [{\citenamefont {Haldane}\ and\ \citenamefont
  {Raghu}(2008)}]{Raghu2008}%
  \BibitemOpen
  \bibfield  {author} {\bibinfo {author} {\bibfnamefont {F.~D.~M.}\
  \bibnamefont {Haldane}}\ and\ \bibinfo {author} {\bibfnamefont
  {S.}~\bibnamefont {Raghu}},\ }\href {\doibase 10.1103/PhysRevLett.100.013904}
  {\bibfield  {journal} {\bibinfo  {journal} {Phys. Rev. Lett.}\ }\textbf
  {\bibinfo {volume} {100}},\ \bibinfo {pages} {013904} (\bibinfo {year}
  {2008})}\BibitemShut {NoStop}%
\bibitem [{\citenamefont {Hafezi}\ \emph {et~al.}(2011)\citenamefont {Hafezi},
  \citenamefont {Demler}, \citenamefont {Lukin},\ and\ \citenamefont
  {Taylor}}]{Hafezi2011}%
  \BibitemOpen
  \bibfield  {author} {\bibinfo {author} {\bibfnamefont {M.}~\bibnamefont
  {Hafezi}}, \bibinfo {author} {\bibfnamefont {E.~A.}\ \bibnamefont {Demler}},
  \bibinfo {author} {\bibfnamefont {M.~D.}\ \bibnamefont {Lukin}}, \ and\
  \bibinfo {author} {\bibfnamefont {J.~M.}\ \bibnamefont {Taylor}},\ }\href
  {\doibase 10.1038/nphys2063} {\bibfield  {journal} {\bibinfo  {journal} {Nat.
  Phys.}\ }\textbf {\bibinfo {volume} {7}},\ \bibinfo {pages} {907} (\bibinfo
  {year} {2011})}\BibitemShut {NoStop}%
\bibitem [{\citenamefont {Cooper}\ \emph {et~al.}(2019)\citenamefont {Cooper},
  \citenamefont {Dalibard},\ and\ \citenamefont {Spielman}}]{Cooper2019}%
  \BibitemOpen
  \bibfield  {author} {\bibinfo {author} {\bibfnamefont {N.~R.}\ \bibnamefont
  {Cooper}}, \bibinfo {author} {\bibfnamefont {J.}~\bibnamefont {Dalibard}}, \
  and\ \bibinfo {author} {\bibfnamefont {I.~B.}\ \bibnamefont {Spielman}},\
  }\href {\doibase 10.1103/RevModPhys.91.015005} {\bibfield  {journal}
  {\bibinfo  {journal} {Rev. Mod. Phys.}\ }\textbf {\bibinfo {volume} {91}},\
  \bibinfo {pages} {015005} (\bibinfo {year} {2019})}\BibitemShut {NoStop}%
\bibitem [{\citenamefont {Schuch}\ \emph {et~al.}(2006)\citenamefont {Schuch},
  \citenamefont {Cirac},\ and\ \citenamefont {Wolf}}]{Schuch2006}%
  \BibitemOpen
  \bibfield  {author} {\bibinfo {author} {\bibfnamefont {N.}~\bibnamefont
  {Schuch}}, \bibinfo {author} {\bibfnamefont {J.~I.}\ \bibnamefont {Cirac}}, \
  and\ \bibinfo {author} {\bibfnamefont {M.~M.}\ \bibnamefont {Wolf}},\ }\href
  {\doibase 10.1007/s00220-006-0049-6} {\bibfield  {journal} {\bibinfo
  {journal} {Commun. Math. Phys.}\ }\textbf {\bibinfo {volume} {267}},\
  \bibinfo {pages} {65} (\bibinfo {year} {2006})}\BibitemShut {NoStop}%
\bibitem [{\citenamefont {Sato}\ \emph {et~al.}(2009)\citenamefont {Sato},
  \citenamefont {Takahashi},\ and\ \citenamefont {Fujimoto}}]{Sato2009}%
  \BibitemOpen
  \bibfield  {author} {\bibinfo {author} {\bibfnamefont {M.}~\bibnamefont
  {Sato}}, \bibinfo {author} {\bibfnamefont {Y.}~\bibnamefont {Takahashi}}, \
  and\ \bibinfo {author} {\bibfnamefont {S.}~\bibnamefont {Fujimoto}},\ }\href
  {\doibase 10.1103/PhysRevLett.103.020401} {\bibfield  {journal} {\bibinfo
  {journal} {Phys. Rev. Lett.}\ }\textbf {\bibinfo {volume} {103}},\ \bibinfo
  {pages} {020401} (\bibinfo {year} {2009})}\BibitemShut {NoStop}%
\bibitem [{\citenamefont {Qi}\ \emph {et~al.}(2010)\citenamefont {Qi},
  \citenamefont {Hughes},\ and\ \citenamefont {Zhang}}]{Qi2010}%
  \BibitemOpen
  \bibfield  {author} {\bibinfo {author} {\bibfnamefont {X.-L.}\ \bibnamefont
  {Qi}}, \bibinfo {author} {\bibfnamefont {T.~L.}\ \bibnamefont {Hughes}}, \
  and\ \bibinfo {author} {\bibfnamefont {S.-C.}\ \bibnamefont {Zhang}},\ }\href
  {\doibase 10.1103/PhysRevB.82.184516} {\bibfield  {journal} {\bibinfo
  {journal} {Phys. Rev. B}\ }\textbf {\bibinfo {volume} {82}},\ \bibinfo
  {pages} {184516} (\bibinfo {year} {2010})}\BibitemShut {NoStop}%
\bibitem [{\citenamefont {Dyson}(1962)}]{Dyson1962}%
  \BibitemOpen
  \bibfield  {author} {\bibinfo {author} {\bibfnamefont {F.~J.}\ \bibnamefont
  {Dyson}},\ }\href {\doibase 10.1063/1.1703863} {\bibfield  {journal}
  {\bibinfo  {journal} {Journal of Mathematical Physics}\ }\textbf {\bibinfo
  {volume} {3}},\ \bibinfo {pages} {1199} (\bibinfo {year} {1962})}\BibitemShut
  {NoStop}%
\bibitem [{\citenamefont {Verbaarschot}(1994)}]{Verbaarschot1994}%
  \BibitemOpen
  \bibfield  {author} {\bibinfo {author} {\bibfnamefont {J.}~\bibnamefont
  {Verbaarschot}},\ }\href {\doibase 10.1103/PhysRevLett.72.2531} {\bibfield
  {journal} {\bibinfo  {journal} {Phys. Rev. Lett.}\ }\textbf {\bibinfo
  {volume} {72}},\ \bibinfo {pages} {2531} (\bibinfo {year}
  {1994})}\BibitemShut {NoStop}%
\bibitem [{\citenamefont {Brouder}\ \emph {et~al.}(2007)\citenamefont
  {Brouder}, \citenamefont {Panati}, \citenamefont {Calandra}, \citenamefont
  {Mourougane},\ and\ \citenamefont {Marzari}}]{Marzari2007}%
  \BibitemOpen
  \bibfield  {author} {\bibinfo {author} {\bibfnamefont {C.}~\bibnamefont
  {Brouder}}, \bibinfo {author} {\bibfnamefont {G.}~\bibnamefont {Panati}},
  \bibinfo {author} {\bibfnamefont {M.}~\bibnamefont {Calandra}}, \bibinfo
  {author} {\bibfnamefont {C.}~\bibnamefont {Mourougane}}, \ and\ \bibinfo
  {author} {\bibfnamefont {N.}~\bibnamefont {Marzari}},\ }\href {\doibase
  10.1103/PhysRevLett.98.046402} {\bibfield  {journal} {\bibinfo  {journal}
  {Phys. Rev. Lett.}\ }\textbf {\bibinfo {volume} {98}},\ \bibinfo {pages}
  {046402} (\bibinfo {year} {2007})}\BibitemShut {NoStop}%
\bibitem [{\citenamefont {Soluyanov}\ and\ \citenamefont
  {Vanderbilt}(2011)}]{Vanderbilt2011}%
  \BibitemOpen
  \bibfield  {author} {\bibinfo {author} {\bibfnamefont {A.~A.}\ \bibnamefont
  {Soluyanov}}\ and\ \bibinfo {author} {\bibfnamefont {D.}~\bibnamefont
  {Vanderbilt}},\ }\href {\doibase 10.1103/PhysRevB.83.035108} {\bibfield
  {journal} {\bibinfo  {journal} {Phys. Rev. B}\ }\textbf {\bibinfo {volume}
  {83}},\ \bibinfo {pages} {035108} (\bibinfo {year} {2011})}\BibitemShut
  {NoStop}%
\bibitem [{\citenamefont {Wen}(1990)}]{Wen1990}%
  \BibitemOpen
  \bibfield  {author} {\bibinfo {author} {\bibfnamefont {X.~G.}\ \bibnamefont
  {Wen}},\ }\href {\doibase 10.1142/S0217979290000139} {\bibfield  {journal}
  {\bibinfo  {journal} {Int. J. Mod. Phys. B}\ }\textbf {\bibinfo {volume}
  {04}},\ \bibinfo {pages} {239} (\bibinfo {year} {1990})}\BibitemShut
  {NoStop}%
\bibitem [{\citenamefont {Haldane}(1983)}]{Haldane1983}%
  \BibitemOpen
  \bibfield  {author} {\bibinfo {author} {\bibfnamefont {F.~D.~M.}\
  \bibnamefont {Haldane}},\ }\href {\doibase 10.1103/PhysRevLett.50.1153}
  {\bibfield  {journal} {\bibinfo  {journal} {Phys. Rev. Lett.}\ }\textbf
  {\bibinfo {volume} {50}},\ \bibinfo {pages} {1153} (\bibinfo {year}
  {1983})}\BibitemShut {NoStop}%
\bibitem [{\citenamefont {Mesaros}\ and\ \citenamefont {Ran}(2013)}]{Ran2013}%
  \BibitemOpen
  \bibfield  {author} {\bibinfo {author} {\bibfnamefont {A.}~\bibnamefont
  {Mesaros}}\ and\ \bibinfo {author} {\bibfnamefont {Y.}~\bibnamefont {Ran}},\
  }\href {\doibase 10.1103/PhysRevB.87.155115} {\bibfield  {journal} {\bibinfo
  {journal} {Phys. Rev. B}\ }\textbf {\bibinfo {volume} {87}},\ \bibinfo
  {pages} {155115} (\bibinfo {year} {2013})}\BibitemShut {NoStop}%
\bibitem [{sym()}]{symc}%
  \BibitemOpen
  \href@noop {} {}\bibinfo {note} {Note that we are actually abusing
  ``symmetry-entriched'' since intrinsic (Gaussian) topological order may
  sometimes be constrained rather than enriched.}\BibitemShut {Stop}%
\bibitem [{\citenamefont {Kitaev}(2001)}]{Kitaev2001}%
  \BibitemOpen
  \bibfield  {author} {\bibinfo {author} {\bibfnamefont {A.~Y.}\ \bibnamefont
  {Kitaev}},\ }\href {https://doi.org/10.1070/1063-7869/44/10S/S29} {\bibfield
  {journal} {\bibinfo  {journal} {Phys. Usp.}\ }\textbf {\bibinfo {volume}
  {44}},\ \bibinfo {pages} {131} (\bibinfo {year} {2001})}\BibitemShut
  {NoStop}%
\bibitem [{\citenamefont {Wen}(2017)}]{Wen2017}%
  \BibitemOpen
  \bibfield  {author} {\bibinfo {author} {\bibfnamefont {X.-G.}\ \bibnamefont
  {Wen}},\ }\href {\doibase 10.1103/RevModPhys.89.041004} {\bibfield  {journal}
  {\bibinfo  {journal} {Rev. Mod. Phys.}\ }\textbf {\bibinfo {volume} {89}},\
  \bibinfo {pages} {041004} (\bibinfo {year} {2017})}\BibitemShut {NoStop}%
\bibitem [{\citenamefont {Gu}\ and\ \citenamefont {Wen}(2014)}]{Gu2014}%
  \BibitemOpen
  \bibfield  {author} {\bibinfo {author} {\bibfnamefont {Z.-C.}\ \bibnamefont
  {Gu}}\ and\ \bibinfo {author} {\bibfnamefont {X.-G.}\ \bibnamefont {Wen}},\
  }\href {\doibase 10.1103/PhysRevB.90.115141} {\bibfield  {journal} {\bibinfo
  {journal} {Phys. Rev. B}\ }\textbf {\bibinfo {volume} {90}},\ \bibinfo
  {pages} {115141} (\bibinfo {year} {2014})}\BibitemShut {NoStop}%
\bibitem [{\citenamefont {Tantivasadakarn}\ and\ \citenamefont
  {Vishwanath}(2018)}]{Tantivasadakarn2018}%
  \BibitemOpen
  \bibfield  {author} {\bibinfo {author} {\bibfnamefont {N.}~\bibnamefont
  {Tantivasadakarn}}\ and\ \bibinfo {author} {\bibfnamefont {A.}~\bibnamefont
  {Vishwanath}},\ }\href {\doibase 10.1103/PhysRevB.98.165104} {\bibfield
  {journal} {\bibinfo  {journal} {Phys. Rev. B}\ }\textbf {\bibinfo {volume}
  {98}},\ \bibinfo {pages} {165104} (\bibinfo {year} {2018})}\BibitemShut
  {NoStop}%
\bibitem [{\citenamefont {Ellison}\ and\ \citenamefont
  {Fidkowski}(2019)}]{Ellison2019}%
  \BibitemOpen
  \bibfield  {author} {\bibinfo {author} {\bibfnamefont {T.~D.}\ \bibnamefont
  {Ellison}}\ and\ \bibinfo {author} {\bibfnamefont {L.}~\bibnamefont
  {Fidkowski}},\ }\href {\doibase 10.1103/PhysRevX.9.011016} {\bibfield
  {journal} {\bibinfo  {journal} {Phys. Rev. X}\ }\textbf {\bibinfo {volume}
  {9}},\ \bibinfo {pages} {011016} (\bibinfo {year} {2019})}\BibitemShut
  {NoStop}%
\bibitem [{\citenamefont {Chen}\ \emph {et~al.}(2021)\citenamefont {Chen},
  \citenamefont {Ellison},\ and\ \citenamefont {Tantivasadakarn}}]{Chen2021}%
  \BibitemOpen
  \bibfield  {author} {\bibinfo {author} {\bibfnamefont {Y.-A.}\ \bibnamefont
  {Chen}}, \bibinfo {author} {\bibfnamefont {T.~D.}\ \bibnamefont {Ellison}}, \
  and\ \bibinfo {author} {\bibfnamefont {N.}~\bibnamefont {Tantivasadakarn}},\
  }\href {\doibase 10.1103/PhysRevResearch.3.013056} {\bibfield  {journal}
  {\bibinfo  {journal} {Phys. Rev. Research}\ }\textbf {\bibinfo {volume}
  {3}},\ \bibinfo {pages} {013056} (\bibinfo {year} {2021})}\BibitemShut
  {NoStop}%
\bibitem [{\citenamefont {Kitaev}(2003)}]{Kitaev2003}%
  \BibitemOpen
  \bibfield  {author} {\bibinfo {author} {\bibfnamefont {A.}~\bibnamefont
  {Kitaev}},\ }\href {\doibase https://doi.org/10.1016/S0003-4916(02)00018-0}
  {\bibfield  {journal} {\bibinfo  {journal} {Ann. Phys.}\ }\textbf {\bibinfo
  {volume} {303}},\ \bibinfo {pages} {2} (\bibinfo {year} {2003})}\BibitemShut
  {NoStop}%
\bibitem [{\citenamefont {Freedman}\ and\ \citenamefont
  {Hastings}(2020)}]{Freedman2020}%
  \BibitemOpen
  \bibfield  {author} {\bibinfo {author} {\bibfnamefont {M.}~\bibnamefont
  {Freedman}}\ and\ \bibinfo {author} {\bibfnamefont {M.~B.}\ \bibnamefont
  {Hastings}},\ }\href {\doibase 10.1007/s00220-020-03735-y} {\bibfield
  {journal} {\bibinfo  {journal} {Comm. Math. Phys.}\ }\textbf {\bibinfo
  {volume} {376}},\ \bibinfo {pages} {1171} (\bibinfo {year}
  {2020})}\BibitemShut {NoStop}%
\bibitem [{\citenamefont {Haah}\ \emph {et~al.}(2018)\citenamefont {Haah},
  \citenamefont {Fidkowski},\ and\ \citenamefont {Hastings}}]{Haah2018}%
  \BibitemOpen
  \bibfield  {author} {\bibinfo {author} {\bibfnamefont {J.}~\bibnamefont
  {Haah}}, \bibinfo {author} {\bibfnamefont {L.}~\bibnamefont {Fidkowski}}, \
  and\ \bibinfo {author} {\bibfnamefont {M.~B.}\ \bibnamefont {Hastings}},\
  }\href@noop {} {\bibfield  {journal} {\bibinfo  {journal} {arXiv:1812.01625}\
  } (\bibinfo {year} {2018})}\BibitemShut {NoStop}%
\bibitem [{\citenamefont {Walker}\ and\ \citenamefont
  {Wang}(2012)}]{Walker2012}%
  \BibitemOpen
  \bibfield  {author} {\bibinfo {author} {\bibfnamefont {K.}~\bibnamefont
  {Walker}}\ and\ \bibinfo {author} {\bibfnamefont {Z.}~\bibnamefont {Wang}},\
  }\href {\doibase 10.1007/s11467-011-0194-z} {\bibfield  {journal} {\bibinfo
  {journal} {Front. Phys.}\ }\textbf {\bibinfo {volume} {7}},\ \bibinfo {pages}
  {150} (\bibinfo {year} {2012})}\BibitemShut {NoStop}%
\bibitem [{\citenamefont {Fu}(2011)}]{Fu2011}%
  \BibitemOpen
  \bibfield  {author} {\bibinfo {author} {\bibfnamefont {L.}~\bibnamefont
  {Fu}},\ }\href {\doibase 10.1103/PhysRevLett.106.106802} {\bibfield
  {journal} {\bibinfo  {journal} {Phys. Rev. Lett.}\ }\textbf {\bibinfo
  {volume} {106}},\ \bibinfo {pages} {106802} (\bibinfo {year}
  {2011})}\BibitemShut {NoStop}%
\bibitem [{\citenamefont {Slager}\ \emph {et~al.}(2013)\citenamefont {Slager},
  \citenamefont {Mesaros}, \citenamefont {Juri\v{c}i\'c},\ and\ \citenamefont
  {Zaanen}}]{Slager2013}%
  \BibitemOpen
  \bibfield  {author} {\bibinfo {author} {\bibfnamefont {R.-J.}\ \bibnamefont
  {Slager}}, \bibinfo {author} {\bibfnamefont {A.}~\bibnamefont {Mesaros}},
  \bibinfo {author} {\bibfnamefont {V.}~\bibnamefont {Juri\v{c}i\'c}}, \ and\
  \bibinfo {author} {\bibfnamefont {J.}~\bibnamefont {Zaanen}},\ }\href
  {http://dx.doi.org/10.1038/nphys2513} {\bibfield  {journal} {\bibinfo
  {journal} {Nat. Phys.}\ }\textbf {\bibinfo {volume} {9}},\ \bibinfo {pages}
  {98} (\bibinfo {year} {2013})}\BibitemShut {NoStop}%
\bibitem [{\citenamefont {Shiozaki}\ and\ \citenamefont
  {Sato}(2014)}]{Shiozaki2014}%
  \BibitemOpen
  \bibfield  {author} {\bibinfo {author} {\bibfnamefont {K.}~\bibnamefont
  {Shiozaki}}\ and\ \bibinfo {author} {\bibfnamefont {M.}~\bibnamefont
  {Sato}},\ }\href {\doibase 10.1103/PhysRevB.90.165114} {\bibfield  {journal}
  {\bibinfo  {journal} {Phys. Rev. B}\ }\textbf {\bibinfo {volume} {90}},\
  \bibinfo {pages} {165114} (\bibinfo {year} {2014})}\BibitemShut {NoStop}%
\bibitem [{\citenamefont {Kruthoff}\ \emph {et~al.}(2017)\citenamefont
  {Kruthoff}, \citenamefont {de~Boer}, \citenamefont {van Wezel}, \citenamefont
  {Kane},\ and\ \citenamefont {Slager}}]{Slager2017}%
  \BibitemOpen
  \bibfield  {author} {\bibinfo {author} {\bibfnamefont {J.}~\bibnamefont
  {Kruthoff}}, \bibinfo {author} {\bibfnamefont {J.}~\bibnamefont {de~Boer}},
  \bibinfo {author} {\bibfnamefont {J.}~\bibnamefont {van Wezel}}, \bibinfo
  {author} {\bibfnamefont {C.~L.}\ \bibnamefont {Kane}}, \ and\ \bibinfo
  {author} {\bibfnamefont {R.-J.}\ \bibnamefont {Slager}},\ }\href {\doibase
  10.1103/PhysRevX.7.041069} {\bibfield  {journal} {\bibinfo  {journal} {Phys.
  Rev. X}\ }\textbf {\bibinfo {volume} {7}},\ \bibinfo {pages} {041069}
  (\bibinfo {year} {2017})}\BibitemShut {NoStop}%
\bibitem [{\citenamefont {Khalaf}\ \emph {et~al.}(2018)\citenamefont {Khalaf},
  \citenamefont {Po}, \citenamefont {Vishwanath},\ and\ \citenamefont
  {Watanabe}}]{Khalaf2018}%
  \BibitemOpen
  \bibfield  {author} {\bibinfo {author} {\bibfnamefont {E.}~\bibnamefont
  {Khalaf}}, \bibinfo {author} {\bibfnamefont {H.~C.}\ \bibnamefont {Po}},
  \bibinfo {author} {\bibfnamefont {A.}~\bibnamefont {Vishwanath}}, \ and\
  \bibinfo {author} {\bibfnamefont {H.}~\bibnamefont {Watanabe}},\ }\href
  {\doibase 10.1103/PhysRevX.8.031070} {\bibfield  {journal} {\bibinfo
  {journal} {Phys. Rev. X}\ }\textbf {\bibinfo {volume} {8}},\ \bibinfo {pages}
  {031070} (\bibinfo {year} {2018})}\BibitemShut {NoStop}%
\bibitem [{\citenamefont {Benalcazar}\ \emph {et~al.}(2017)\citenamefont
  {Benalcazar}, \citenamefont {Bernevig},\ and\ \citenamefont
  {Hughes}}]{Benalcazar2017}%
  \BibitemOpen
  \bibfield  {author} {\bibinfo {author} {\bibfnamefont {W.~A.}\ \bibnamefont
  {Benalcazar}}, \bibinfo {author} {\bibfnamefont {B.~A.}\ \bibnamefont
  {Bernevig}}, \ and\ \bibinfo {author} {\bibfnamefont {T.~L.}\ \bibnamefont
  {Hughes}},\ }\href {\doibase 10.1126/science.aah6442} {\bibfield  {journal}
  {\bibinfo  {journal} {Science}\ }\textbf {\bibinfo {volume} {357}},\ \bibinfo
  {pages} {61} (\bibinfo {year} {2017})}\BibitemShut {NoStop}%
\bibitem [{\citenamefont {Schindler}\ \emph {et~al.}(2018)\citenamefont
  {Schindler}, \citenamefont {Cook}, \citenamefont {Vergniory}, \citenamefont
  {Wang}, \citenamefont {Parkin}, \citenamefont {Bernevig},\ and\ \citenamefont
  {Neupert}}]{Schindler2018}%
  \BibitemOpen
  \bibfield  {author} {\bibinfo {author} {\bibfnamefont {F.}~\bibnamefont
  {Schindler}}, \bibinfo {author} {\bibfnamefont {A.~M.}\ \bibnamefont {Cook}},
  \bibinfo {author} {\bibfnamefont {M.~G.}\ \bibnamefont {Vergniory}}, \bibinfo
  {author} {\bibfnamefont {Z.}~\bibnamefont {Wang}}, \bibinfo {author}
  {\bibfnamefont {S.~S.~P.}\ \bibnamefont {Parkin}}, \bibinfo {author}
  {\bibfnamefont {B.~A.}\ \bibnamefont {Bernevig}}, \ and\ \bibinfo {author}
  {\bibfnamefont {T.}~\bibnamefont {Neupert}},\ }\href {\doibase
  10.1126/sciadv.aat0346} {\bibfield  {journal} {\bibinfo  {journal} {Sci.
  Adv.}\ }\textbf {\bibinfo {volume} {4}} (\bibinfo {year} {2018}),\
  10.1126/sciadv.aat0346}\BibitemShut {NoStop}%
\bibitem [{\citenamefont {Trifunovic}\ and\ \citenamefont
  {Brouwer}(2019)}]{Trifunovic2019}%
  \BibitemOpen
  \bibfield  {author} {\bibinfo {author} {\bibfnamefont {L.}~\bibnamefont
  {Trifunovic}}\ and\ \bibinfo {author} {\bibfnamefont {P.~W.}\ \bibnamefont
  {Brouwer}},\ }\href {\doibase 10.1103/PhysRevX.9.011012} {\bibfield
  {journal} {\bibinfo  {journal} {Phys. Rev. X}\ }\textbf {\bibinfo {volume}
  {9}},\ \bibinfo {pages} {011012} (\bibinfo {year} {2019})}\BibitemShut
  {NoStop}%
\bibitem [{\citenamefont {Bardyn}\ \emph {et~al.}(2013)\citenamefont {Bardyn},
  \citenamefont {Baranov}, \citenamefont {Kraus}, \citenamefont {Rico},
  \citenamefont {\.Imamo\v{g}lu}, \citenamefont {Zoller},\ and\ \citenamefont
  {Diehl}}]{Diehl2013}%
  \BibitemOpen
  \bibfield  {author} {\bibinfo {author} {\bibfnamefont {C.-E.}\ \bibnamefont
  {Bardyn}}, \bibinfo {author} {\bibfnamefont {M.~A.}\ \bibnamefont {Baranov}},
  \bibinfo {author} {\bibfnamefont {C.~V.}\ \bibnamefont {Kraus}}, \bibinfo
  {author} {\bibfnamefont {E.}~\bibnamefont {Rico}}, \bibinfo {author}
  {\bibfnamefont {A.}~\bibnamefont {\.Imamo\v{g}lu}}, \bibinfo {author}
  {\bibfnamefont {P.}~\bibnamefont {Zoller}}, \ and\ \bibinfo {author}
  {\bibfnamefont {S.}~\bibnamefont {Diehl}},\ }\href
  {http://stacks.iop.org/1367-2630/15/i=8/a=085001} {\bibfield  {journal}
  {\bibinfo  {journal} {New J. Phys.}\ }\textbf {\bibinfo {volume} {15}},\
  \bibinfo {pages} {085001} (\bibinfo {year} {2013})}\BibitemShut {NoStop}%
\bibitem [{\citenamefont {Budich}\ and\ \citenamefont
  {Diehl}(2015)}]{Budich2015}%
  \BibitemOpen
  \bibfield  {author} {\bibinfo {author} {\bibfnamefont {J.~C.}\ \bibnamefont
  {Budich}}\ and\ \bibinfo {author} {\bibfnamefont {S.}~\bibnamefont {Diehl}},\
  }\href {\doibase 10.1103/PhysRevB.91.165140} {\bibfield  {journal} {\bibinfo
  {journal} {Phys. Rev. B}\ }\textbf {\bibinfo {volume} {91}},\ \bibinfo
  {pages} {165140} (\bibinfo {year} {2015})}\BibitemShut {NoStop}%
\bibitem [{\citenamefont {Mink}\ \emph {et~al.}(2019)\citenamefont {Mink},
  \citenamefont {Fleischhauer},\ and\ \citenamefont {Unanyan}}]{Mink2019}%
  \BibitemOpen
  \bibfield  {author} {\bibinfo {author} {\bibfnamefont {C.~D.}\ \bibnamefont
  {Mink}}, \bibinfo {author} {\bibfnamefont {M.}~\bibnamefont {Fleischhauer}},
  \ and\ \bibinfo {author} {\bibfnamefont {R.}~\bibnamefont {Unanyan}},\ }\href
  {\doibase 10.1103/PhysRevB.100.014305} {\bibfield  {journal} {\bibinfo
  {journal} {Phys. Rev. B}\ }\textbf {\bibinfo {volume} {100}},\ \bibinfo
  {pages} {014305} (\bibinfo {year} {2019})}\BibitemShut {NoStop}%
\bibitem [{\citenamefont {Altland}\ \emph {et~al.}(2021)\citenamefont
  {Altland}, \citenamefont {Fleischhauer},\ and\ \citenamefont
  {Diehl}}]{Altland2021}%
  \BibitemOpen
  \bibfield  {author} {\bibinfo {author} {\bibfnamefont {A.}~\bibnamefont
  {Altland}}, \bibinfo {author} {\bibfnamefont {M.}~\bibnamefont
  {Fleischhauer}}, \ and\ \bibinfo {author} {\bibfnamefont {S.}~\bibnamefont
  {Diehl}},\ }\href {\doibase 10.1103/PhysRevX.11.021037} {\bibfield  {journal}
  {\bibinfo  {journal} {Phys. Rev. X}\ }\textbf {\bibinfo {volume} {11}},\
  \bibinfo {pages} {021037} (\bibinfo {year} {2021})}\BibitemShut {NoStop}%
\bibitem [{\citenamefont {Shi}\ \emph {et~al.}(2018)\citenamefont {Shi},
  \citenamefont {Demler},\ and\ \citenamefont {Cirac}}]{Shi2018}%
  \BibitemOpen
  \bibfield  {author} {\bibinfo {author} {\bibfnamefont {T.}~\bibnamefont
  {Shi}}, \bibinfo {author} {\bibfnamefont {E.}~\bibnamefont {Demler}}, \ and\
  \bibinfo {author} {\bibfnamefont {J.~I.}\ \bibnamefont {Cirac}},\ }\href
  {\doibase https://doi.org/10.1016/j.aop.2017.11.014} {\bibfield  {journal}
  {\bibinfo  {journal} {Ann. Phys.}\ }\textbf {\bibinfo {volume} {390}},\
  \bibinfo {pages} {245} (\bibinfo {year} {2018})}\BibitemShut {NoStop}%
\bibitem [{\citenamefont {Guaita}\ \emph {et~al.}(2019)\citenamefont {Guaita},
  \citenamefont {Hackl}, \citenamefont {Shi}, \citenamefont {Hubig},
  \citenamefont {Demler},\ and\ \citenamefont {Cirac}}]{Guaita2019}%
  \BibitemOpen
  \bibfield  {author} {\bibinfo {author} {\bibfnamefont {T.}~\bibnamefont
  {Guaita}}, \bibinfo {author} {\bibfnamefont {L.}~\bibnamefont {Hackl}},
  \bibinfo {author} {\bibfnamefont {T.}~\bibnamefont {Shi}}, \bibinfo {author}
  {\bibfnamefont {C.}~\bibnamefont {Hubig}}, \bibinfo {author} {\bibfnamefont
  {E.}~\bibnamefont {Demler}}, \ and\ \bibinfo {author} {\bibfnamefont {J.~I.}\
  \bibnamefont {Cirac}},\ }\href {\doibase 10.1103/PhysRevB.100.094529}
  {\bibfield  {journal} {\bibinfo  {journal} {Phys. Rev. B}\ }\textbf {\bibinfo
  {volume} {100}},\ \bibinfo {pages} {094529} (\bibinfo {year}
  {2019})}\BibitemShut {NoStop}%
\bibitem [{\citenamefont {Hackl}\ \emph {et~al.}(2020)\citenamefont {Hackl},
  \citenamefont {Guaita}, \citenamefont {Shi}, \citenamefont {Haegeman},
  \citenamefont {Demler},\ and\ \citenamefont {Cirac}}]{Hackl2020}%
  \BibitemOpen
  \bibfield  {author} {\bibinfo {author} {\bibfnamefont {L.}~\bibnamefont
  {Hackl}}, \bibinfo {author} {\bibfnamefont {T.}~\bibnamefont {Guaita}},
  \bibinfo {author} {\bibfnamefont {T.}~\bibnamefont {Shi}}, \bibinfo {author}
  {\bibfnamefont {J.}~\bibnamefont {Haegeman}}, \bibinfo {author}
  {\bibfnamefont {E.}~\bibnamefont {Demler}}, \ and\ \bibinfo {author}
  {\bibfnamefont {J.~I.}\ \bibnamefont {Cirac}},\ }\href {\doibase
  10.21468/SciPostPhys.9.4.048} {\bibfield  {journal} {\bibinfo  {journal}
  {SciPost Phys.}\ }\textbf {\bibinfo {volume} {9}},\ \bibinfo {pages} {48}
  (\bibinfo {year} {2020})}\BibitemShut {NoStop}%
\bibitem [{Note2()}]{Note2}%
  \BibitemOpen
  \bibinfo {note} {A subtle point here is that $2\protect \mathbb {Z}$ number
  for topological fGSs might correspond to $\protect \mathbb {Z}$ number for
  topological fGOs, as is the case for class DIII in 3D and class CI in 7D.
  Nevertheless, the group structure is always isomorphic to $\protect \mathbb
  {Z}$.}\BibitemShut {Stop}%
\end{thebibliography}%

\clearpage
\begin{center}
\textbf{\large Supplemental Materials}
\end{center}
\setcounter{equation}{0}
\setcounter{figure}{0}
\setcounter{table}{0}
\makeatletter
\renewcommand{\theequation}{S\arabic{equation}}
\renewcommand{\thefigure}{S\arabic{figure}}
\renewcommand{\bibnumfmt}[1]{[S#1]}

We derive the explicit symmetry (either unitary or anti-unitary) constraints on the covariance matrices for GSs and the representation matrices for GOs. In particular, we confirm the consistency with the widely used Hamiltonian notation in the literature. We also provide the full periodic table for topological fGSs and fGOs, as well as the refined ones based on disentanglability.

\section{Symmetry constraints on and topological classifications of GSs and GOs}

\subsection{Unitary and anti-unitary symmetries}
A pure fGS is fully characterized by its covariance matrix
\begin{equation}
(\Gamma_{\rm f})_{jj'} = \frac{i}{2} \braket{\Psi_{\rm f}| [\hat\gamma_j,\hat\gamma_{j'}] |\Psi_{\rm f}} .
\end{equation}
We consider symmetry operators $\hat U_{\rm s}$ that we assume to be Gaussian unitaries or anti-unitaries, meaning that they transform the modes linearly:
\begin{equation}
\hat U_{\rm s}^\dag \hat{\gamma}_j \hat U_{\rm s} = \sum_{j'} (V_{\rm s})_{jj'} \hat{\gamma}_{j'} .
\end{equation}

Imposing the symmetry on a fGS corresponds to requiring $[\hat U_s, |\Psi_{\rm f}\rangle\langle\Psi_{\rm f}|]=0$. At the level of the covariance matrix this is equivalent to $i/2 \braket{\Psi_{\rm f}|\hat U^\dag_{\rm s} [\hat\gamma_j,\hat\gamma_{j'}] \hat U_{\rm s}|\Psi_{\rm f}} =i/2 \braket{\Psi_{\rm f}| [\hat\gamma_j,\hat\gamma_{j'}] |\Psi_{\rm f}}$ for unitary symmetries and  $i/2 \braket{\Psi_{\rm f}|\hat U^\dag_{\rm s} [\hat\gamma_j,\hat\gamma_{j'}] \hat U_{\rm s}|\Psi_{\rm f}}=i/2 \braket{\Psi_{\rm f}| [\hat\gamma_j,\hat\gamma_{j'}] |\Psi_{\rm f}}^*$ for anti-unitary symmetries. Considering that $\Gamma_{\rm f}$ is imaginary, we have
\begin{equation}
V_{\rm s}\Gamma_{\rm f} V_{\rm s}^\dag = \pm \Gamma_{\rm f}
\end{equation}
where the $+$ / $-$ holds for unitary / anti-unitary symmetries.

A pure bGS is fully characterized by its displacement vector
\begin{equation}
(\boldsymbol{\Delta}_{\rm b})_{j} = \braket{\Psi_{\rm b}| \hat\xi_j |\Psi_{\rm b}} ,
\end{equation}
and covariance matrix
\begin{equation}
(\Gamma_{\rm b})_{jj'} = \frac{1}{2} \braket{\Psi_{\rm b}| \{\delta\hat\xi_j,\delta\hat\xi_{j'}\} |\Psi_{\rm b}} ,
\end{equation}
where $\delta\hat\xi_j=\hat\xi_j-(\boldsymbol{\Delta}_{\rm b})_{j}$.

Invariance under the symmetry operators $\hat U_{\rm s}$, defined similarly to before, is given by
\begin{align}
    V_{\rm s} \boldsymbol{\Delta}_{\rm b} &= \boldsymbol{\Delta}_{\rm b} \\
    V_{\rm s}\Gamma_{\rm b} V_{\rm s}^\dag &= \Gamma_{\rm b} \,.
\end{align}
Due to the fact that $\boldsymbol{\Delta}_{\rm b}$ and $\Gamma_{\rm b}$ are real, in the bosonic case there is no difference between unitary and anti-unitary symmetries. 

One can continuously deform $\boldsymbol{\Delta}_{\rm b}$ into $\boldsymbol{0}$ in a symmetric manner simply via $\boldsymbol{\Delta}_{\rm b}(\lambda)=(1-\lambda)\boldsymbol{\Delta}_{\rm b}$. This is realized by on-site displacement operations and thus does not alter the short-range nature of the bGS. In what follows we therefore assume $\boldsymbol{\Delta}_{\rm b}=\boldsymbol{0}$ and in the case of bGOs we do not consider displacements.

Let us now consider the effect of imposing symmetries on a unitary GO $\hat U_{\rm f/b}$. We have to impose the commutation relation $[\hat U_{\rm f/b},\hat U_{\rm s}]=0$. This coincides with requiring 
\begin{equation}
    [V_{\rm f/b},V_{\rm s}]=0\,.
\end{equation}
As $V_{\rm f/b}$ is always real, there is no difference between unitary and anti-unitary symmetries. This result implies the consistency of symmetry constraint in Eq.~(\ref{Kfb}) in the main text, especially for the case of fermions and anti-unitary symmetries: while $\Gamma_{\rm f}$ anti-commutes with $V_{\rm s}$, $K_{\rm f}$ turns out to commute with $V_{\rm s}$ and so does $V_{\rm f}$.

We now review some unitary and anti-unitary symmetries. We define them in the case of fermions, but analogous results hold, where applicable, also for bosons, simply by replacing $\hat c_j$ with $\hat a_j$ and $\hat \gamma_j$ with $\hat \xi_j$.

\emph{${\rm U}(1)$ particle-number symmetry ---} In the case of the unitary particle-number symmetry $\hat U_{\rm s}=\hat \Phi$, for Gaussian states it is sufficient to impose the discrete subgroup $\mathbb{Z}_4$ generated by $\Phi: \hat c_{\boldsymbol{r}s}\mapsto -i \hat c_{\boldsymbol{r}s}, \quad \hat c^\dag_{\boldsymbol{r}s}\mapsto i \hat c^\dag_{\boldsymbol{r}s}$. In the Majorana representation this leads to
\begin{equation}
\hat\Phi^\dag\hat{\boldsymbol{\gamma}}_{\boldsymbol{r}s} \hat\Phi = i\sigma_y \hat{\boldsymbol{\gamma}}_{\boldsymbol{r}s} \,,
\end{equation}
where $\hat{\boldsymbol{\gamma}}_{\boldsymbol{r}s}\equiv \left(\hat\gamma_{\boldsymbol{r}+s},\hat\gamma_{\boldsymbol{r}-s}\right)^{\rm T}$.

\emph{${\rm SU}(2)$ spin-rotation symmetry ---} In the case of the unitary spin rotation symmetry $\hat U_{\rm s}=\hat R$, for Gaussian states it is sufficient to impose the $\pi$-rotation symmetry along $x$ and $z$ directions:
\begin{equation}
\hat R_{x,z}^\dag \begin{pmatrix} \hat c^\dag_{\boldsymbol{r}\uparrow\tilde s} \\ \hat c^\dag_{\boldsymbol{r}\downarrow\tilde s} \end{pmatrix} \hat R_{x,z} = i\sigma_{x,z} \begin{pmatrix} \hat c^\dag_{\boldsymbol{r}\uparrow\tilde s} \\ \hat c^\dag_{\boldsymbol{r}\downarrow\tilde s} \end{pmatrix}.
\end{equation}
In the Majorana representation this leads to
\begin{equation}
\hat R_{x,z}^\dag\hat{\boldsymbol{\gamma}}_{\boldsymbol{r}\tilde s} \hat R_{x,z} = (i\sigma_y\otimes \sigma_{x,z})\,\hat{\boldsymbol{\gamma}}_{\boldsymbol{r}\tilde s}\,,
\end{equation}
where $\hat{\boldsymbol{\gamma}}_{\boldsymbol{r}\tilde s}\equiv \left(\hat\gamma_{\boldsymbol{r}+\uparrow\tilde s},\hat\gamma_{\boldsymbol{r}+\downarrow\tilde s},\hat\gamma_{\boldsymbol{r}-\uparrow\tilde s},\hat\gamma_{\boldsymbol{r}-\downarrow\tilde s}\right)^{\rm T}$ and $\tilde s$ indicates the internal degrees of freedom other than spin.

\emph{Time-reversal symmetry ---} In the case of spinless fermions we can always find a basis such that the anti-unitary time-reversal symmetry (TRS) $\hat U_{\rm s}=\hat \Theta$ leaves the operators $\hat c_{\boldsymbol{r}s}$ invariant. That is, in the Majorana basis,
\begin{equation}
\hat\Theta^\dag \hat{\boldsymbol{\gamma}}_{\boldsymbol{r}s} \hat\Theta = \sigma_z\, \hat{\boldsymbol{\gamma}}_{\boldsymbol{r}s}.
\end{equation}

In the case of spin-$1/2$ fermions, the TRS additionally flips the spin according to
\begin{equation}
\hat \Theta^\dag \begin{pmatrix} \hat c^\dag_{\boldsymbol{r}\uparrow\tilde s} \\ \hat c^\dag_{\boldsymbol{r}\downarrow\tilde s} \end{pmatrix}  \hat \Theta = i\sigma_{y} \begin{pmatrix} \hat c^\dag_{\boldsymbol{r}\uparrow\tilde s} \\ \hat c^\dag_{\boldsymbol{r}\downarrow\tilde s} \end{pmatrix},
\end{equation}
and we therefore have
\begin{equation}
\hat\Theta^\dag\hat{\boldsymbol{\gamma}}_{\boldsymbol{r}\tilde s} \hat \Theta = (\sigma_z\otimes i\sigma_y)\hat{\boldsymbol{\gamma}}_{\boldsymbol{r}\tilde s}\,.
\end{equation}

In summary:
\begin{align}
    \mbox{Particle-number (U(1)):}& \quad V_\Phi = \openone_\Lambda \otimes i\sigma_y\otimes\openone_n  \nonumber\\
    \mbox{Spin-rotation (SU(2)):}& \quad V_R^{x,z} = \openone_\Lambda \otimes i\sigma_y\otimes \sigma_{x,z} \otimes\openone_{\tilde n} \nonumber\\
    \mbox{Spinless TRS:}& \quad V_\Theta = \openone_\Lambda\otimes\sigma_z \otimes\openone_n \nonumber\\
    \mbox{Spin-$1/2$ TRS:}& \quad V_\Theta = \openone_\Lambda\otimes \sigma_z\otimes i\sigma_y \otimes\openone_{\tilde n} \nonumber,
\end{align}
where $\tilde n= n/2$.

\subsection{Classification of states with Hamiltonian-based AZ classes}

We now show that the constraints deriving from imposing the physical symmetries discussed above on fermionic Gaussian states are equivalent to imposing certain emergent symmetries at the level of the matrix $i\Gamma$. These emergent symmetries can be understood in terms of the Hamiltonian-based AZ classes.

In the Hamiltonian-based formulation, symmetries can be classified as time-reversal, particle-hole or sub-lattice symmetries acting on Bloch/Bogoliubov-de Gennes (BdG) Hamiltonians in the following way:
\begin{equation}
\begin{split}
V_{\rm T} h(\boldsymbol{k})^* V_{\rm T}^\dag &= h(-\boldsymbol{k}),\;\;\;\; V_{\rm T}V_{\rm T}^*=\pm\openone,  \\
V_{\rm C} h(\boldsymbol{k})^* V_{\rm C}^\dag &= -h(-\boldsymbol{k}),\;\;\;\; V_{\rm C}V_{\rm C}^*=\pm\openone, \\
V_{\rm SL} h(\boldsymbol{k}) V_{\rm SL}^\dag &= -h(\boldsymbol{k}),\;\;\;\;V_{\rm SL}^2=\openone.
\end{split}
\label{TCSL}
\end{equation}
The different combinations of these symmetries, and whether they square to $+\openone$ (involutory) or to $-\openone$ (anti-involutory), lead to the ten AZ symmetry classes, as summarized in Table~\ref{tab:AZ}. Note that if the Hamiltonian satisfies both a TRS and a particle-hole symmetry (PHS), which always commute with each other, it will also satisfy a sub-lattice symmetry, which can be constructed out of the product of $V_{\rm T}$ and $V_{\rm C}$.

\begin{table}[h]
\caption{Altland-Zirnbauer (AZ) classes in terms of symmetries in the Hamiltonian-based formalism.}
\label{tab:AZ}
\begin{center}
\begin{tabular}{c|ccc|c}
\hline\hline
AZ & TRS & PHS & SLS & Classifying space \\
\hline
A    & 0 & 0 & 0 & $\mathcal{C}_0$ \\
AIII & 0 & 0 & 1 & $\mathcal{C}_1$ \\
\hline
AI   & $+$ & 0 & 0 & $\mathcal{R}_0$ \\
BDI  & $+$ & $+$ & 1 & $\mathcal{R}_1$ \\
D    & 0 & $+$ & 0 & $\mathcal{R}_2$ \\
DIII & $-$ & $+$ & 1 & $\mathcal{R}_3$ \\
AII  & $-$ & 0 & 0 & $\mathcal{R}_4$ \\
CII  & $-$ & $-$ & 1 & $\mathcal{R}_5$ \\
C    & 0 & $-$ & 0 & $\mathcal{R}_6$ \\
CI   & $+$ & $-$ & 1 & $\mathcal{R}_7$ \\
\hline\hline
\end{tabular}
\end{center}
\end{table}

For fermions, we have that the matrix $i\Gamma_{\rm f}(\boldsymbol{k})$ is Hermitian and involutory and can thus be regarded as a flattened Hamiltonian. It also has to fulfill the condition 
\begin{equation}
    [i\Gamma_{\rm f}(\boldsymbol{k})]^*=-i\Gamma_{\rm f}(-\boldsymbol{k}),
    \label{eq:supp-phs}
\end{equation}
which in the Hamiltonian-based formalism corresponds to an involutory PHS. In the case of no physical symmetries there no further constraints, leading to class D.

In the case of spinless TRS alone, we have the condition $\{\Gamma_{\rm f}(\boldsymbol{k}),\sigma_z\otimes\openone_n\}=0$. At the Hamiltonian-based level this can be seen as the result of $i\Gamma_{\rm f}(\boldsymbol{k})$ satisfying both the PHS~\eqref{eq:supp-phs} and the TRS
\begin{equation}
    (\sigma_z\otimes\openone_n) [i\Gamma_{\rm f}(\boldsymbol{k})]^* (\sigma_z\otimes\openone_n)=i\Gamma_{\rm f}(-\boldsymbol{k}),
    \label{BDITRS}
\end{equation}
both of which square to $+\openone$. This means that $i\Gamma_{\rm f}(\boldsymbol{k})$ belongs to class BDI. The general form of $i\Gamma_{\rm f}(\boldsymbol{k})$ reads 
\begin{equation}
i\Gamma_{\rm f}(\boldsymbol{k})=\begin{pmatrix} 0 & q(\boldsymbol{k}) \\ q(\boldsymbol{k})^\dag & 0 \end{pmatrix},
\label{BDIH}
\end{equation}
where $q(\boldsymbol{k})$ is a unitary satisfying $q(\boldsymbol{k})^*=q(-\boldsymbol{k})$.

In the case of spin-$1/2$ TRS alone, we have the condition $\{\Gamma_{\rm f}(\boldsymbol{k}),\sigma_z\otimes i\sigma_y\otimes\openone_{\tilde n}\}=0$. This can again be understood as 
the result of $i\Gamma_{\rm f}(\boldsymbol{k})$ satisfying both the PHS~\eqref{eq:supp-phs} and the TRS
\begin{equation}
    (\sigma_z\otimes i\sigma_y\otimes\openone_{\tilde n}) [i\Gamma_{\rm f}(\boldsymbol{k})]^* (\sigma_z\otimes i\sigma_y\otimes\openone_{\tilde n})^\dag=i\Gamma_{\rm f}(-\boldsymbol{k}),
    \label{eq:supp-spin-trs}
\end{equation}
where now the TRS squares to $-\openone$. This means that $i\Gamma_{\rm f}(\boldsymbol{k})$ belongs to class DIII. The general form of $i\Gamma_{\rm f}(\boldsymbol{k})$ reads
\begin{equation}
\mathbb{S}i\Gamma_{\rm f}(\boldsymbol{k})\mathbb{S}=\frac{1}{2}\begin{pmatrix} i[q(\boldsymbol{k})-q(\boldsymbol{k})^\dag] & [q(\boldsymbol{k})+q(\boldsymbol{k})^\dag]\sigma_z  \\ \sigma_z[q(\boldsymbol{k})+q(\boldsymbol{k})^\dag] & -i\sigma_z[q(\boldsymbol{k})-q(\boldsymbol{k})^\dag]\sigma_z \end{pmatrix},
\label{DIIIH}
\end{equation}
where $\sigma_z$ is a simplified notation for $\sigma_z\otimes\openone_{\tilde n}$ (this simplification will be adopted hereafter, sometimes with $\tilde n$ replaced by $n$ or $\sigma_\mu$ replaced by $\sigma_0\otimes\sigma_\mu$), $\mathbb{S}=(\sum_{\mu=0,x,y,z}\sigma_\mu\otimes\sigma_\mu)\otimes\openone_{\tilde n}/2$ swaps the Majorana and spin degrees of freedom and $q(\boldsymbol{k})$ is an $n\times n$ unitary satisfying $q(\boldsymbol{k})^{\rm T}=-q(-\boldsymbol{k})$.

In the case of ${\rm U}(1)$ particle-number symmetry alone, the condition $[\Gamma_{\rm f}(\boldsymbol{k}),i\sigma_y\otimes\openone_n]=0$, toghether with Eq.~\eqref{eq:supp-phs}, implies that $i\Gamma_{\rm f}(\boldsymbol{k})$ must have the form
\begin{align}
i\Gamma_{\rm f}(\boldsymbol{k})&=\frac{1}{2}\begin{pmatrix} h(\boldsymbol{k}) - h(-\boldsymbol{k})^* & ih(\boldsymbol{k}) + ih(-\boldsymbol{k})^* \\ - ih(\boldsymbol{k}) - ih(-\boldsymbol{k})^* & h(\boldsymbol{k}) - h(-\boldsymbol{k})^* \end{pmatrix} \nonumber\\
&= \frac{\sigma_0-\sigma_y}{2}\otimes h(\boldsymbol{k}) - \frac{\sigma_0+\sigma_y}{2}\otimes h(-\boldsymbol{k})^*\,,
\label{eq:supp-iGammma-as-function-of-H}
\end{align}
where $h(\boldsymbol{k})$ is a flat $n\times n$ Hermitian matrix with no symmetry constraint, which belongs to class A.

In case we impose ${\rm U}(1)$ particle-number symmetry and spinless TRS, then $i\Gamma_{\rm f}(\boldsymbol{k})$ will have to be of the form~\eqref{eq:supp-iGammma-as-function-of-H} with the additional constraint, coming from the TRS, that $h(\boldsymbol{k})$ satisfies
\begin{equation}
    h(\boldsymbol{k})^*=h(-\boldsymbol{k}).
    \label{eq:supp-trs}
\end{equation}
$h(\boldsymbol{k})$ is therefore a Hamiltonian satisfying an involutory TRS, leading to class AI.

In case we impose ${\rm U}(1)$ particle-number symmetry and spin-$1/2$ TRS, then $i\Gamma_{\rm f}(\boldsymbol{k})$ will have to be of the form~\eqref{eq:supp-iGammma-as-function-of-H} where now $h(\boldsymbol{k})$ satisfies
\begin{equation}
    (i\sigma_y\otimes\openone_{\tilde n})h(\boldsymbol{k})^*(i\sigma_y\otimes\openone_{\tilde n})^\dag=h(-\boldsymbol{k}).
    \label{AIIH}
\end{equation}
Therefore $h(\boldsymbol{k})$ now satisfies an anti-involutory TRS, leading to class AII.

In the case of ${\rm SU}(2)$ spin-rotation symmetry alone, the Hermitian and involutory matrix $i\Gamma_{\rm f}$ has to satisfy Eq.~\eqref{eq:supp-phs} and additionally $[i\Gamma_{\rm f}(\boldsymbol{k}),\sigma_y\otimes\sigma_x\otimes\openone_{\tilde n}]=0$ and $[i\Gamma_{\rm f}(\boldsymbol{k}),\sigma_y\otimes\sigma_z\otimes\openone_{\tilde n}]=0$. Imposing these constraints we obtain that $i\Gamma_{\rm f}$ should be of the form 
\begin{align}
&i\Gamma_{\rm f}(\boldsymbol{k})=\nonumber\\
&\frac{1}{2}\begin{pmatrix} h(\boldsymbol{k}) - h(-\boldsymbol{k})^* & i[h(\boldsymbol{k}) + h(-\boldsymbol{k})^*]\sigma_z \\ - i\sigma_z [h(\boldsymbol{k}) + h(-\boldsymbol{k})^*] & \sigma_z [h(\boldsymbol{k}) - h(-\boldsymbol{k})^*]\sigma_z \end{pmatrix},
\label{eq:supp-iGammma-as-function-of-Q}
\end{align}
where $h(\boldsymbol{k})$ is a flat $n\times n$ Hermitian matrix satisfying
\begin{equation}
    (i\sigma_y\otimes\openone_{\tilde n})h(\boldsymbol{k})^*(i\sigma_y\otimes\openone_{\tilde n})^\dag=-h(-\boldsymbol{k}).
    \label{eq:supp-anti-inv-phs}
\end{equation}
We therefore have a Hamiltonian satisfying an anti-involutory PHS, leading to class C.

In case we impose ${\rm SU}(2)$ spin-rotation symmetry and spin-$1/2$ TRS, we have that $i\Gamma_{\rm f}$ should be of the form~\eqref{eq:supp-iGammma-as-function-of-Q} where $h(\boldsymbol{k})$ satisfies the anti-involutory PHS~\eqref{eq:supp-anti-inv-phs}. Additionally the TRS implies that
\begin{equation}
    h(\boldsymbol{k})^*=h(-\boldsymbol{k}),
\end{equation}
which from the Hamiltonian-based point of view is an involutory TRS, leading to class CI. The general form of $h(\boldsymbol{k})$ reads 
\begin{equation}
h(\boldsymbol{k})=\frac{1}{2}\begin{pmatrix} \tilde q(\boldsymbol{k})+\tilde q(\boldsymbol{k})^\dag & i[\tilde q(\boldsymbol{k})-\tilde q(\boldsymbol{k})^\dag] \\ i[\tilde q(\boldsymbol{k})-\tilde q(\boldsymbol{k})^\dag] & -\tilde q(\boldsymbol{k})-\tilde q(\boldsymbol{k})^\dag \end{pmatrix},
\label{CIH}
\end{equation}
where $\tilde q(\boldsymbol{k})$ is an $\tilde n\times \tilde n$ unitary satisfying $\tilde q(\boldsymbol{k})^{\rm T}=\tilde q(-\boldsymbol{k})$.

If, on the other hand, we impose the ${\rm SU}(2)$ rotation symmetry on another internal degree of freedom different from the spin, then $i\Gamma_{\rm f}$ will be of the form~\eqref{eq:supp-iGammma-as-function-of-Q} (where $\sigma_z$ should be understood as $\sigma_0\otimes\sigma_z\otimes\openone_{\tilde n}$) with $h(\boldsymbol{k})$ satisfying the anti-involutory PHS~\eqref{eq:supp-anti-inv-phs}, where now $i\sigma_y$ acts on this other degree of freedom: 
\begin{equation}
    (\sigma_0\otimes i\sigma_y\otimes\openone_{\tilde n/2})h(\boldsymbol{k})^*(\sigma_0\otimes i\sigma_y\otimes\openone_{\tilde n/2})^\dag=-h(-\boldsymbol{k}).
\end{equation}
The TRS~\eqref{eq:supp-spin-trs} will then act on the spin indices of $h(\boldsymbol{k})$ as 
\begin{equation}
    (i\sigma_y\otimes\openone_{\tilde n})h(\boldsymbol{k})^*(i\sigma_y\otimes\openone_{\tilde n})^\dag=h(-\boldsymbol{k}),
\end{equation}
giving this time an anti-involutory TRS. We are then in class CII. The general form of $h(\boldsymbol{k})$ reads with respect to the spin indices
\begin{equation}
h(\boldsymbol{k})=\frac{1}{2}\begin{pmatrix}i[q(\boldsymbol{k})-q(\boldsymbol{k})^\dag] & -[q(\boldsymbol{k})+q(\boldsymbol{k})^\dag]\sigma_y \\
-\sigma_y[q(\boldsymbol{k})+q(\boldsymbol{k})^\dag] & -i\sigma_y [q(\boldsymbol{k})-q(\boldsymbol{k})^\dag]\sigma_y \end{pmatrix},
\label{CIIH}
\end{equation}
where $q(\boldsymbol{k})$ is a unitary satisfying $(i\sigma_y\otimes\openone_{\tilde n/2})q(\boldsymbol{k})^*(i\sigma_y\otimes\openone_{\tilde n/2})^\dag=q(-\boldsymbol{k})$. We emphasize again that here $i\sigma_y$ acts on the other degree of freedom.

Finally, if we consider a spin-$1/2$ system with TRS and we impose a further ${\rm U}(1)$ symmetry, namely a spin-rotation symmetry around the $z$-axis, then $i\Gamma_{\rm f}$ must have the form of Eq.~(\ref{DIIIH}), where $q(\boldsymbol{k})$ is given by
\begin{equation}
q(\boldsymbol{k})=\frac{\sigma_0-\sigma_y}{2}\otimes\tilde q(\boldsymbol{k}) - \frac{\sigma_0+\sigma_y}{2}\otimes\tilde q(-\boldsymbol{k})^{\rm T},
\label{AIIIH}
\end{equation}
and $\tilde q(\boldsymbol{k})$ is an arbitrary $\tilde n\times \tilde n$ unitary. Alternatively, $i\Gamma_{\rm f}$ can be generally expressed as Eq.~(\ref{eq:supp-iGammma-as-function-of-Q}), where $h(\boldsymbol{k})$ given by Eq.~(\ref{CIH}) with $\tilde q(\boldsymbol{k})$ being arbitrary. Arbitrary unitaries belong to class AIII.

\subsection{Classification of operations with Hamiltonian-based AZ classes}
We will use the Hermitianization technique to map the constraints on unitary representation matrices to conditions which, like before, can be classified in terms of the Hamiltonian-based AZ symmetry classes. In some cases we will be in situations where we have additional order 2 symmetries on top of the ones summarized in Table~\ref{tab:AZ}. To classify these cases we will use the formalism introduced in Ref.~\cite{Shiozaki2014}. The results will be double checked by the general forms of the representation matrices.

\subsubsection{fGOs}
We will first treat the case of fermions. In this case, $V_{\rm f}(\boldsymbol{k})$ is unitary and satisfies $V_{\rm f}(\boldsymbol{k})^*=V_{\rm f}(\boldsymbol{-k})$. Using the Hermitianization technique, we will consider
\begin{equation}
    X(\boldsymbol{k})=\begin{pmatrix} 0 & V_{\rm f}(\boldsymbol{k}) \\ V_{\rm f}(\boldsymbol{k})^\dag & 0 \end{pmatrix},
    \label{eq:supp-hermitianization}
\end{equation}
where we now have that $X(\boldsymbol{k})$ is Hermitian, involutory ($X^2=\openone$) and satisifies
\begin{align}
     X(\boldsymbol{k})^* &= X(-\boldsymbol{k}), \\
     (\sigma_z \otimes \openone_{2n}) X(\boldsymbol{k})^* (\sigma_z \otimes \openone_{2n}) &= -X(-\boldsymbol{k}), \label{eq:supp-X-phs}
\end{align}
where $2n$ is the dimension of $V_{\rm f}(\boldsymbol{k})$. We can therefore regard $X$ as a flat Hamiltonian satisfying involutory TRS and PHS. For this reason, fGOs with no physical symmetries fall into class BDI with classifying space $\mathcal{R}_1$.

In the case of spinless TRS alone we have the condition $[\sigma_z\otimes\openone_n,V_{\rm f}(\boldsymbol{k})]=0$, which at the level of the Hermitianized operator $X$ means
\begin{equation}
    (\sigma_0\otimes\sigma_z\otimes\openone_n) X(\boldsymbol{k}) (\sigma_0\otimes\sigma_z\otimes\openone_n) = X(\boldsymbol{k}).
\end{equation}
This is an additional unitary and involutory symmetry that commutes with both the TRS and PHS of $X$. We are therefore in the case $s=1$, $t=0$ ($U^+_{++}$) of Sec. III C of Ref.~\cite{Shiozaki2014}. This means the $K$-group is given by 
\begin{equation}
    K(s=1,t=0;d)=\pi_d(\mathcal{R}_1^2). 
\end{equation}
This result is consistent with the following general form:
\begin{equation}
V_{\rm f}(\boldsymbol{k})=\begin{pmatrix} u_1(\boldsymbol{k}) & 0 \\ 0 & u_2(\boldsymbol{k}) \end{pmatrix},
\label{BDIV}
\end{equation}
where $u_1(\boldsymbol{k})$ and $u_2(\boldsymbol{k})$ are two independent $n\times n$ unitaries satisfying $u_{1,2}(\boldsymbol{k})^*=u_{1,2}(-\boldsymbol{k})$.

In the case of spin-$1/2$ TRS alone we have the condition $[\sigma_z\otimes i\sigma_y\otimes\openone_{\tilde n},V_{\rm f}(\boldsymbol{k})]=0$, which at the level of the Hermitianized operator $X$ means
\begin{equation}
    (\sigma_0\otimes\sigma_z\otimes i\sigma_y\otimes\openone_{\tilde n}) X(\boldsymbol{k}) (\sigma_0\otimes\sigma_z\otimes i\sigma_y\otimes\openone_{\tilde n}) = X(\boldsymbol{k}).
\end{equation}
This is an additional unitary and anti-involutory symmetry that commutes with both the TRS and PHS of $X$. We are therefore in the case $s=1$, $t=2$ ($U^-_{++}$) of Sec. III C of Ref.~\cite{Shiozaki2014}. This means the $K$-group is given by 
\begin{equation}
    K(s=1,t=2;d)=\pi_d(\mathcal{C}_1).
\end{equation}
This result is consistent with the following general form:
\begin{align}
&\mathbb{S}V_{\rm f}(\boldsymbol{k})\mathbb{S}=\nonumber \\
&\frac{1}{2}\begin{pmatrix} u(\boldsymbol{k}) + u(-\boldsymbol{k})^* & i[u(\boldsymbol{k}) - u(-\boldsymbol{k})^*] \sigma_z \\ - i\sigma_z [u(\boldsymbol{k}) - u(-\boldsymbol{k})^*] & \sigma_z [u(\boldsymbol{k}) + u(-\boldsymbol{k})^*] \sigma_z \end{pmatrix} ,
\label{eq:supp-V-as-function-of-Utilde}
\end{align}
where $\mathbb{S}$ is the same as that in Eq.~(\ref{DIIIH}) and $u(\boldsymbol{k})$ is an arbitrary $n \times n$ unitary.

In the case of ${\rm U}(1)$ particle-number symmetry alone, we have the condition on $[V_{\rm f}(\boldsymbol{k}),i\sigma_y\otimes\openone_n]=0$. Similarly to the case of fGSs, this, combined with $V_{\rm f}(\boldsymbol{k})^*=V_{\rm f}(\boldsymbol{-k})$ implies that $V_{\rm f}(\boldsymbol{k})$ must have the form
\begin{align}
V_{\rm f}(\boldsymbol{k})&=\frac{1}{2}\begin{pmatrix} u(\boldsymbol{k}) + u(-\boldsymbol{k})^* & iu(\boldsymbol{k}) - iu(-\boldsymbol{k})^* \\ - iu(\boldsymbol{k}) + iu(-\boldsymbol{k})^* & u(\boldsymbol{k}) + u(-\boldsymbol{k})^* \end{pmatrix} \nonumber\\
&= \frac{\sigma_0-\sigma_y}{2}\otimes u(\boldsymbol{k}) + \frac{\sigma_0+\sigma_y}{2}\otimes u(-\boldsymbol{k})^*\,,
\label{eq:supp-V-as-function-of-U}
\end{align}
where $u(\boldsymbol{k})$ is an arbitrary $n\times n$ unitary matrix. Let us again reduce ourselves back to the Hamiltonian-based formalism through the Hermitianization technique. In this case we have to consider the object
\begin{equation}
    Y(\boldsymbol{k})=\begin{pmatrix} 0 & u(\boldsymbol{k}) \\ u(\boldsymbol{k})^\dag & 0 \end{pmatrix},
    \label{eq:supp-Y}
\end{equation}
which is an involutory Hermitian matrix satisfying
\begin{align}
     (\sigma_z \otimes \openone_{n}) Y(\boldsymbol{k}) (\sigma_z \otimes \openone_{n}) &= -Y(\boldsymbol{k}) \label{eq:supp-Y-chiral}.
\end{align}
This can be understood as an involutory sub-lattice symmetry. There are no further symmetry constraints and we are therefore in class AIII with classifying space $\mathcal{C}_1$. 

In case of ${\rm U}(1)$ particle-number symmetry and spinless TRS, then $V_{\rm f}(\boldsymbol{k})$ will still take the form~\eqref{eq:supp-V-as-function-of-U}, but now the TRS imposes the additional constraint
\begin{align}
     u(\boldsymbol{k})^* &= u(-\boldsymbol{k}).
     \label{AIV}
\end{align}
At the Hamiltonian level therefore we have the following symmetries on $Y(\boldsymbol{k})$:
\begin{align}
    (\sigma_z \otimes \openone_{n}) Y(\boldsymbol{k})^* (\sigma_z \otimes \openone_{n}) &= -Y(-\boldsymbol{k}),\\
    Y(\boldsymbol{k})^* &= Y(-\boldsymbol{k}),
\end{align}
which can be seen as involutory PHS and TRS respectively, leading to class BDI, with classifying space $\mathcal{R}_1$.

In case of ${\rm U}(1)$ particle-number symmetry and spin-$1/2$ TRS, we similarly find that $V_{\rm f}(\boldsymbol{k})$ takes the form~\eqref{eq:supp-V-as-function-of-U}, but now the TRS imposes the constraint
\begin{align}
     (i\sigma_y \otimes \openone_{\tilde n}) u(\boldsymbol{k})^* (i\sigma_y \otimes \openone_{\tilde n})^\dag &= u(-\boldsymbol{k}).
     \label{AIIV}
\end{align}
At the Hamiltonian level therefore we have
\begin{align}
    (\sigma_z \otimes i\sigma_y \otimes \openone_{\tilde n}) Y(\boldsymbol{k})^* (\sigma_z \otimes i\sigma_y \otimes \openone_{\tilde n})^\dag &= -Y(-\boldsymbol{k}), \label{eq:supp-phs-Y}\\
    (\sigma_0 \otimes i\sigma_y \otimes \openone_{\tilde n}) Y(\boldsymbol{k})^* (\sigma_0 \otimes i\sigma_y \otimes \openone_{\tilde n})^\dag &= Y(-\boldsymbol{k}),
     \label{eq:supp-trs-Y}
\end{align}
which are now anti-involutory PHS and TRS, leading to class CII with classifying space $\mathcal{R}_5$.

In the case of ${\rm SU}(2)$ spin-rotation symmetry alone, we have the constraints $[V_{\rm f}(\boldsymbol{k}),\sigma_y\otimes\sigma_x\otimes\openone_{\tilde n}]=0$ and $[V_{\rm f}(\boldsymbol{k}),\sigma_y\otimes\sigma_z\otimes\openone_{\tilde n}]=0$.
Similarly to the case of fGSs, this, combined with $V_{\rm f}(\boldsymbol{k})^*=V_{\rm f}(\boldsymbol{-k})$ implies that $V_{\rm f}(\boldsymbol{k})$ must have the form of the rhs of  Eq.~(\ref{eq:supp-V-as-function-of-Utilde}):
\begin{align}
&V_{\rm f}(\boldsymbol{k})=\nonumber \\
&\frac{1}{2}\begin{pmatrix} u(\boldsymbol{k}) + u(-\boldsymbol{k})^* & i[u(\boldsymbol{k}) - u(-\boldsymbol{k})^*] \sigma_z \\ - i\sigma_z [u(\boldsymbol{k}) - u(-\boldsymbol{k})^*] & \sigma_z [u(\boldsymbol{k}) + u(-\boldsymbol{k})^*] \sigma_z \end{pmatrix},
\label{CVI}
\end{align} 
where $u(\boldsymbol{k})$ is a unitary matrix, satisfying 
\begin{equation}
    (i\sigma_y \otimes \openone_{\tilde n}) u(\boldsymbol{k})^*(i\sigma_y \otimes \openone_{\tilde n})^\dag = u(-\boldsymbol{k}).
    \label{CV}
\end{equation}
This is the same situation as in the case of ${\rm U}(1)$ particle-number symmetry and spin-$1/2$ TRS, which can be seen as equivalent to imposing the anti-involutory PHS~\eqref{eq:supp-phs-Y} and TRS~\eqref{eq:supp-trs-Y} to the Hermitianized martix~\eqref{eq:supp-Y}. Therefore we are again in class CII with classifying space $\mathcal{R}_5$.

In case we impose ${\rm SU}(2)$ spin-rotation symmetry and spin-$1/2$ TRS, we have again that $V_{\rm f}(\boldsymbol{k})$ takes the form of Eq.~(\ref{CVI}) and that the Hermitianized matrix~\eqref{eq:supp-Y} satisfies the PHS~\eqref{eq:supp-phs-Y} and TRS~\eqref{eq:supp-trs-Y}. Additionally, the TRS implies that
\begin{equation}
    (\sigma_0\otimes i\sigma_y \otimes \openone_{\tilde n}) Y(\boldsymbol{k})(\sigma_0\otimes i\sigma_y \otimes \openone_{\tilde n})^\dag = Y(\boldsymbol{k}).
\end{equation}
This is an additional anti-involutory unitary symmetry that commutes with both the PHS and TRS of $Y(\boldsymbol{k})$. This means we are in the case $s=5$, $t=2$ ($U^-_{++}$) of Sec. III C of Ref.~\cite{Shiozaki2014}, where the $K$-group is given by
\begin{equation}
    K(s=5,t=2;d)=\pi_d(\mathcal{C}_1).
\end{equation}
This result is consistent with the following general form:
\begin{equation}
\begin{split}
V_{\rm f}(\boldsymbol{k})&=\frac{\sigma_0\otimes\sigma_0- \sigma_z\otimes\sigma_y}{2}\otimes\tilde{u}(\boldsymbol{k}) \\
&+ \frac{\sigma_0\otimes\sigma_0 + \sigma_z\otimes\sigma_y}{2}\otimes\tilde{u}(-\boldsymbol{k})^* ,
\end{split}
\label{CIV}
\end{equation}
where $\tilde{u}(\boldsymbol{k})$ is an arbitrary $\tilde n \times \tilde n$ unitary.

If we impose the ${\rm SU}(2)$ rotation symmetry on an internal degree of freedom different from the spin, then $V_{\rm f}(\boldsymbol{k})$ will be of the form Eq.~(\ref{CVI}) (where $\sigma_z$ should be understood as $\sigma_0\otimes\sigma_z\otimes\openone_{\tilde n/2}$) with $Y(\boldsymbol{k})$ satisfying the PHS~\eqref{eq:supp-phs-Y} and TRS~\eqref{eq:supp-trs-Y}, where now $i\sigma_y$ acts on this other degree of freedom. The TRS will then act on the spin indices of $Y(\boldsymbol{k})$ as 
\begin{equation}
    (\sigma_0\otimes i\sigma_y \otimes \openone_{\tilde n}) Y(\boldsymbol{k})^*(\sigma_0\otimes i\sigma_y \otimes \openone_{\tilde n})^\dag = Y(-\boldsymbol{k}).
\end{equation}
giving an additional anti-involutory anti-unitary symmetry that commutes with both the PHS and TRS of $Y(\boldsymbol{k})$. This means we are in the case $s=5$, $t=0$ ($A^-_{++}$) of Sec. III C of Ref.~\cite{Shiozaki2014}, where the $K$-group is given by
\begin{equation}
    K(s=5,t=0;d)=\pi_d(\mathcal{R}_5^2). 
\end{equation}
This result is consistent with $V_{\rm f}(\boldsymbol{k})$ having the form of Eq.~\eqref{CVI} with $u(\boldsymbol{k})$ of the general form:
\begin{equation}
    \begin{split}
        u&(\boldsymbol{k})=\\
        &\frac{1}{2}\begin{pmatrix} \tilde u_1(\boldsymbol{k})+\tilde u_2(\boldsymbol{k}) & i[\tilde u_1(\boldsymbol{k})-\tilde u_2(\boldsymbol{k})]\sigma_y \\
        -i\sigma_y [\tilde u_1(\boldsymbol{k})-\tilde u_2(\boldsymbol{k})] & \sigma_y [\tilde u_1(\boldsymbol{k})+\tilde u_2(\boldsymbol{k})] \sigma_y \end{pmatrix},
    \end{split}
\label{CIIV}
\end{equation}
where $\tilde u_1(\boldsymbol{k})$ and $\tilde u_2(\boldsymbol{k})$ are two independent $\tilde n\times\tilde n$ unitaries satisfying $(i\sigma_y\otimes \openone_{\tilde n /2}) \tilde u_{1,2}(\boldsymbol{k})^*(i\sigma_y\otimes \openone_{\tilde n /2})^\dag =\tilde u_{1,2}(-\boldsymbol{k})$.

Finally, if we consider a spin-$1/2$ system with TRS and we impose a further ${\rm U}(1)$ spin-rotation symmetry around the $z$-axis, then one can show that $V_{\rm f}(\boldsymbol{k})$ must have the form
\begin{equation}
\begin{split}
&V_{\rm f}(\boldsymbol{k}) = \\
&\frac{1}{4}\left[ (\sigma_0\otimes\sigma_0 + \sigma_y\otimes\sigma_z + \sigma_z\otimes\sigma_y + \sigma_x\otimes\sigma_x)\otimes \tilde u_1(\boldsymbol{k})\right.\\
&+(\sigma_0\otimes\sigma_0 - \sigma_y\otimes\sigma_z - \sigma_z\otimes\sigma_y + \sigma_x\otimes\sigma_x)\otimes \tilde u_1(-\boldsymbol{k})^* \\
&+(\sigma_0\otimes\sigma_0 - \sigma_y\otimes\sigma_z + \sigma_z\otimes\sigma_y - \sigma_x\otimes\sigma_x)\otimes \tilde u_2(\boldsymbol{k})\\
&\left.+(\sigma_0\otimes\sigma_0 + \sigma_y\otimes\sigma_z - \sigma_z\otimes\sigma_y - \sigma_x\otimes\sigma_x)\otimes \tilde u_2(-\boldsymbol{k})^* \right],
\end{split}
\label{AIIIV}
\end{equation}
where $\tilde u_1(\boldsymbol{k})$ and $\tilde u_2(\boldsymbol{k})$ are two independent arbitrary $\tilde n\times \tilde n$ unitaries. Therefore, the $K$-group is given by $\pi_d(\mathcal{C}_1^2)$. 

\subsubsection{bGOs}
Let us now turn to bosons. As discussed in the main text, the symplectic matrix $V_{\rm b}$ can always be continuously unitarized. This is based on considering the polar decomposition $V_{\rm b}(\boldsymbol{k}) = W_{\rm b}(\boldsymbol{k}) P_{\rm b}(\boldsymbol{k})$, where $W_{\rm b}(\boldsymbol{k})$ is unitary and $P_{\rm b}(\boldsymbol{k})$ is Hermitian and positive definite. Substituting this into Eq.~(\ref{unisym}) in the main text, we obtain
\begin{equation}
[-\sigma W_{\rm b}(\boldsymbol{k})\sigma][- \sigma P_{\rm b}(\boldsymbol{k})\sigma] =W_{\rm b}(\boldsymbol{k}) P_{\rm b}(\boldsymbol{k})^{-1},
\end{equation}
where $\sigma\equiv i\sigma_y\otimes\openone_n$ is the symplectic matrix. Recalling the uniqueness of polar decomposition, this means that $W_{\rm b}(\boldsymbol{k})$ and $P_{\rm b}(\boldsymbol{k})$ are also symplectic. Finally, $P_{\rm b}(\boldsymbol{k})$ can be continuously deformed to the identity according to Eq.~\eqref{exptri}.

We then reduce ourselves to classifying $W_{\rm b}$ which is unitary and fulfills
\begin{align}
    W_{\rm b}(\boldsymbol{k})^*=W_{\rm b}(\boldsymbol{-k}),\\
    W_{\rm b}(\boldsymbol{k})\sigma W_{\rm b}(\boldsymbol{k})^\dag=\sigma, \label{eq:supp-symplectic-condition}.
\end{align}
Considering that $W_{\rm b}(\boldsymbol{k})$ is unitary, the condition~\eqref{eq:supp-symplectic-condition} is equivalent to $[W_{\rm b}(\boldsymbol{k}),\sigma]=0$. We are therefore in a situation completely equivalent to the one of fGOs with a ${\rm U}(1)$ particle-number symmetry. Just like in that case, we have that $W_{\rm b}(\boldsymbol{k})$ must have the form of Eq.~(\ref{eq:supp-V-as-function-of-U}):
\begin{align}
W_{\rm b}(\boldsymbol{k})&=
\frac{\sigma_0-\sigma_y}{2}\otimes u(\boldsymbol{k}) + \frac{\sigma_0+\sigma_y}{2}\otimes u(-\boldsymbol{k})^*\,,
\label{eq:supp-W-as-function-of-U}
\end{align}
where $u(\boldsymbol{k})$ is a unitary matrix. As before, we can see this as imposing a sub-lattice symmetry on the Hamiltonian matrix (\ref{eq:supp-Y}), which leads to class AIII with classifying space $\mathcal{C}_1$.

In case of a spinless TRS alone, we have that this can be seen as imposing involutory PHS and TRS on $Y(\boldsymbol{k})$ (as in the case of fGOs with ${\rm U}(1)$ symmetry and spinless TRS). This leads to class BDI and classifying space $\mathcal{R}_1$.

In case of a spin-$1/2$ TRS alone, we have that this can be seen as imposing anti-involutory PHS and TRS on $Y(\boldsymbol{k})$ (as in the case of fGOs with ${\rm U}(1)$ symmetry and spin-$1/2$ TRS). This leads to class CII and classifying space $\mathcal{R}_5$.

Imposing a ${\rm U}(1)$ particle number symmetry just corresponds to enforcing $[V_{\rm b}(\boldsymbol{k}),\sigma]=0$ directly, without having to first unitarize $V_{\rm b}(\boldsymbol{k})$. Because of~\eqref{eq:supp-symplectic-condition} this means that $V_{\rm b}(\boldsymbol{k})$ has to already be unitary. This constraint therefore does not change the classification of bGOs.

Imposing ${\rm SU}(2)$ rotation symmetry on a matrix $W_{\rm b}(\boldsymbol{k})$ of the form~\eqref{eq:supp-W-as-function-of-U} corresponds to imposing it trivially on $u(\boldsymbol{k})$, i.e., $[u(\boldsymbol{k}),\sigma_x\otimes\openone_{\tilde n}]=[u(\boldsymbol{k}),\sigma_z\otimes\openone_{\tilde n}]=0$. This means $u(\boldsymbol{k})=\sigma_0 \otimes \tilde u(\boldsymbol{k})$ for an arbitrary unitary $\tilde u(\boldsymbol{k})$. We therefore remain in class AIII with classifying space $\mathcal{C}_1$.

Finally if we consider a system with spin-$1/2$ TRS and we impose a spin-rotation symmetry around the $z$-axis only, we have that $W_{\rm b}(\boldsymbol{k})$ must be of the form
\begin{equation}
\begin{split}
W_{\rm b}(\boldsymbol{k}) &= \frac{\sigma_0\otimes\sigma_0 + \sigma_z\otimes\sigma_y}{2}\otimes  \tilde u(\boldsymbol{k}) \\
&+ \frac{\sigma_0\otimes\sigma_0 - \sigma_z\otimes\sigma_y}{2}\otimes \tilde u(-\boldsymbol{k})^*,
\end{split}
\end{equation}
for an arbitrary $\tilde n\times \tilde n$ unitary $\tilde u(\boldsymbol{k})$, which leads again to class AIII with classifying space $\mathcal{C}_1$.

\section{Periodic tables}
After a brief review of the expressions of the topological invariants (cf. Ref.~\cite{Ryu2016}), we first present the full periodic table for fGSs and fGOs as well as the group homomorphism from the latter to the former. Then we refine the periodic tables according to the disentanglability as well as the genuinely dynamical topology.

\subsection{Topological invariants}
There are two different types of $\mathbb{Z}$ (or $2\mathbb{Z}$) topological numbers in the periodic table. The first one is associated to an involutory Hermitian matrix $h(\boldsymbol{k})$ in even dimensions $d=2m$ ($m\in\mathbb{N}$). This is the $m$th Chern number: 
\begin{equation}
{\rm Ch}_m = \frac{1}{m!}\left(\frac{i}{2\pi}\right)^m \int_{T^d} \Tr\mathcal{F}^m,
\label{Ch}
\end{equation}
where $\mathcal{F}$ is the Berry curvature built from the Berry connection via $\mathcal{F} \equiv d\mathcal{A}+ \mathcal{A}\wedge\mathcal{A}$ with
\begin{equation}
(\mathcal{A}(\boldsymbol{k}))_{\alpha\beta}\equiv\boldsymbol{\psi}_\alpha(\boldsymbol{k})^\dag d\boldsymbol{\psi}_\beta(\boldsymbol{k}) =\sum^d_{\mu=1} \boldsymbol{\psi}_\alpha(\boldsymbol{k})^\dag \partial_{k_\mu} \boldsymbol{\psi}_\beta(\boldsymbol{k}) dk_\mu,
\label{FA}
\end{equation}
Here $\boldsymbol{\psi}_\alpha(\boldsymbol{k})$'s are the eigenvectors of $h(\boldsymbol{k})$ with eigenvalue $-1$. Alternatively, we can express the Chern number (\ref{Ch}) directly in terms of $h(\boldsymbol{k})$ as
\begin{equation}
{\rm Ch}_m = -\frac{1}{2^{2m+1}m!}\left(\frac{i}{2\pi}\right)^m \int_{T^d} \Tr[h(dh)^{2m}].
\label{Ch2}
\end{equation}

The second-type integer topological number is associated with a unitary matrix $q(\boldsymbol{k})$ in odd dimensions $d=2m+1$. This is the $m$th winding number: 
\begin{equation}
\omega_m = \frac{(-)^m m!}{(2m+1)!} \left(\frac{i}{2\pi}\right)^{m+1} \int_{T^d} \Tr[(q^\dag dq)^{2m+1}]. 
\label{wd}
\end{equation}
Alternatively, the winding number can be expressed as twice of the Chern-Simons form
\begin{equation}
{\rm CS}_m = \frac{1}{m!} \left(\frac{i}{2\pi}\right)^{m+1} \int_{T^d} \int^1_0 dt \Tr (\mathcal{A}\mathcal{F}^m_t)
\label{CS}
\end{equation}
with $\mathcal{F}_t \equiv t d\mathcal{A} + t^2 \mathcal{A}\wedge\mathcal{A}$ under a special gauge. Here $\mathcal{A}$ is defined in the same form as Eq.~(\ref{FA}) with $\boldsymbol{\psi}_\alpha(\boldsymbol{k})$'s being the eigenstates of $\sigma_+\otimes q(\boldsymbol{k}) + \sigma_-\otimes q(\boldsymbol{k})^\dag$ with eigenvalue $-1$. The special gauge is chosen such that 
\begin{equation}
\boldsymbol{\psi}_\alpha(\boldsymbol{k})= \frac{1}{\sqrt{2}} \begin{pmatrix}\boldsymbol{\phi}_\alpha \\ -q(\boldsymbol{k})^\dag\boldsymbol{\phi}_\alpha\end{pmatrix}, 
\label{spg}
\end{equation}
where $\phi_\alpha$'s form an orthonormal basis. In general, the Chern-Simons form is well-defined up to an integer due to the gauge freedom and is not necessarily quantized.

As for the $\mathbb{Z}_2$ index, there are also two different types of formulas depending on $d$. If $d=2m$ is even, the $\mathbb{Z}_2$ index is given by the Fu-Kane formula: 
\begin{equation}
\begin{split}
{\rm FK}_m &= \frac{1}{m!}\left(\frac{i}{2\pi}\right)^m \int_{\frac{1}{2}T^d} \Tr\mathcal{F}^m  \\
&-  \frac{1}{(m-1)!} \left(\frac{i}{2\pi}\right)^m \int_{\partial (\frac{1}{2}T^d)} \int^1_0 dt \Tr (\mathcal{A}\mathcal{F}^{m-1}_t),
\label{FK}
\end{split}
\end{equation}
where $\frac{1}{2}T^d=\{\boldsymbol{k}: k_1\in[0,\pi], k_\mu\in[-\pi,\pi]\;\forall\mu>1\}$ refers to half of the Brillouin zone and it boundary reads $\partial(\frac{1}{2}T^d)=\{\boldsymbol{k}: k_1=0,\pi; k_\mu\in[-\pi,\pi]\;\forall\mu>1\}$. Recalling that the Chern-Simons form (\ref{CS}) is gauge dependent, we have to set some constraints on the gauge at $\partial(\frac{1}{2}T^d)$ such that ${\rm FK}_m$ (\ref{FK}) is well-defined up to an even integer. For classes AI, AII, DIII and CI, the constraint is given by
\begin{equation}
\boldsymbol{\psi}_\alpha(-\boldsymbol{k})^\dag V_{\rm T}\boldsymbol{\psi}_\beta(\boldsymbol{k})^*|_{\boldsymbol{k}\in \frac{1}{2}T^d} = {\rm const.},
\label{TRSc}
\end{equation}
where $V_{\rm T}$ is the emergent TRS on $h(\boldsymbol{k})$ (cf Eq.~(\ref{TCSL})) and the constant, although being $\boldsymbol{k}$-independent, may still depend on $\alpha,\beta$. For classes D, C, BDI and CII, the constraint is given by
\begin{equation}
\begin{split}
&\int_{\partial(\frac{1}{2}T^d)} \Tr[ ( X^\dag d X)^{2m-1}]=0,\\
&X(\boldsymbol{k})=
\begin{bmatrix}
\uparrow & ... & \uparrow  & \uparrow & ... & \uparrow  \\
\boldsymbol{\psi}_1(\boldsymbol{k}) & ... & \boldsymbol{\psi}_n(\boldsymbol{k}) & V_{\rm C}\boldsymbol{\psi}_1(-\boldsymbol{k})^* & ... &  V_{\rm C}\boldsymbol{\psi}_n(-\boldsymbol{k})^* \\
\downarrow & ... & \downarrow & \downarrow & ... & \downarrow  \\
\end{bmatrix},
\label{PHSc}
\end{split}
\end{equation}
where $V_{\rm C}$ is the emergent PHS on $h(\boldsymbol{k})$ (cf Eq.~(\ref{TCSL})).

In contrast, if $d=2m+1$ is odd, we can compute the Chern-Simons form (\ref{CS}). For classes AI, D, AII and C, the gauge freedom introduces an integer ambiguity and a half-integer implies a nontrivial $\mathbb{Z}_2$ index. For the remaining four classes, however, we have to constrain the gauge degree of freedom by imposing Eq.~(\ref{TRSc}) / Eq.~(\ref{PHSc}) for classes DIII and CI / BDI and CII over the whole Brillouin zone (i.e., replacing $\partial (\frac{1}{2}T^d)$ by $T^d$), such that the Chern-Simons form is well-defined up to an even integer. In this case, an odd integer implies a nontrivial $\mathbb{Z}_2$ index.

\begin{table*}[tbp]
\caption{Topological classifications of fGSs / fGOs in the AZ classes. Just like Fig.~\ref{fig1} in the main text, here non-disentanglable topological fGSs and genuinely dynamical topological fGOs are marked in red and blue, respectively. In particular, light red / blue is used to indicate that only a subgroup is non-disentanglable / genuinely dynamical.} 
\begin{center}
\begin{tabular}{ccccccccc}
\hline\hline
\;\;\;\;AZ\;\;\;\; & \;\;\;\;\;\;$d=0$\;\;\;\;\;\; & \;\;\;\;\;\;\;\;\;\;1\;\;\;\;\;\;\;\;\;\; & \;\;\;\;\;\;\;\;\;\;2\;\;\;\;\;\;\;\;\;\; & \;\;\;\;\;\;\;\;\;\;3\;\;\;\;\;\;\;\;\;\; & \;\;\;\;\;\;\;\;\;\;4\;\;\;\;\;\;\;\;\;\; & \;\;\;\;\;\;\;\;\;\;5\;\;\;\;\;\;\;\;\;\; & \;\;\;\;\;\;\;\;\;\;6\;\;\;\;\;\;\;\;\;\; & \;\;\;\;\;\;\;\;\;\;7\;\;\;\;\;\;\;\;\;\; \\
\hline
A & \colorbox{red!25!white}{$\mathbb{Z}$} / 0 & 0 / \colorbox{blue!25!white}{$\mathbb{Z}$} & \colorbox{red!25!white}{$\mathbb{Z}$} / 0 & 0 / \colorbox{blue!25!white}{$\mathbb{Z}$} & \colorbox{red!25!white}{$\mathbb{Z}$} / 0 & 0 / \colorbox{blue!25!white}{$\mathbb{Z}$} & \colorbox{red!25!white}{$\mathbb{Z}$} / 0 & 0 / \colorbox{blue!25!white}{$\mathbb{Z}$} \\
AIII & 0 / 0 & $\mathbb{Z}$ / \colorbox{blue!10!white}{$\mathbb{Z}^2$} & 0 / 0 & $\mathbb{Z}$ / \colorbox{blue!10!white}{$\mathbb{Z}^2$} & 0 / 0 & $\mathbb{Z}$ / \colorbox{blue!10!white}{$\mathbb{Z}^2$} & 0 / 0 & $\mathbb{Z}$ / \colorbox{blue!10!white}{$\mathbb{Z}^2$} \\
\hline
AI &  \colorbox{red!25!white}{$\mathbb{Z}$} / \colorbox{blue!25!white}{$\mathbb{Z}_2$}  & 0 / \colorbox{blue!25!white}{$\mathbb{Z}$} & 0 / 0 & 0 / 0 & \colorbox{red!25!white}{$2\mathbb{Z}$} / 0 & 0 / \colorbox{blue!25!white}{$2\mathbb{Z}$} & \colorbox{red!25!white}{$\mathbb{Z}_2$} / 0 & \colorbox{red!25!white}{$\mathbb{Z}_2$} / \colorbox{blue!25!white}{$\mathbb{Z}_2$} \\
BDI & $\mathbb{Z}_2$ / \colorbox{blue!10!white}{$\mathbb{Z}_2^2$} & $\mathbb{Z}$ / \colorbox{blue!10!white}{$\mathbb{Z}^2$} & 0 / 0 & 0 / 0 & 0 / 0 & $2\mathbb{Z}$ /  \colorbox{blue!10!white}{$2\mathbb{Z}^2$} & 0 / 0 & $\mathbb{Z}_2$ /  \colorbox{blue!10!white}{$\mathbb{Z}_2^2$} \\
D & $\mathbb{Z}_2$ / $\mathbb{Z}_2$  & $\mathbb{Z}_2$ / \colorbox{blue!10!white}{$\mathbb{Z}$}  & \colorbox{red!25!white}{$\mathbb{Z}$} / 0 & 0 / 0 & 0 / 0 & 0 / \colorbox{blue!25!white}{$2\mathbb{Z}$} & \colorbox{red!25!white}{$2\mathbb{Z}$} / 0 & 0 / \colorbox{blue!25!white}{$\mathbb{Z}_2$} \\
DIII & 0 / 0 & $\mathbb{Z}_2$ / \colorbox{blue!10!white}{$\mathbb{Z}$} & \colorbox{red!25!white}{$\mathbb{Z}_2$} / 0 & \colorbox{red!10!white}{$\mathbb{Z}$} / $\mathbb{Z}$ & 0 / 0 & 0 / \colorbox{blue!25!white}{$\mathbb{Z}$} & 0 / 0 & $2\mathbb{Z}$ / $\mathbb{Z}$ \\
AII & \colorbox{red!25!white}{$2\mathbb{Z}$} / 0 & 0 / \colorbox{blue!25!white}{$2\mathbb{Z}$} & \colorbox{red!25!white}{$\mathbb{Z}_2$} / 0 & \colorbox{red!25!white}{$\mathbb{Z}_2$} / \colorbox{blue!25!white}{$\mathbb{Z}_2$} & \colorbox{red!25!white}{$\mathbb{Z}$} / \colorbox{blue!25!white}{$\mathbb{Z}_2$} & 0 / \colorbox{blue!25!white}{$\mathbb{Z}$} & 0 / 0 & 0 / 0 \\
CII & 0 / 0 & $2\mathbb{Z}$ / \colorbox{blue!10!white}{$2\mathbb{Z}^2$} & 0 / 0 & $\mathbb{Z}_2$ / \colorbox{blue!10!white}{$\mathbb{Z}_2^2$} & $\mathbb{Z}_2$ / \colorbox{blue!10!white}{$\mathbb{Z}_2^2$} & $\mathbb{Z}$ / \colorbox{blue!10!white}{$\mathbb{Z}^2$} & 0 / 0 & 0 / 0 \\
C & 0 / 0 & 0 / \colorbox{blue!25!white}{$2\mathbb{Z}$} & \colorbox{red!25!white}{$2\mathbb{Z}$} / 0 & $0$ / \colorbox{blue!25!white}{$\mathbb{Z}_2$} & $\mathbb{Z}_2$ / $\mathbb{Z}_2$ & $\mathbb{Z}_2$ / \colorbox{blue!10!white}{$\mathbb{Z}$} &  \colorbox{red!25!white}{$\mathbb{Z}$} / 0 & 0 / 0 \\
CI & 0 / 0 & 0 / \colorbox{blue!25!white}{$\mathbb{Z}$} & 0 / 0 & $2\mathbb{Z}$ / $\mathbb{Z}$ & 0 / 0 & $\mathbb{Z}_2$ / \colorbox{blue!10!white}{$\mathbb{Z}$}  & \colorbox{red!25!white}{$\mathbb{Z}_2$} / 0 & \colorbox{red!10!white}{$\mathbb{Z}$} / $\mathbb{Z}$ \\
\hline\hline
\end{tabular}
\end{center}
\label{tableS2}
\end{table*}

\subsection{Full periodic table}
Having in mind the general forms of the covariance and representation matrices, we can readily identify the corresponding classifying spaces and $K$-groups. The image of the group homomorphism can then be determined by the subgroup that consists of all the covariance matrices in the following form:
\begin{equation}
\Gamma_{\rm f}(\boldsymbol{k}) = V_{\rm f}(\boldsymbol{k}) \Gamma_0 V_{\rm f}(\boldsymbol{k})^\dag,
\label{GVG0}
\end{equation}
where $\Gamma_0$ corresponds to a fixed trivial state. The full results are presented in Table~\ref{tableS2}. As mentioned in the main text, the homomorphism is always surjective (trivial) for the chiral symmetry (Wigner-Dyson) classes, but becomes rather complicated for the BdG classes. Remarkably, it turns out that there are four nontrivial bijective homomorphisms in the BdG classes: class D in 0D, class C in 4D, class CI in 3D and class DIII in 7D.

\subsubsection{Chiral symmetry classes}
Let us first consider the chiral symmetry classes, which include AIII, BDI and CII. As shown previously, the covariance matrix of a fGS is uniquely determined by a unitary $q(\boldsymbol{k})$, which is arbitrary for class AIII (cf. Eq.~(\ref{AIIIH})) and satisfies $q(\boldsymbol{k})^* = q(-\boldsymbol{k})$ / $(\sigma_y\otimes\tilde\openone) q(\boldsymbol{k})^*(\sigma_y\otimes\tilde\openone) = q(-\boldsymbol{k})$ for class BDI / CII (cf. Eq.~(\ref{BDIH}) / Eq.~(\ref{CIIH})). Here (and after) for simplicity, we do not distinguish $q(\boldsymbol{k})$ and $\tilde q(\boldsymbol{k})$ as well as $\openone_n$ with different $n$. On the other hand, the representation matrix is uniquely determined by two independent unitaries $u_{1,2}(\boldsymbol{k})$ satisfying the same symmetry constraints (cf. Eqs.~(\ref{AIIIV}), (\ref{BDIV}) and (\ref{CIIV})). This is why the classification of fGOs is simply a double of that of fGSs. 

By properly choosing $\Gamma_0$ in Eq.~(\ref{GVG0}), we can relate $q(\boldsymbol{k})$ to $u_{1,2}(\boldsymbol{k})$ via 
\begin{equation}
q(\boldsymbol{k}) = u_1(\boldsymbol{k})u_2(\boldsymbol{k})^\dag.
\end{equation}
This implies all the topological fGSs in the chiral symmetry classes are disentanglable, and all those topological fGOs with $u_1(\boldsymbol{k})$ and $u_2(\boldsymbol{k})$ deformable (under the symmetry constraint) into each other are genuinely dynamical.

\subsubsection{Wigner-Dyson classes}
We turn to consider the Wigner-Dyson classes, which include A, AI and AII. As shown previously, the covariance matrix of a fGS is uniquely determined by a Hermitian matrix $h(\boldsymbol{k})$ (cf. Eq.~(\ref{eq:supp-iGammma-as-function-of-H})), which is involutory for class A and further satisfies $h(\boldsymbol{k})^* = h(-\boldsymbol{k})$ / $(\sigma_y\otimes\tilde\openone) h(\boldsymbol{k})^*(\sigma_y\otimes\tilde\openone) = h(-\boldsymbol{k})$ for class AI / AII (cf. Eq.~(\ref{eq:supp-trs}) / Eq.~(\ref{AIIH})). In contrast, the representation matrix is uniquely determined by a unitary $u(\boldsymbol{k})$, which is arbitrary for class A (cf. Eq.~(\ref{eq:supp-V-as-function-of-U})) and $u(\boldsymbol{k})^* = u(-\boldsymbol{k})$ / $(\sigma_y\otimes\tilde\openone) u(\boldsymbol{k})^*(\sigma_y\otimes\tilde\openone) = u(-\boldsymbol{k})$ for class AI / AII (cf. Eq.~(\ref{AIV}) / Eq.~(\ref{AIIV})). This is why the classification of fGOs coincides with the state classification of AIII, BDI and CII.

It is already clear from the classifications that the only possibilities of topological fGSs in the Wigner-Dyson classes being disentanglable are class AI in 7D and class AII in 3D, where both fGSs and fGOs are characterized by $\mathbb{Z}_2$. Note that any group homomorphism from a torsion group, such as $\mathbb{Z}_2$, to $\mathbb{Z}$ is trivial. This observation rules out the possibility of disentanglable topological fGSs in class AI in 0D and class AII in 4D. The $\mathbb{Z}_2$ index for TRS fGSs in odd dimensions are 
given by the Chern-Simons form (\ref{CS}), which is equal to half of the winding number (\ref{wd}) of the sewing matrix
\begin{equation}
(w(\boldsymbol{k}))_{\alpha\beta}\equiv \boldsymbol{\psi}_\alpha(-\boldsymbol{k})^\dag V_{\rm T}\boldsymbol{\psi}_\beta(\boldsymbol{k})^*,
\label{sewmtrx}
\end{equation}
where $\boldsymbol{\psi}_\alpha(\boldsymbol{k})$'s are the eigenvectors of $h(\boldsymbol{k})$ with eigenvalue $-1$ and $V_{\rm T}=\openone$ / $i\sigma_y\otimes\tilde\openone$ for class AI / AII. Now consider a TRS fGS which is disentangled by $\hat U_{\rm f}$ represented by $V_{\rm f}(\boldsymbol{k})$. This implies $\boldsymbol{\psi}_\alpha(\boldsymbol{k})$'s can be related to some $\boldsymbol{k}$-independent TRS basis $\boldsymbol{\phi}_\alpha$'s via
\begin{equation}
\boldsymbol{\psi}_\alpha(\boldsymbol{k})= u(\boldsymbol{k})\boldsymbol{\phi}_\alpha,
\end{equation}
where $u(\boldsymbol{k})$ also satisfies the TRS. No matter whether $u(\boldsymbol{k})$ is topological or not, we can check that the corresponding sewing matrix (\ref{sewmtrx}) is $\boldsymbol{k}$-independent and thus its winding number vanishes, implying a trivial $\mathbb{Z}_2$ index. Therefore, all the topological fGSs in the Wigner-Dyson classes are non-disentanglable.

\subsubsection{BdG classes without TRS}
We move on to the most complicated BdG classes, which include D, DIII, C and CI. Let us first consider classes D and C without TRS. As shown in Eq.~(\ref{eq:supp-phs}) / Eq.~(\ref{eq:supp-anti-inv-phs}), the covariance matrix is uniquely determined by a Hermitian involutory matrix $h(\boldsymbol{k})$, which further satisfies $h(\boldsymbol{k})^*= -h(-\boldsymbol{k})$ / $(\sigma_y\otimes\tilde\openone)h(\boldsymbol{k})^*(\sigma_y\otimes\tilde\openone)= -h(-\boldsymbol{k})$ for class D / C. In contrast, the representation matrix of a fGO is uniquely determined by a unitary that satisfies exactly the same symmetry constraint as class AI / AII (cf. the fundamental constraint $V_{\rm f}(\boldsymbol{k})^*=V_{\rm f}(-\boldsymbol{k})$ / Eq.~(\ref{CV})). This is why the classification of fGOs in class D / C is the same as that in class AI / AII, which coincides with the state classification for class BDI / CII.

According to the classification, we know that the topological fGSs in class D / C are possibly disentanglable only in the dimensions where class BDI / CII also have nontrivial fGSs (and fGOs). If the dimension is even ($d=0$ / $4$), the $\mathbb{Z}_2$ index of classes D and BDI / C and CII are computed by the same Fu-Kane formula (\ref{FK}) under the same gauge constraint (\ref{PHSc}). If the dimension is odd ($d=1$ / $5$), the PHS-protected $\mathbb{Z}_2$ index of the fGSs in class BDI / CII is nontrivial if and only if the winding number is odd, as can be understood from the fact that the winding number is twice of the Chern-Simons form for a specific gauge (\ref{spg}). In short, we have a surjective inclusion (by forgetting the TRS) of the topological classes of BDI in D and those of CII in C in these dimensions. 
As demonstrated previously, the former is always disentanglable, so is the latter.

\subsubsection{BdG classes with TRS}
The remaining two classes are DIII and CI. As shown in Eq.~(\ref{DIIIH}) / Eq.~(\ref{CIH}), the covariance matrix is uniquely determined by a unitary $q(\boldsymbol{k})$, which satisfies $q(\boldsymbol{k})^{\rm T}=-q(-\boldsymbol{k})$ / $q(\boldsymbol{k})^{\rm T}=q(-\boldsymbol{k})$ for class DIII / CI. In contrast, the representation matrix of a fGO is uniquely determined by a unitary $u(\boldsymbol{k})$ without any constraint (cf. Eqs.~(\ref{eq:supp-V-as-function-of-Utilde}) and (\ref{CIV})). This is why the classification of fGOs in class DIII / CI is the same as that in class A, which coincides with the state classification for class AIII.

The classification suggests that the fGSs in class DIII / CI could be generally disentanglable except for $d=2$ / $6$. To see whether this is the case, we note that all the disentanglable fGSs are associated with $q(\boldsymbol{k})$ that takes the following form: 
\begin{equation}
q(\boldsymbol{k}) =\left\{\begin{array}{ll} u(\boldsymbol{k})(\sigma_y\otimes \tilde\openone)u(-\boldsymbol{k})^{\rm T}, & {\rm DIII};  \\ u(\boldsymbol{k})u(-\boldsymbol{k})^{\rm T}, & {\rm CI}. \end{array} \right.
\label{BdGqu}
\end{equation}
In 1D / 5D, the $\mathbb{Z}_2$ index for class DIII / CI is determined by the Chern-Simons form under the gauge constraint that the sewing matrix is $\boldsymbol{k}$-independent (cf. Eq.~(\ref{TRSc})). One can readily write down a valid solution as
\begin{equation}
\boldsymbol{\psi}_\alpha(\boldsymbol{k})=\frac{1}{\sqrt{2}}
\begin{pmatrix} v(\boldsymbol{k})\boldsymbol{\phi}_\alpha \\ -u(-\boldsymbol{k})^*\boldsymbol{\phi}_\alpha \end{pmatrix},
\end{equation}
where $v(\boldsymbol{k})= u(\boldsymbol{k})(\sigma_y\otimes\tilde\openone)$ for class DIII / $v(\boldsymbol{k})= u(\boldsymbol{k})$ for class CI. The $\mathbb{Z}_2$ index thus turns out to be the parity (even or odd) of the winding number of $u(\boldsymbol{k})$. Since the $\mathbb{Z}_2$ index can be nontrivial for odd winding numbers, we know that the topological fGSs in class DIII in 1D / CI in 5D are disentanglable. In other nontrivial dimensions, where the state topological invariants are simply the winding number of $q(\boldsymbol{k})$ ($d=3,7$), we find that while the topological fGSs in class DII in 7D / CI in 3D are all disentanglable, only a subgroup $2\mathbb{Z}$ characterized by even winding numbers is disentanglable in 3D / 7D. This is because the winding number of $u(\boldsymbol{k})$ coincides with that of $u(-\boldsymbol{k})^{\rm T}$ for $d\equiv 3\mod 4$ (cf. Eq.~(\ref{wd})), so the winding number of $q(\boldsymbol{k})$ in Eq.~(\ref{BdGqu}) is always even.

\subsection{Refined periodic tables}
We can separate Table~\ref{tableS2} into three refined periodic tables according to the disentanglability. In Table~\ref{tableS3} we present the periodic table for disentanglable topological fGSs, which is nothing but the non-shaded part in Table~\ref{tableS2} and, by definition, coincides with that for state-like topological fGOs \footnote{A subtle point here is that $2\mathbb{Z}$ number for topological fGSs might correspond to $\mathbb{Z}$ number for topological fGOs, as is the case for class DIII in 3D and class CI in 7D. Nevertheless, the group structure is always isomorphic to $\mathbb{Z}$.}. Here we consistently use the term ``disentanglable" in 0D to refer to those topological fGSs that cannot be generated by any topological fGOs from some trivial reference states.

\begin{table}[tbp]
\caption{Disentanglable topological fGSs / state-like topological fGOs. Intrinsic and symmetry-enriched Gaussian topological orders are marked in purple and orange, respectively. In particular, class BDI in 1D marked by light orange only has a $\mathbb{Z}_2$ subgroup that remains topological in the absence of symmetries.} 
\begin{center}
\begin{tabular}{ccccccccc}
\hline\hline
AZ & $d=0$ & \;\;\;\;1\;\;\;\; & \;\;\;\;2\;\;\;\; & \;\;\;\;3\;\;\;\; & \;\;\;\;4\;\;\;\; & \;\;\;\;5\;\;\;\; & \;\;\;\;6\;\;\;\; & \;\;\;\;7\;\;\;\; \\
\hline
A & 0 & 0 & 0 & 0 & 0 & 0 & 0 & 0 \\
AIII & 0 & $\mathbb{Z}$ & 0 & $\mathbb{Z}$ & 0 & $\mathbb{Z}$ & 0 & $\mathbb{Z}$ \\
\hline
AI & 0  & 0 & 0 & 0 & 0 & 0 & 0 & 0 \\
BDI & \colorbox{orange!30!white}{$\mathbb{Z}_2$} & \colorbox{orange!15!white}{$\mathbb{Z}$} & 0 & 0 & 0 & $2\mathbb{Z}$ & 0 & $\mathbb{Z}_2$ \\
D & \colorbox{purple!30!white}{$\mathbb{Z}_2$}  & \colorbox{purple!30!white}{$\mathbb{Z}_2$}  & 0 & 0 & 0 & 0 & 0 & 0 \\
DIII & 0 & $\mathbb{Z}_2$ & 0 & $2\mathbb{Z}$ & 0 & 0 & 0 & $2\mathbb{Z}$ \\
AII & 0 & 0 & 0 & 0 & 0 & 0 & 0 & 0 \\
CII & 0 & $2\mathbb{Z}$ & 0 & $\mathbb{Z}_2$ & $\mathbb{Z}_2$ & $\mathbb{Z}$ & 0 & 0 \\
C & 0 & 0 & 0 & 0 & $\mathbb{Z}_2$ & $\mathbb{Z}_2$ & 0 & 0 \\
CI & 0 & 0 & 0 & $2\mathbb{Z}$ & 0 & $\mathbb{Z}_2$  & 0 & $2\mathbb{Z}$ \\
\hline\hline
\end{tabular}
\end{center}
\label{tableS3}
\end{table}

There remains a problem of distinguishing symmetry-protected and -enriched orders for the nontrivial AZ classes other than D. This can be done by studying the $K$-group homomorphism to class D by forgetting the symmetry. This homomorphism is obviously trivial in $d\ge2$D so it suffices to consider $d=0,1$. In $0$D, the only nontrivial class is BDI, which can certainly be intrinsically nontrivial (i.e., symmetry-enriched) since the $\mathbb{Z}_2$ number is also characterized by the Pfaffian of the covariance matrix, just like class D. In $1$D, it turns out that only those topological fGSs in class BDI with odd winding numbers exhibit symmetry-enriched orders. Otherwise, we can always deform the fGS into a product state of two copies with spin up and down, which exhibits a trivial $\mathbb{Z}_2$ number by construction.

Let us move on to non-disentanglable topological fGSs. This can be simply obtained by dividing the disentanglable topological classes from the full periodic table for fGSs. Mathematically speaking, this is the quotient group of the full $K$-group for fGSs with respect to the subgroup consisting of disentanglable topological classes.

\begin{table}[tbp]
\caption{Non-disentanglable topological fGSs. Colors share the same meanings as those in Table~\ref{tableS3}.} 
\begin{center}
\begin{tabular}{ccccccccc}
\hline\hline
AZ & $d=0$ & \;\;\;\;1\;\;\;\; & \;\;\;\;2\;\;\;\; & \;\;\;\;3\;\;\;\; & \;\;\;\;4\;\;\;\; & \;\;\;\;5\;\;\;\; & \;\;\;\;6\;\;\;\; & \;\;\;\;7\;\;\;\; \\
\hline
A & $\mathbb{Z}$ & 0 & \colorbox{orange!30!white}{$\mathbb{Z}$} & 0 & $\mathbb{Z}$ & 0 & \colorbox{orange!30!white}{$\mathbb{Z}$} & 0 \\
AIII & 0 & 0 & 0 & 0 & 0 & 0 & 0 & 0 \\
\hline
AI & $\mathbb{Z}$  & 0 & 0 & 0 & $2\mathbb{Z}$ & 0 & $\mathbb{Z}_2$ & $\mathbb{Z}_2$ \\
BDI & 0  & 0 & 0 & 0 & 0 & 0 & 0 & 0 \\
D & 0  & 0  & \colorbox{purple!30!white}{$\mathbb{Z}$} & 0 & 0 & 0 & \colorbox{purple!30!white}{$2\mathbb{Z}$} & 0 \\
DIII & 0 & 0 & $\mathbb{Z}_2$ & $\mathbb{Z}_2$ & 0 & 0 & 0 & 0 \\
AII & $2\mathbb{Z}$ & 0 & $\mathbb{Z}_2$ & $\mathbb{Z}_2$ & $\mathbb{Z}$ & 0 & 0 & 0 \\
CII & 0  & 0 & 0 & 0 & 0 & 0 & 0 & 0 \\
C & 0 & 0 & \colorbox{orange!30!white}{$2\mathbb{Z}$} & 0 & 0 & 0 & \colorbox{orange!30!white}{$\mathbb{Z}$} & 0 \\
CI & 0 & 0 & 0 & 0 & 0 & 0  & $\mathbb{Z}_2$ & $\mathbb{Z}_2$ \\
\hline\hline
\end{tabular}
\end{center}
\label{tableS4}
\end{table}

To distinguish symmetry-protected and -enriched orders, we first note that, in the absence of symmetries (class D), topological fGSs appear only in 2D and 6D, and the corresponding topological (Chern) numbers are integers. This observation already allows us to conclude that all the topological fGSs characterized by $\mathbb{Z}_2$ are symmetry-protected. In contrast, those topological fGSs characterized by integers are symmetry-enriched. Moreover, the group homomorphism to class D is a multiplication by $2$, as can be understood from Eq.~(\ref{eq:supp-iGammma-as-function-of-H}) and the fact that $h(\boldsymbol{k})$ and $-h(-\boldsymbol{k})^*$ share the same Chern number in 2D and 6D (space inversion $\boldsymbol{k}\to -\boldsymbol{k}$ does not change the Chern number, while the complex conjugate inverses the Chern number in $d\equiv 2\mod 4$D; cf. Eq.~(\ref{Ch2})).

We finally turn to genuinely dynamical topological fGOs. Similar to non-disentanglable fGSs, the topological classes can be simply obtained by dividing the state-like ones from the full periodic table for fGOs. This is the quotient group of the full $K$-group for fGOs with respect to the subgroup consisting of state-like topological classes.

\begin{table}[tbp]
\caption{Genuinely dynamical topological fGOs. Colors share the same meanings as those in Table~\ref{tableS3}.} 
\begin{center}
\begin{tabular}{ccccccccc}
\hline\hline
AZ & $d=0$ & \;\;\;\;1\;\;\;\; & \;\;\;\;2\;\;\;\; & \;\;\;\;3\;\;\;\; & \;\;\;\;4\;\;\;\; & \;\;\;\;5\;\;\;\; & \;\;\;\;6\;\;\;\; & \;\;\;\;7\;\;\;\; \\
\hline
A & 0 & \colorbox{orange!30!white}{$\mathbb{Z}$} & 0 & $\mathbb{Z}$ & 0 &  \colorbox{orange!30!white}{$\mathbb{Z}$} & 0 & \colorbox{orange!15!white}{$\mathbb{Z}$} \\
AIII & 0 &   \colorbox{orange!30!white}{$\mathbb{Z}$} & 0 & $\mathbb{Z}$ & 0 &  \colorbox{orange!30!white}{$\mathbb{Z}$} & 0 & $\mathbb{Z}$ \\
\hline
AI & $\mathbb{Z}_2$  &  \colorbox{orange!30!white}{$\mathbb{Z}$}  & 0 & 0 & 0 &  \colorbox{orange!30!white}{$2\mathbb{Z}$} & 0 & $\mathbb{Z}_2$ \\
BDI & $\mathbb{Z}_2$ &  \colorbox{orange!30!white}{$\mathbb{Z}$} & 0 & 0 & 0 &  \colorbox{orange!30!white}{$2\mathbb{Z}$} & 0 & $\mathbb{Z}_2$ \\
D & 0  & \colorbox{purple!30!white}{$2\mathbb{Z}$}  & 0 & 0 & 0 & \colorbox{purple!30!white}{$2\mathbb{Z}$} & 0 & \colorbox{purple!30!white}{$\mathbb{Z}_2$} \\
DIII & 0 &  \colorbox{orange!30!white}{$2\mathbb{Z}$} & 0 & 0 & 0 &  \colorbox{orange!30!white}{$\mathbb{Z}$} & 0 & 0 \\
AII & 0 & \colorbox{orange!30!white}{$2\mathbb{Z}$} & 0 & $\mathbb{Z}_2$ & $\mathbb{Z}_2$ &  \colorbox{orange!30!white}{$\mathbb{Z}$} & 0 & 0 \\
CII & 0 &  \colorbox{orange!30!white}{$2\mathbb{Z}$} & 0 & $\mathbb{Z}_2$ & $\mathbb{Z}_2$ &  \colorbox{orange!30!white}{$\mathbb{Z}$} & 0 & 0 \\
C & 0 &  \colorbox{orange!30!white}{$2\mathbb{Z}$} & 0 & $\mathbb{Z}_2$ & 0 &  \colorbox{orange!30!white}{$2\mathbb{Z}$} & 0 & 0 \\
CI & 0 &  \colorbox{orange!30!white}{$\mathbb{Z}$} & 0 & 0 & 0 &  \colorbox{orange!30!white}{$2\mathbb{Z}$}  & 0 & 0 \\
\hline\hline
\end{tabular}
\end{center}
\label{tableS5}
\end{table}

It turns out that, in the absence of symmetries, genuinely dynamical topological fGOs appear in $d=1,5$ and $7$D. In particular, in 1D and 5D, all the topological invariants are integer winding numbers. It follows that all the genuinely dynamical topological fGOs in these two dimensions are symmetry-enriched (more precisely, symmetry constrained). The homomorphism to class D is a multiplication by $2$ for classes A, AI, BDI, DIII, AII and C, and a multiplication by $4$ for classes AIII, CII and CI, respectively. This can be seen by counting the number of topological equivalent blocks in the general forms of representation matrices.

Things become much trickier in 7D, where general fGOs are classified by $\mathbb{Z}_2$. One can show that a representative genuinely dynamical fGO in class AI or BDI takes the form $\sigma_0\otimes v(\boldsymbol{k})$, where $v(\boldsymbol{k})=v(-\boldsymbol{k})^*$ exhibits a nontrivial $\mathbb{Z}_2$ index. Therefore,  $\sigma_0\otimes v(\boldsymbol{k})= v(\boldsymbol{k})\oplus v(\boldsymbol{k})$ is trivial by construction, implying that the $\mathbb{Z}_2$ indices for class AI and BDI are both symmetry-protected. What remains unclear is whether the homomorphism from class A or AIII to D could be nontrivial. To simplify the problem, we make use of the Bott periodicity and consider the effective dimension $d=-1$, where, instead of the wave vector, we have a periodic space-like parameter $x$ \cite{Teo2010}. Now the fundamental symmetry constraint on the representation matrix is $V(x)=V(x+2\pi)=V(x)^*$ and the $\mathbb{Z}_2$ index can be understood from the fact $\pi_1({\rm O}(N))=\mathbb{Z}_2$ for $N\ge3$. Noting that the embedding of $\pi_1({\rm O}(2))=\mathbb{Z}$ into $\pi_1({\rm O}(N))=\mathbb{Z}_2$ is surjective and a generator of the former can be realized in class A:
\begin{equation}
\begin{pmatrix} \cos x & \sin x \\ -\sin x & \cos x \end{pmatrix} = \frac{\sigma_0-\sigma_y}{2}e^{-ix}
+ \frac{\sigma_0+\sigma_y}{2}e^{ix},
\end{equation}
we know that fGOs in class A with odd winding numbers are symmetry-enriched, while those with even winding numbers are symmetry-protected. On the other hand, genuinely dynamical fGOs in class AIII always have even winding numbers upon being embedded in class A (if we choose the ${\rm U}(1)$ symmetry to be the spin-rotation symmetry in $z$ direction), and are thus symmetry-protected.

\end{document}